\documentclass[aps,reprint,twocolumn,superscriptaddress,floats,10pt,floatfix]{revtex4-2}
\usepackage[dvipdfmx]{graphicx}
\usepackage{siunitx}
\usepackage{amsfonts}
\usepackage{amsmath}
\usepackage{amssymb}
\usepackage{bm}
\usepackage{graphicx}
\usepackage{dcolumn}
\usepackage{mciteplus}
\usepackage{euscript}
\usepackage{multirow}
\usepackage{color}
\usepackage{textcomp}
\usepackage{gensymb}
\usepackage{float}
\usepackage{enumitem}
\usepackage{xcolor}
\usepackage{sepfootnotes}
\usepackage{mhchem}
\usepackage{outlines}
\usepackage[hidelinks]{hyperref}

\graphicspath{{./}{../figs/}}

\DeclareSIUnit{\atomicunit}{a.u.}

\newcommand{\para}[1]{\paragraph*{#1}}

\newcommand{\req}[1]{Eq.~\eqref{#1}}

\newcommand{\rfig}[1]{Fig.~\ref{#1}}
\newcommand{\rFig}[1]{Figure~\ref{#1}}
\newcommand{\rtbl}[1]{Table~\ref{#1}}

\newcommand{\ef}{E_{\rm F}}

\newcommand\abinitio{\emph{ab initio}}
\newcommand\etal{\textit{et al.}}
\newcommand{\mub}{\mu_\mathrm{B}}

\newcommand\tc{T_\mathrm{C}}

\newcommand{\st}[1]{\lvert{#1}\rangle}

\graphicspath{{./}{../figs/}{../psd/}}

\newcommand{\rmnsn}{R\text{Mn}_6\text{Sn}_6}

\newcommand{\yymnsn}{\text{YMn}_6\text{Sn}_6}

\newcommand{\gdmnsn}{\text{GdMn}_6\text{Sn}_6}
\newcommand{\tbmnsn}{\text{TbMn}_6\text{Sn}_6}
\newcommand{\dymnsn}{\text{DyMn}_6\text{Sn}_6}
\newcommand{\homnsn}{\text{HoMn}_6\text{Sn}_6}
\newcommand{\ermnsn}{\text{ErMn}_6\text{Sn}_6}
\newcommand{\tsr}{T_\text{SR}}
\newcommand{\jrm}{J_{R\text{M}}}
\newcommand{\avg}[1]{\langle{#1}\rangle}
\newcommand{\Oe}{\mathcal{O}}

\newcommand{\rY}{\mathcal{Y}}

\mathchardef\mhyphen="2D

\definecolor{forestgreen}{rgb}{0.13, 0.55, 0.13}

\begin{document}

\title{Interplay between magnetism and band topology in Kagome magnets $R$Mn$_6$Sn$_6$}

\author{Y. Lee}
\affiliation{Ames Laboratory, U.S.~Department of Energy, Ames, Iowa 50011, USA}

\author{R. Skomski}
\thanks{Deceased}
\affiliation{Department of Physics and Astronomy, Nebraska Center for Materials and Nanoscience, University of Nebraska, Lincoln, Nebraska 68588, USA}

\author{X. Wang}
\affiliation{Sophyics Technology, LLC, McLean, Virginia 22102, USA}

\author{P. P. Orth}
\affiliation{Ames Laboratory, U.S.~Department of Energy, Ames, Iowa 50011, USA}
\affiliation{Department of Physics and Astronomy, Iowa State University, Ames, Iowa 50011, USA}

\author{Y. Ren}
\affiliation{Department of Materials Science and Engineering, University of Washington, Seattle, Washington 98195, USA}

\author{Byungkyun Kang}
\affiliation{College of Arts and Sciences, University of Delaware, Newark, Delaware 19716, USA}

\author{A. K. Pathak}
\affiliation{Department of Physics, SUNY Buffalo State, Buffalo, New York 14222, USA}

\author{A.~Kutepov}
\affiliation{5 Carriage Lane, Roxbury, Connecticut 06783, USA}

\author{B.~N.~Harmon}
\affiliation{Ames Laboratory, U.S.~Department of Energy, Ames, Iowa 50011, USA}
\affiliation{Department of Physics and Astronomy, Iowa State University, Ames, Iowa 50011, USA}

\author{R.~J.~McQueeney}
\affiliation{Ames Laboratory, U.S.~Department of Energy, Ames, Iowa 50011, USA}
\affiliation{Department of Physics and Astronomy, Iowa State University, Ames, Iowa 50011, USA}

\author{I.~I.~Mazin}
\affiliation{Department of Physics and Astronomy, George Mason University, Fairfax, Virginia 22030, USA}
\affiliation{Quantum Science and Engineering Center, George Mason University, Fairfax, Virginia 22030, USA}

\author{Liqin Ke}
\email{liqinke@ameslab.gov}
\affiliation{Ames Laboratory, U.S.~Department of Energy, Ames, Iowa 50011, USA}

\date{\today}

\begin{abstract}
Kagome-lattice magnets $R$Mn$_6$Sn$_6$ recently emerged as a new platform to exploit the interplay between magnetism and topological electronic states.
Some of the most exciting features of this family are the dramatic dependence of the easy magnetization direction on the rare-earth specie, despite other magnetic and electronic properties being essentially unchanged, and the kagome geometry of the Mn planes that in principle can generate flat bands and Dirac points;
gapping of the Dirac points by spin-orbit coupling has been suggested recently to be  responsible for the observed anomalous Hall response in the member TbMn$_6$Sn$_6$.
In this paper, we address both issues with \textit{ab initio} calculations.
We have discovered the significant role played by higher-order crystal-field parameters and rare-earth magnetic anisotropy constants in these systems.
We demonstrate that the microscopic origin of rare-earth magnetic anisotropy can also be quantified and understood at various levels: \textit{ab initio}, phenomenological, and analytical.
In particular, using a simple and physically transparent analytical model based on perturbation theory, we are able to explain, with full quantitative agreement, the evolution of rare-earth magnetic anisotropy across the series.
We analyze in detail the topological properties of Mn-dominated bands and demonstrate how they emerge from the multiorbital planar kagome model.
We further show that, despite this fact, most of the topological features at the Brillouin zone corner K are strongly 3D, and therefore cannot explain the observed quasi-2D anomalous Hall effect, while the most pronounced quasi-2D dispersion are too far removed from the Fermi level.
By employing self-consistent calculations with \textit{ab initio} many-body approaches, we demonstrate that the exchange-correlation effects beyond the density functional theory for itinerant Mn-$d$ electrons do not significantly alter the obtained electronic and magnetic structure.
Therefore, we conclude that, contrary to previous claims, the most pronounced 2D kagome-derived topological band features bear little relevance to transport in $R$Mn$_6$Sn$_6$, albeit they may possibly be brought to focus by electron or hole doping.

\end{abstract}

\keywords{spin-orbit coupling, Rare-earth, magnetocrystalline anisotropy, kagome lattice, Dirac crossing, spin-reorientation}

\maketitle

\section{Introduction}
Two-dimensional (2D) kagome-lattices of $3d$ ions have initially attracted considerable attention due to their exceptionally strong magnetic frustration.
The first experimental realizations were in systems featuring correlated Mott insulators based, for instance, on Cu$^{2+}$, with strong nearest neighbor antiferromagnetic exchange.
These materials were investigated for potential spin liquid behavior~\cite{Norman2016} and fluctuation-driven phenomena such as unconventional superconductivity~\cite{Mazin2014}.
A relatively newer development is metallic kagome materials with unusual magnetic and topological properties~\cite{Ghimire2020}.
In particular, a 2D single-orbital kagome model exhibits such features as flat band and Dirac crossing (DC).
As we discuss later in the paper, the same features survive in the 2D five-orbital nearest-neighbor hopping kagome planes, but not all of them retain their 2D character in real 3D materials like the family considered in this paper.
Spin-polarized DCs may be gapped by the spin-orbit coupling (SOC) in quasi-2D ferromagnetic (FM) metals,  resulting in Chern gaps~\cite{xu2015prl,tang2011prl,neupert2011prl}.
When these topological electronic states are near the Fermi level, large Berry curvatures are manifested, resulting in novel quantum properties such as the quantum anomalous Hall effect.

\begin{figure}[htb]
  \includegraphics[width=0.8\linewidth,clip]{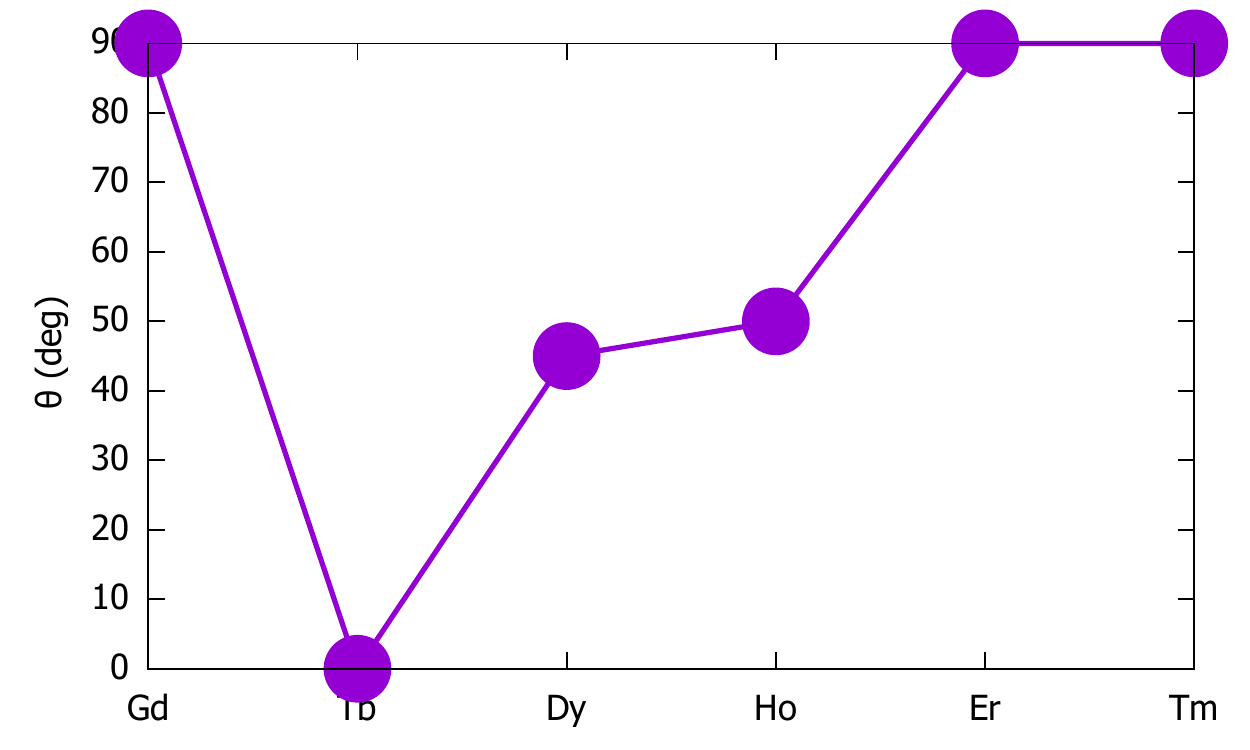}
  \caption{Easy-axis angle $\theta$, with respect to the crystallographic $c$ direction, in $\rmnsn$ at low temperature, with $R=$ Gd, Tb, Dy, Ho, Er, and Tm~\cite{idrissi1991jlcm,venturini1991jmmm,dirken1991jac,venturini1993jac,venturini1996jac,lefevre2003jac}.
At low temperatures, the anisotropy is easy-axis when $R=$ Tb, easy-plane when $R=$ Gd, Er, and Tm, and easy-cone with $\theta=40$--$50$ degrees when $R=$ Dy and Ho.  }
\label{fig:inset}
\end{figure}

An especially popular lately family of FM kagome metals is $R$Mn$_6$Sn$_6$, with the rare earth $R=$ Gd, Tb, Dy, Ho, or Er (the structure also forms with nonmagnetic rare earths but in that case the lack of the transferred FM interaction between the Mn layers bridged by a magnetic rare earth leads to complex antiferromagnetic spiral structures).
Intriguingly, and importantly, all of them form collinear ferrimagnets, but the direction of the ordered moments varies, seemingly randomly, as shown in \rfig{fig:inset}, from material to material.
Given that SOC, as well as such properties as an anomalous Hall effect (AHE) or magnetooptical Kerr effect (MOKE), are intimately related to the direction of magnetization, understanding this interesting variation of the magnetocrystalline anisotropy (MA) is of utmost importance. 

Another hot topic, prominently featured in the recent literature~\cite{yin2020n}, is the possibility of Chern topological magnetism.
In principle, Chern physics can be triggered by the DCs genetically related to the kagome geometry.
In that case, the size of the Chern gap is determined by the orbital characters of corresponding bands, as well as the size of the spin projection along the direction normal to the kagome layer~\cite{xu2015prl}.
The prerequisites are (i) out-of-plane spin alignment, which is necessary for generating the Chern gap; (ii)
minimal $k_z$ dispersion of the relevant DCs; and (iii) proximity of the DC in question to the Fermi level.

The first condition is satisfied in, and only in, the Tb compound in the $\rmnsn$ family.
This has motivated intense research of this compound~\cite{yin2020n,xu2022topological,gao2021apl,riberolles2022prx,mielkeiii2022cp,GMU}.
The main challenge here is establishing a connection between surface probes such as tunneling and bulk properties controlling effects like AHE and MOKE.
Recently, Yin and coworkers, using tunneling spectroscopy, identified a feature that could be interpreted in terms of a DC located $\sim 130$ meV above the Fermi level, and conjectured that this DC is a source of the observed bulk AHE.
The intriguing observation depends on these quasi-2D DCs lying close to the Fermi level, and warrants a closer inspection, which is done in a companion paper~\cite{GMU}.

In this work, we investigate the electronic structures and intrinsic magnetic properties of $\rmnsn$ with $R$ = Gd, Tb, Dy, Ho, and Er.
Besides the excellent agreement of magnetic results with existing experiments, our $\abinitio$ calculations also uncover the higher-order nature of crystal field (CF) parameters and MA constants in these systems.
We further demonstrate that this discovery can be understood qualitatively in the phenomenological model and quantitatively within a simple analytical model based on the CF at the rare-earth site, which is also calculated from first principles.
We then address the topological aspect of the electronic structure, paying particular attention to the DCs, their location and origin, and their potential impact upon the bulk topological properties, and how they can be affected by spin-reorientation, surface effects, and electron correlation.

\section{\textit{Ab initio} methods}

The density functional theory (DFT) calculations are performed using a full-potential linear augmented plane wave (FP-LAPW) method, as implemented in \textsc{Wien2k}~\cite{WIEN2k}.
The generalized gradient approximation of Perdew, Burke, and Ernzerhof~\cite{perdew1996} is used for the correlation and exchange potentials.
Unless specified, low-temperature experimental lattice parameters~\cite{idrissi1991jlcm} are adopted in all bulk calculations.
SOC is included using a second variational method.

The strongly correlated $R$-$4f$ electrons are treated using the DFT+$U$ method with the fully-localized-limit (FLL) double-counting scheme and the so-called open-core approach.
The ground states of the heavy-$R$ $4f$ shell are generally expected to satisfy Hund's rules due to the dominance of SOC over CF.
However, it is well-known that DFT+$U$ can have many metastable solutions, and worse still, the ground state may appear as a metastable state in DFT+$U$.
Therefore, the initial orbital occupancy of $4f$ states should be controlled to ensure that the self-consistent electron configurations satisfy Hund's rules.
This is the only constraint we enforced in our DFT+$U$ calculations.
As long as it is enforced (which itself requires that $U$ cannot be too small), we found that the calculated magnetic properties are not very sensitive to the $U$ value.
Therefore, in this work, we only present DFT+$U$ results with $U=$ \SI{0.52}{Ry}, which falls within the typical range of $U$ values used for $R$-$4f$ elements.
Without adjusting $U$ parameters for each $R$ element, our calculations, as demonstrated later, can capture the essence of anisotropy evolution in this entire series of compounds.
In contrast to DFT+$U$, the open-core approach incorporates the occupied $4f$ electrons as core states.
This approximation is reasonable when describing band structures near the Fermi level with minimal contributions from the $4f$ electrons.
Additionally, the open-core approach allows us to examine the contributions of non-$4f$ electrons to MA.

We also explore the effects of electronic correlation of non-$4f$ electrons beyond DFT by employing the quasiparticle self-consistent GW (QSGW) method~\cite{van-schilfgaarde2006prl,kotani2007prb}, which is based on many-body theory.
While the QSGW method represents a simplification (from a technical standpoint) of the more general fully self-consistent GW (scGW) approximation~\cite{hedin1965pr}, it is typically more accurate~\cite{kutepov2017prb,grumet2018prb}.
In the QSGW, the fully frequency-dependent self-energy of the scGW method is replaced with a static (frequency-independent) self-energy.
However, this replacement is performed in a special way to ensure the so-called $Z$-factor cancellation~\cite{kotani2007prb}.
This cancellation guarantees that the QSGW method, unlike the scGW method, satisfies the Ward Identity in an important long-wave and zero-bosonic frequency limit.
Compared to the DFT approximation, the QSGW method considerably improves the calculated electronic structure in both simple $sp$-materials~\cite{kotani2007prb,deguchi2016jsap} and materials with strong electron correlations involving $d$- or $f$-electrons~\cite{kotani2007prb,chantis2007prb,chantis2008prb,lee2020prbr,ke2021ncm}.
Unlike the DFT+$U$ approximation, the QSGW method has the advantage of being fully \textit{ab-initio}, without any adjustable parameters~\cite{ke2013prb}.
In this study, we employ the QSGW method~\cite{kotani2014jpsj} to investigate the non-$4f$ band structures near the Fermi level in these compounds.

\section{Magnetic ordering and exchange coupling}

\begin{figure}[htb]
  \includegraphics[width=0.8\linewidth,clip]{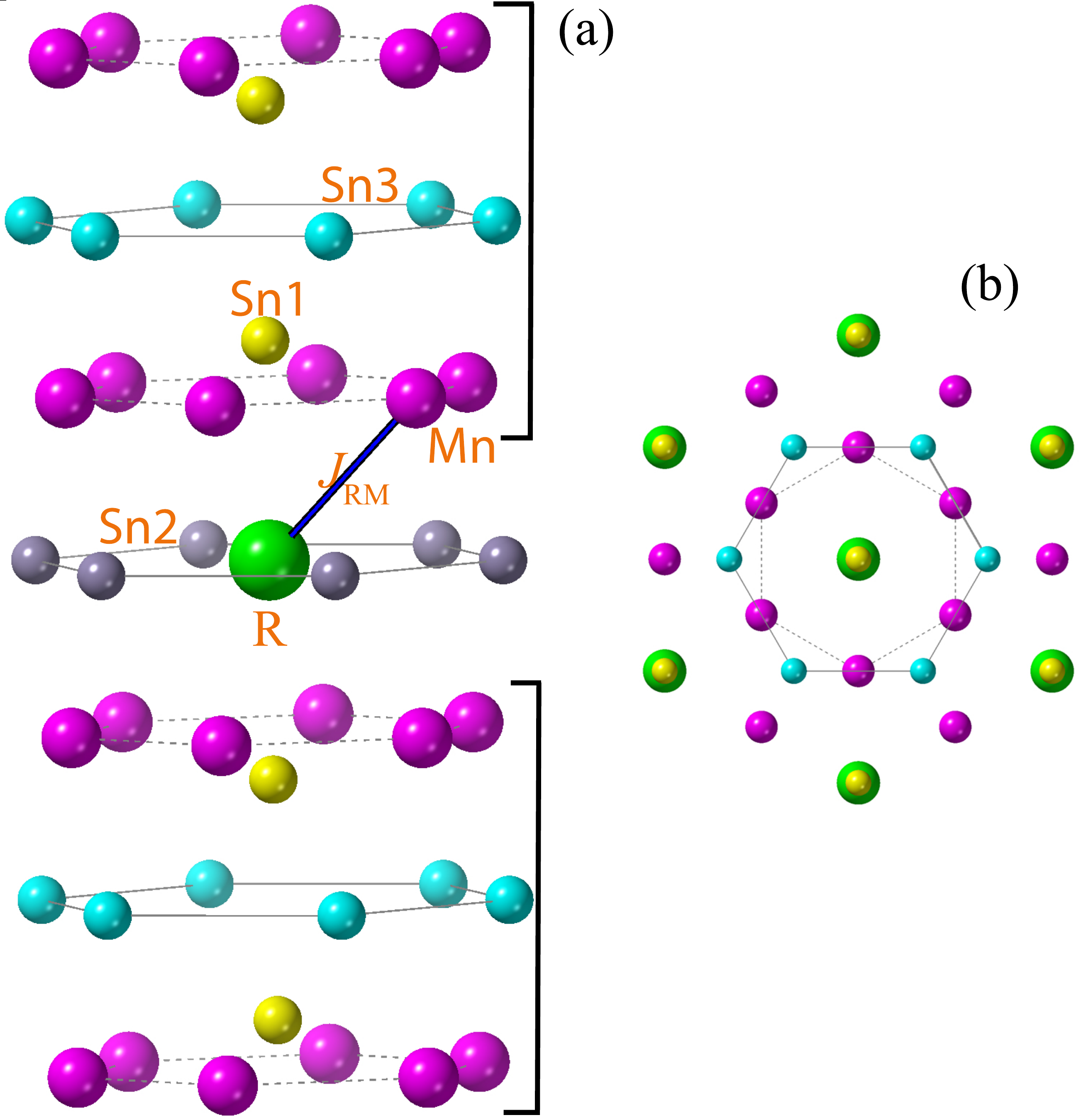}
  \caption{
Crystal structure of $\rmnsn$ (a) and its top view (b).
  Atomic layers are stacked in the order of [Mn-Sn$_1$-Sn$_3$-Sn$_1$-Mn]-[$R$-Sn$_2$]-[Mn-Sn$_1$-Sn$_3$-Sn$_1$-Mn] along the $c$ axis.
  The kagome Mn bilayers that sandwich the Sn$_3$ layer, denoted by the square brackets, are ferromagnetically strongly coupled, while the coupling between two Mn-bilayer blocks is weak.
  The antiferromagnetic coupling between the heavy rare-earth atoms $R$ and neighboring Mn atoms, $J_{RM}$, is crucial to maintain the ferromagnetic Mn ordering in $\rmnsn$ at low temperature.
}
\label{fig:xtal}
\end{figure}

$\rmnsn$ with heavy $R$ elements crystallizes in the hexagonal HfFe$_6$Ge$_6$-type ($P6/mmm$, space group no.~191) structure, as shown in~\rfig{fig:xtal}. 
$R$ atoms ($D_{6h}, \text{or } 6/mmm$) forms a triangular lattice with each $R$ atom neighboring with six Sn$_2$ atoms in the basal plane.
The nearest neighbor of $R$ atoms is the Sn$_1$ atoms, which are along the axial direction and pushed slightly off the Mn kagome plane by $R$ atoms.
The six Mn atoms ($2mm$) in the unit cell form two FM kagome layers that sandwich the Sn$_3$ honeycomb layer and are ferromagnetically coupled via the Mn-Sn$_3$-Mn superexchange~\cite{riberolles2022prx}.
Mn sublattices prefer easy-plane spin orientation.
The couplings between neighboring Mn-bilayers blocks across the $R$-Sn$_1$ layer are weaker or even antiferromagnetic (AFM), depending on the $R$ element type.
As a result, the AFM $R$-Mn exchange coupling $J_{RM}$ and $R$ magnetic anisotropy are essential to determine the overall magnetic structure and band topology.

The lattice parameters and atomic coordinates slightly vary with different element types of $R$.
To separate the chemical and structural effects on magnetic properties, we also perform calculations for all $\rmnsn$ compounds using the lattice parameters of $\gdmnsn$.

\subsection{Spin and orbital magnetic moments}
\rtbl{tbl:ms_morb_mtot} summarizes the magnetic moments and their components in $\rmnsn$ calculated in DFT+$U$ and compared with experimental values and the corresponding values expected for $4f$ shells from Hund's rules.
Reported experimental spin-reorientation temperatures $\tsr$ and Curie temperatures $\tc$ are also listed for comparison.
The calculations adopt the experimental low-temperature collinear magnetic structure and corresponding easy directions.

The deviation of the spin and orbital magnetic moments of $R$ from the integer values expected for $4f$ electrons, as dictated by Hund's rules, is attributed to the contributions from $R$-$5d$ electrons.
The $R$-$5d$ states are primarily spin-polarized by the neighboring 12 magnetic Mn atoms through $3d$-$5d$ hybridization.
They are further polarized by the on-site $4f$ moment.
The Mn-$3d$ spin aligns antiferromagnetically with the $R$-$5d$ spin, which is parallel with the $R$-$4f$ spin, resulting in $R$-Mn ferrimagnetic (FI) ordering in $\rmnsn$ for heavy $R$ atoms.
Without considering the variation of structural parameters with $R$, we found that the induced $5d$ spin moment of various $R$ atoms, calculated using the $\gdmnsn$ crystal structure parameters, can be approximately written as
\begin{equation}
  m_{R\textrm{-}5d}^{s}=12 \alpha\, m_{\text{Mn-}3d}^s + \beta\, m_{R\textrm{-}4f}^{s},   
\end{equation}
with $\alpha \approx 0.007$ and $\beta \approx 0.02$.
When considering the $R$ dependence of structural parameters, we found that the $5d$ spin moment decreases with the $4f$ spin moment by approximately 40\% as $R$ progresses from Gd to Er.

The calculated magnetic moments, as summarized in \rtbl{tbl:ms_morb_mtot}, show good overall agreement with previously reported experimental values.
Mn moments are calculated to have values of 2.38--2.42~$\mub$/Mn, consistent with the reported experimental values of 2.11--2.5~$\mub$/Mn in various $\rmnsn$ compounds.
For the magnetic moment of $R$ atoms, experimental magnetic moments agree reasonably well with the calculated ones, suggesting that the orbital occupancy of $4f$ electrons in these compounds respects Hund's rule as expected for heavy $R$ atoms.
The calculated value of $m_\text{Tb}=9.23~\mub$/Tb is nearly identical to the very recent experimental value measured by Mielke and co-workers~\cite{mielkeiii2022cp} at 2~K.
The calculated $m_\text{Dy}$ also agrees well with neutron diffraction measurements~\cite{mielkeiii2022cp,malaman1999jmmm,idrissi1991jlcm}.
For other $R$ elements, the calculated $m_R$ values are somewhat larger than reported experimental ones. 
For example, Ho in $\homnsn$ has the largest difference between the calculated and experimental values, 10.14 and 8.43~$\mub$/Ho, respectively.
However, the calculated overall magnetization agree better with experiments; Clatterbuck $\etal$~\cite{clatterbuck1999jmmm} estimated the net magnetic moment of $\homnsn$ from the magnetization curve at \SI{10}{\K} and obtained 3.26~$\mub$/f.u., agreeing fairly well with the calculated value of 3.0~$\mub$/f.u.
Furthermore, larger experimental Ho moment measured by the neutron diffraction had been reported in doped $\homnsn$ compounds~\cite{lefevre2003jac}, e.g., with $m_\text{Ho}= 9.53~\mub$/Ho in HoMn$_{6}$Sn$_{5}$In.
The difference between the experiments and theory may be relevant to the easy-cone orientation and the fact that we also partition magnetization into interstitial and Sn sites, which are slightly AFM with respect to Mn.

\begin{widetext}
\begin{table*}[t]
  \caption{ The spin magnetic moment $m_{R}^s$ and orbital magnetic moment $m_{R}^l$ of $R$ atom (in $\mub$/$R$), the total magnetic moment of Mn atom $m_\text{Mn}$ (in $\mub$/Mn), and magnetization $M$ (in $\mub$/cell) in $\rmnsn$ and compared to experiments. $R$-$4f$ orbitals are treated within DFT+$U$.
    The calculated $m_\text{Mn}$, consisting of $\sim1\%$ orbital magnetic moment, is antiparallel with $R$ moment.
    Sn atoms have a moment of $\sim0.11~\mub$/Sn, and the interstitial has a moment of $\sim0.5~\mub$/f.u.; both align antiparallelly with respect to the Mn moments.
    Electron occupancy in the minority $R$-$4f$ channel $n^{\downarrow}_{f}$, spin magnetic moment $m^s_{4f}$, orbital magnetic moment $m^l_{4f}$, and total magnetic moment $m_{4f}$ of $R$-$4f$ electrons, according to Hund's rules, are also shown. On-site spin and orbital magnetic moments are in units of $\mub$/atom. Experimental spin-reorientation temperature $\tsr$ (in K) and Curie temperature $\tc$ (in K) values are also listed.}
  \begin{tabular*}{\linewidth}{cc@{\extracolsep{\fill}}ccccccrrrrrrlrcccr}
  \hline\hline
\multirow{2}{*}{$R$} & \multirow{2}{*}{$Z$} &   & \multicolumn{4}{c}{Hund's Rules} & & \multicolumn{5}{c}{Calculations} & & \multicolumn{6}{c}{Experiments} \\  \cline{4-7}  \cline{9-13} \cline{15-20}
  &  &   & $n^{\downarrow}_{4f}$ & $m_{4f}^s$ & $m_{4f}^l$ & $m_{4f}$ & & $m_{R}^s$ & $m_{R}^l$ & $m_{R}$ & $m_\text{Mn}$ & $M$ & & $m_R$  & $m_\text{Mn}$ & $M$ & $\tsr$ & $\tc$ & References\\
\hline
Gd & 64 & & 0 & 7 & 0 & 7   & & 7.33 & -0.02 &  7.31 & 2.38 &  5.83 &  & 6.5  & 2.5  & 8.5        &           & 435--445 & [\onlinecite{kimura2006jac,venturini1991jmmm,clatterbuck1999jmmm}]   \\
Tb & 65 & & 1 & 6 & 3 & 9   & & 6.26 & 2.96  &  9.23 & 2.42 &  4.10 &  & 9.2  & 2.39 & 5.77       & 310--330  & 423--450 & [\onlinecite{mielkeiii2022cp,kimura2006jac,clatterbuck1999jmmm,venturini1991jmmm}] \\
Dy & 66 & & 2 & 5 & 5 & 10  & & 5.21 & 4.96  & 10.18 & 2.40 &  3.05 &  & 9.97 & 2.11 & 2.69       & 270--320  & 393--410 & [\onlinecite{kimura2006jac,venturini1991jmmm,clatterbuck1999jmmm}] \\
Ho & 67 & & 3 & 4 & 6 & 10  & & 4.17 & 5.97  & 10.14 & 2.39 &  3.07 &  & 8.43 & 2.39 & 3.26--5.91 & 175--200  & 376--400 & [\onlinecite{kimura2006jac,venturini1991jmmm}] \\
Er & 68 & & 4 & 3 & 6 & 9   & & 3.19 & 5.93  &  9.12 & 2.38 &  4.03 &  & 8.40 & 2.21 & 4.86       &    75     & 340--352 & [\onlinecite{venturini1991jmmm,clatterbuck1999jmmm,yazdi2012jmmm}] \\
\hline\hline
\end{tabular*}
\label{tbl:ms_morb_mtot}
\end{table*}
\end{widetext}

\subsection{Intersublattice $R$-Mn exchange coupling}
\begin{figure}[ht]
\centering
\begin{tabular}{c}
    \includegraphics[width=.55\linewidth,clip]{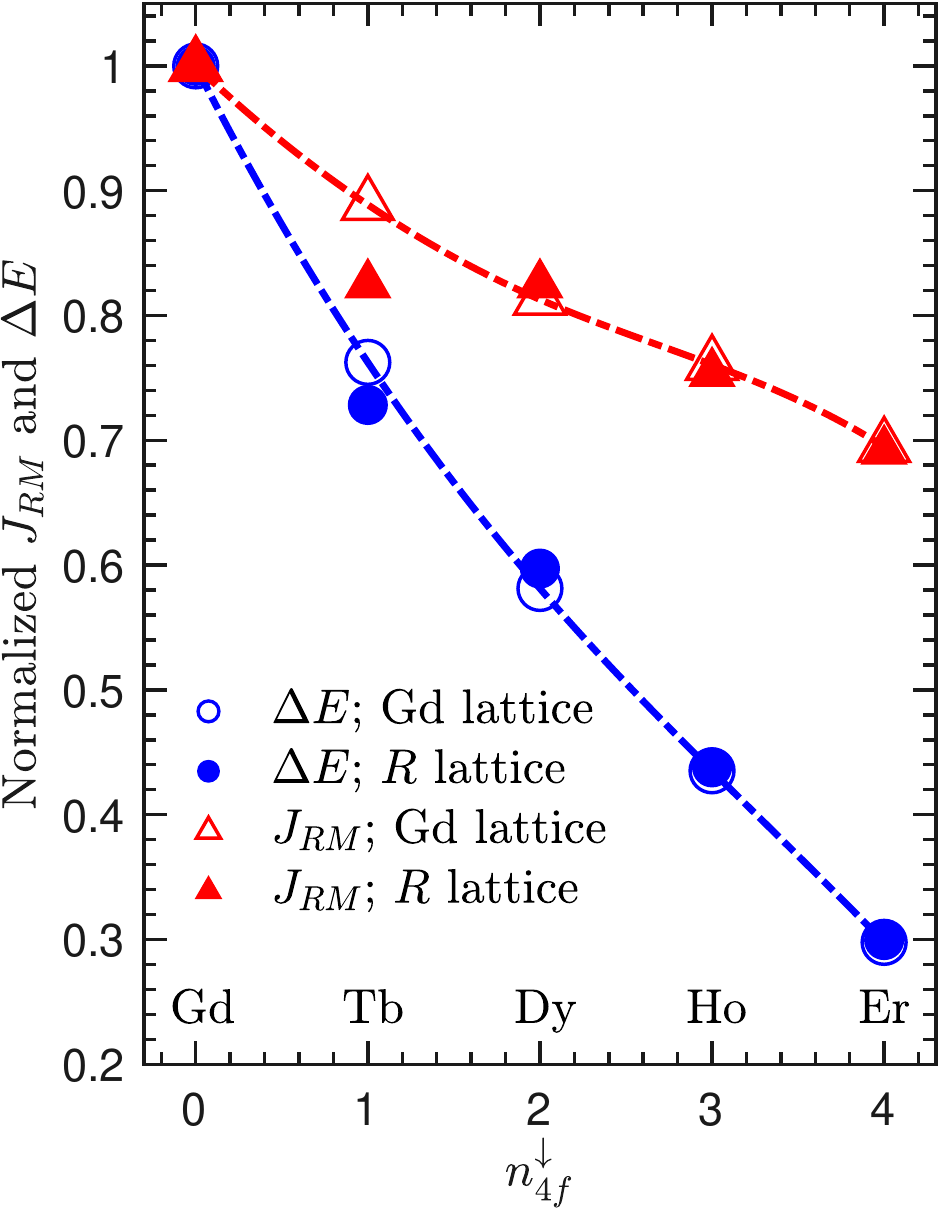} \\  
\end{tabular}%
\caption{ Normalized intersublattice $R$-Mn exchange coupling parameter $J_{R{\rm M}}$ and magnetic energy $\Delta E$ as functions of the electron occupancy in $R$-$4f$ minority spin channel $n_{4f}^{\downarrow}$ in $\rmnsn$ with $R=$~Gd, Tb, Dy, Ho, and Er.
$\Delta E=E_\text{FM}-E_\text{AFM}$ is calculated as the energy difference between the FM and AFM spin configurations of $R$ and Mn sublattices.
To separate the structural and chemical effects, calculations using the lattice parameters of $\gdmnsn$ are also carried out and denoted as open triangles and circles.
The values of $\Delta E$ and $J_{RM}$ are normalized with respect to the values of Gd compounds, $\Delta E^\text{Gd} = 299$ meV and $J_{RM}^\text{Gd} = 2.96$ meV.
}
\label{fig:degennes}
\end{figure}

The intersublattice magnetic couplings between $R$ and Mn sublattice play an essential role in aligning the FM Mn-bilayers and stabilizing long-range Mn ordering.
It  also affects $\tsr$ as a larger $\jrm$ suppresses the thermal activation of $4f$ electrons into excited multiplet, which ultimately makes the thermal average of the $4f$ charge cloud more spherical and isotropic.
We estimate  the $R$-Mn coupling $\jrm$ by mapping the total energies of FM and FI $R$-Mn spin configurations into a Heisenberg model defined as
\begin{equation}
    H_{R\text{M}}= \sum_{i\in R,j\in \text{Mn}}\, \jrm {\bf S}_i \cdot {\bf S}_j
\end{equation}
Here, $S_i=|{\bf S}_i| = m_i^s/2$ and $m_i^s$ is the spin magnetic moment on site $i$. A positive $\jrm$ corresponds to the AFM $R$-Mn coupling.

\rFig{fig:degennes} shows the $R$-Mn magnetic energy $\Delta E$ and exchange parameter $\jrm$, normalized with respect to the values of $\gdmnsn$, as functions of the electron occupancy in the minority $R$-$4f$ spin channel.
The $R$-Mn magnetic interaction energy $\Delta E=E_\text{FM}-E_\text{AFM}=24\jrm S_{R}S_\text{Mn}$ is calculated as the energy difference between the FM and AFM spin configurations of the $R$ and Mn sublattices.
The $R$-Mn intersublattice couplings are AFM for all $R$ elements, consistent with experiments.
The corresponding magnetic energy $\Delta E$ and exchange parameter $\jrm$ decrease by $\sim70\%$ and  $\sim30\%$, respectively, when $R$ goes from Gd to Er.
The abnormality of $\jrm$ at $R=$Tb is related to the structural change, considering that the calculations that use the Gd lattice parameters give a smooth curve, as shown in \rfig{fig:degennes}.

In addition to the decrease in the $R$ spin moment, the reduction of the $R$-Mn exchange energy from Gd to Er is also caused by the weakening of the $\jrm$. A similar decrease of $\jrm$ with increasing atomic number has also been observed in other rare-earth transition-metal alloys. This trend is especially pronounced in the light rare-earth series~\cite{verhoef1990jmmm,liu1994jmmm,duc1993jmmm}. However, the mechanism behind the decreasing $\jrm$ is not apparent, as one might assume that $\jrm$ should remain the same considering the similarities of band structures throughout the series. The exchange coupling between the $R$-$4f$ spin and Mn-$3d$ spin primarily occurs through the $R$-$5d$ electrons. The decrease in $\jrm$ with increasing atomic number may be due to the lanthanide contraction, which reduces the overlap between the $4f$ and $5d$ charge densities~\cite{radwanski1986pssb,belorizky1987jap}. The change in lattice parameters can also affect the $4f$-$5d$ overlap and $5d$-$3d$ hybridization, thus influencing $\jrm$, as demonstrated by the abnormality of $\jrm$ at $R=$Tb in \rfig{fig:degennes}.

\section{Magnetocrystalline anisotropy}
MA in $\rmnsn$ consists of contributions from both the $R$ and Mn sublattices.
These contributions have different temperature dependencies and dominate at lower and higher temperatures, respectively.
MA becomes essential in maintaining long-range magnetic ordering in low-dimensional materials or bulk materials composed of weakly coupled magnetic layers, in accordance with the Mermin-Wagner theorem~\cite{mermin1966prl,mkhitaryan2021prb}.

Although the easy directions of $\rmnsn$ have been well established experimentally, the anisotropy amplitudes, the entire $E(\theta)$ profile, the constituent sublattice contributions, and the underlying microscopic origin of these anisotropies remain largely unknown.
In this section, we demonstrate the evolution of the easy axis in this series of compounds can be well described theoretically.
Moreover, by decomposing the anisotropy into sublattice contributions, we discover the significant role played by higher-order CF parameters and MA constants in these systems.
In this section, we demonstrate that the MA mechanism in $\rmnsn$ can be quantified and understood at various levels: \textit{ab initio}, phenomenological, and analytical.

\subsection{\textit{Ab initio} calculations}

\begin{figure}[thb]
\centering
\begin{tabular}{c}
\includegraphics[width=.99\linewidth,clip]{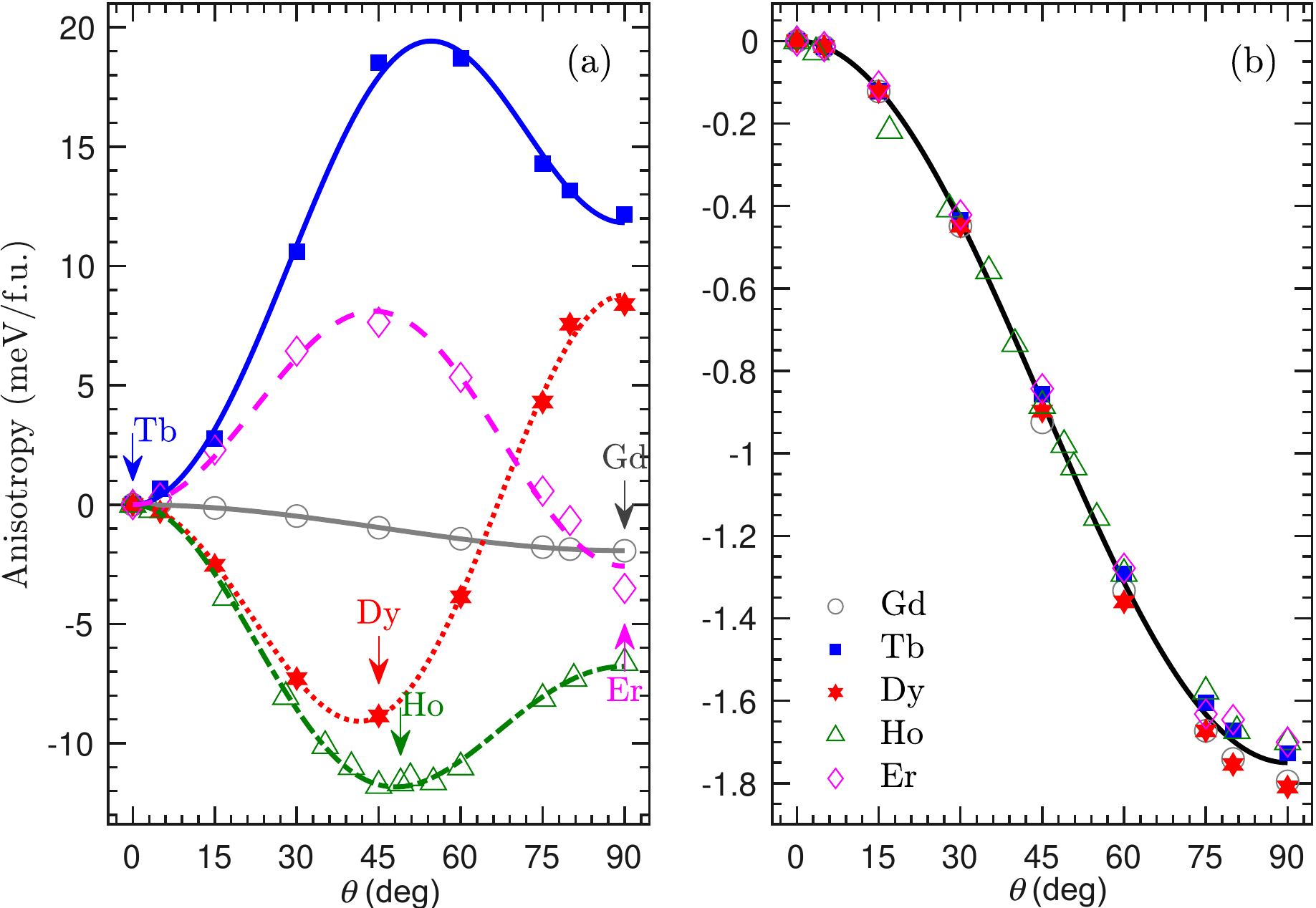}
\end{tabular}%
\caption{
  Variation of magnetic energy (in meV/f.u.) as a function of spin-axis rotation in $\rmnsn$, with $R=$ Gd, Tb, Dy, Ho, and Er, calculated (a) with and (b) without $R$-$4f$ contributions. $\theta$ is the angle between the spin direction and the out-of-plane direction.
  The experimental easy directions for each compound are denoted by arrows in panel (a).
  The lines are fittings of $E(\theta)=K_1 \sin^2\theta + K_2 \sin^4\theta + K_3 \sin^6\theta$ in panel (a) and $E(\theta)=K_1 \sin^2\theta$ in panel (b), respectively.
} 
\label{fig:mae_from_mn_sublattice}
\end{figure}

At lower temperatures, as shown in \rfig{fig:inset}, experiments found that $\tbmnsn$ has an easy-axis anisotropy and $\ermnsn$ has an easy-plane anisotropy, while the $\homnsn$ and $\dymnsn$ have an easy-cone anisotropy with the quantization axis along the $\theta=40$--$50\degree$ directions.
\rFig{fig:mae_from_mn_sublattice}(a) shows the calculated total energies $E(\theta)$ as functions of spin-quantization direction, characterized by the angle $\theta$ deviated from the $c$ axis.
The calculated large easy-axis anisotropy in $\tbmnsn$ is comparable to the experimental value of 23.1 meV/f.u.~estimated  from recent inelastic neutron scattering (INS) measurements~\cite{riberolles2022prx}.
This value is also comparable to the well-studied SmCo$_5$ magnet~\cite{kirchmayr1978jmmm}.
The calculated easy directions for all five compounds agree well with experiments.
$\gdmnsn$ shows a cosine-like $E(\theta)$ dependence, and the amplitude is one order of magnitude smaller than other $\rmnsn$ compounds.
In contrast, all four other compounds show a nonmonotonic dependence of $E$ on $\theta$ with an energy minimum or maximum near 45\degree, suggesting substantial higher-order CF parameters (CFP) and MA constants.

Mn sublattice contribution dominates MA at temperatures above $\tsr$; experiments~\cite{idrissi1991jlcm, venturini1991jmmm,malaman1999jmmm} found that all compounds have an easy axis within the basal plane with $\tsr<T<\tc$.
Here, we theoretically confirm the easy-plane contribution of Mn sublattice by calculating the MA contributions from non-$4f$ electrons.
This is achieved by treating $R$-$4f$ electrons in the open-core approach, in which $R$-$4f$ charges are treated as spherical and do not contribute to MA.

\rFig{fig:mae_from_mn_sublattice}(b) shows the non-$4f$ contributions to MA.
Unlike the total MA, the non-$4f$ MA energy (MAE) can be perfectly fitted as $E(\theta) = K_1 \sin^2\theta$, without higher-order terms ($K_2$ and $K_3$), as generally expected.
Moreover, remarkably, all compounds have a similar amplitude as calculated in $\gdmnsn$.
Overall, the non-$4f$ MAE is generally weaker than the $R$-Mn exchange coupling in $\rmnsn$, which maintains a collinear spin configuration between $R$ and Mn sublattices.
As a result, at lower temperatures, the easy direction is dictated by the $R$ sublattice.
Furthermore, it is worth noting that although we often associate the non-$4f$ MA contribution with the Mn sublattice, in fact it is a combined effect of the Mn-$3d$ spin polarization and the large Sn-$4p$ SOC. 
This MA mechanism is rather general in many systems that consist of strongly spin-polarized atoms and large-SOC heavier atoms, such as permanent magnet FePt~\cite{ke2019prb}, topologcial materials MnBi$_2$Te$_4$~\cite{li2020prl}, and magnetic 2D van der Waals matrials CrI$_3$~\cite{gordon2021jpda}.
Mn sublattice MA can be further resolved into single-ion and two-ion~\cite{ke2019prb} (anisotropic exchange) contributions.
Ghimire $\etal$ found that the MA in $\yymnsn$ consists of an easy-axis single-ion MA and a stronger easy-plane anisotropic exchange, resulting in an overall easy-plane MA~\cite{nirmal2020sciadv}.

The mechanism of the easy-cone MA in $\dymnsn$ and $\homnsn$ is not well understood.
It has been argued that the easy-cone directions in  $\dymnsn$ and $\homnsn$ result from the competition between easy-plane Mn anisotropy and easy-axis (weaker than those of Tb) anisotropy from the Dy or Ho sublattice~\cite{idrissi1991jlcm,malaman1999jmmm}.
However, considering the Mn sublattice contribution is much smaller than the total MAE, as shown in our calculations, we argue that Dy and Ho MAE themselves prefer the easy direction off the $z$-axis.
To verify, we turn off the SOC on Mn and Sn sites in $\homnsn$ and find that the calculated easy direction remains the same.
Thus, we conclude that the easy-cone axis results from the dominant Dy or Ho MA itself instead of the competition between  easy-axis $R$ MA and easy-plane Mn MA.
This can be verified by future measurements of the easy directions of Dy or Ho compounds in other $R$166 compounds with a nonmagnetic transition metal sublattice, such as V.

While the easy directions calculated in DFT agrees well with experiments for all $\rmnsn$ compounds we studied here, it is desirable to understand the evolution of rare-earth anisotropy further.
In the following two sections, we elucidate the microscopic origin of this computational $R$ anisotropy using simple and physically transparent analytical models.

\subsection{Rare-earth anisotropy I: Phenomenological crystal-field model}

\begin{figure}[thb]
\centering
\begin{tabular}{c}
  \includegraphics[width=.99\linewidth,clip]{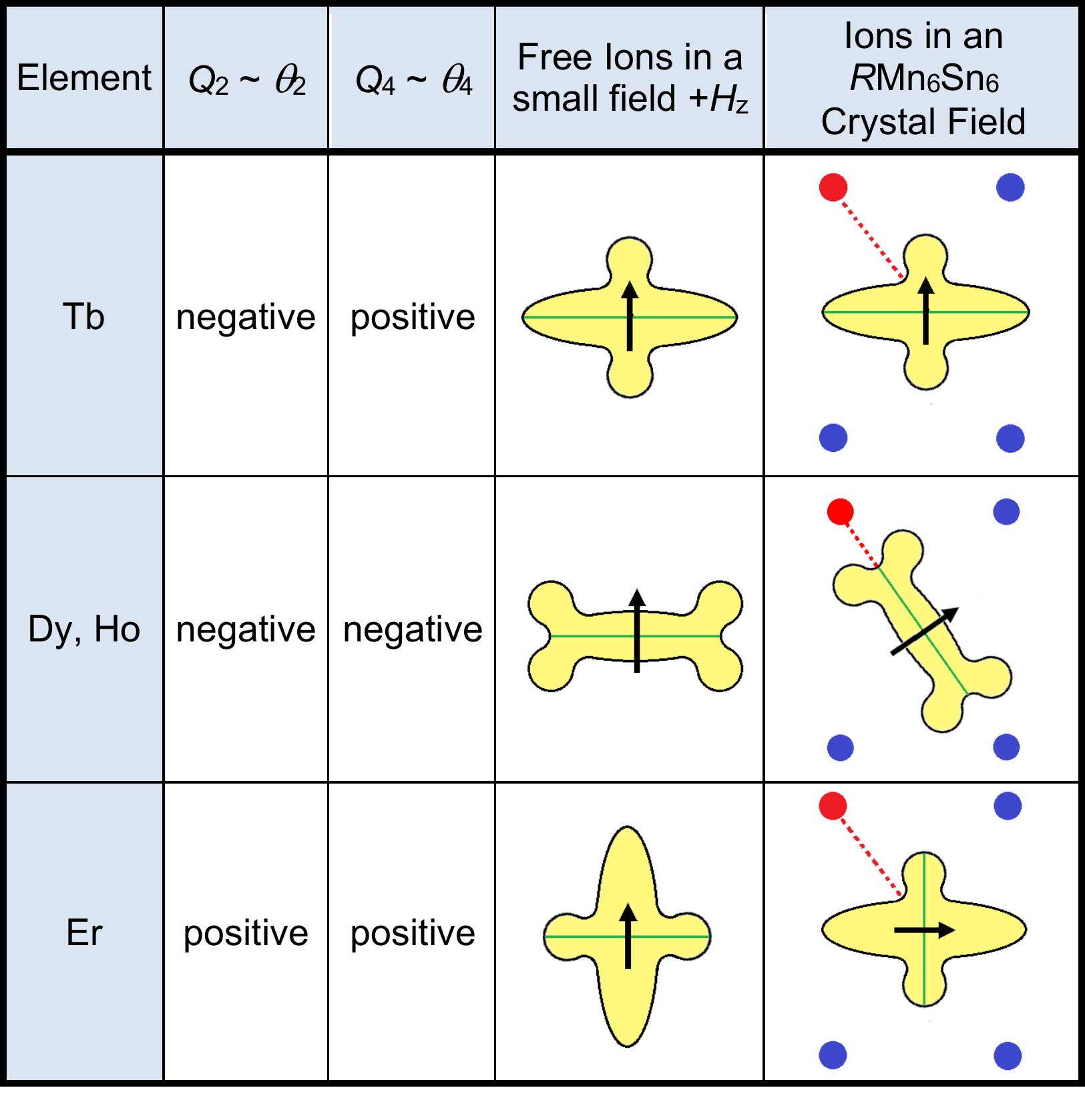} \\  
\end{tabular}%
\caption{Crystal-field origin of easy-axis (Tb), easy-cone (Dy, Ho), and easy-plane (Er) anisotropies in $\rmnsn$.
The magnetization of free ions can point in any direction, so a small magnetic field $H= +H_z \hat{e}_z$  has been added to create a unique spin direction.
In the crystal, symbolized by Mn ligands (blue, red), the spin direction is determined by the electrostatic interaction between the rare-earth $4f$ shell (yellow) and the Mn atoms.
Crystal-field charges are negative, so the crystal-field interaction is repulsive.
The right column focuses on the fourth-order interaction ($Q_4 \to K_2$), the dashed red line showing how the repulsive interaction with Mn stabilizes the spin structure.
The green lines are the equators of the uniaxial $4f$ charge distribution, which is always perpendicular to the spin direction (arrows).}
\label{fig:mae_4f_shapes}
\end{figure}

The dominant rare-earth contribution to MA reflects the CF interaction of the $4f$ electrons.
This interaction was first described in terms of electrostatic interaction in insulators~\cite{bethe1929adp}, but the theory also applies to covalent solids and metals, where it is often called ligand-field theory~\cite{ballhausen1962book,newman1989rpp}.
Up to fourth order, the CF interaction of hexagonal crystals is described by the CFP $A_2^0$ and $A_4^0$~\cite{bethe1929adp,ballhausen1962book,hutchings1964ssp,skomski1999book,skomski2008book}.
The anisotropy energy is, up to fourth order,

\begin{equation}
E_a =  K_1 \sin^2(\theta) + K_2\sin^4(\theta),
\end{equation}
where
\begin{eqnarray}
  K_1 &=& - \frac{3}{2}  A_2^0 Q_2 - 5A_4^0 Q_4, \\
  K_2 &=&   \frac{35}{8} A_4^0 Q_4.
 \label{eq:k1_k2}
\end{eqnarray}
In these equations, the $Q_l=\Theta_l \avg{r^l}_{4f} \Oe_l^0$ are the electrostatic multipole moments of the rare-earth $4f$ shells; quadrupole moment $Q_2=a_J \avg{r^2}_{4f} \Oe_2^0$ and hexadecapole moment $Q_4 = b_J \avg{r^4}_{4f}\Oe_4^0$.
Here, the Stevens coefficients $a_J=\Theta_2$ and $b_J=\Theta_4$, the operator equivalents $\Oe_l^0$, and the rare-earth radii $\avg{r^l}_{4f}$ are well-known~\cite{hutchings1964ssp,taylor1972book}, and low-temperature values of
$Q_2$ and $Q_4$ have been tabulated in Ref.~[\onlinecite{skomski1999book}].
The distinguishing behavior of $\rmnsn$ is the large fourth-order CFP ($A_4^0$) and anisotropy ($K_2$) and the corresponding big energy minimum or maximum near 45\degree.

In isostructural compounds, $A_2^0$ and $A_4^0$ exhibit little change across the lanthanide series, because they reflect the crystalline environment of the rare-earth atoms.
The fact that $\tbmnsn$ ($Q_2<0$ and $Q_4>0$) has the largest easy-axis anisotropy among the series suggests $A_2^0>0$ and $A_4^0<0$ (Indeed, we also confirmed $A_2^0>0$ and $A_4^0<0$ in DFT; see supplementary).
The striking differences in \rfig{fig:mae_from_mn_sublattice}(a) reflect the multipole moments.
Physically, the $4f$ electrons mostly confined within the Muffin-Tin sphere and the hybridization between $4f$ and ligands are small; the domination of $4f$ SOC over the weak CF yields a rigid coupling between the spin and the orbital moments of the $R$ atom, so that the magnetic anisotropy is mainly determined by the electrostatic interaction of the $R$-$4f$ charge clouds with the crystalline environment~\cite{skomski1999book,skomski2008book}.
The charge distribution of the Gd-$4f$ electrons is spherical (half-filled $4f$ shell), but other lanthanides have aspherical charge distributions and exhibit nonzero anisotropy contributions.
This asphericity provides a qualitative explanation of the curves in \rfig{fig:mae_from_mn_sublattice}(a).
Lowest-order interactions ($Q_2$) determine the basic spin orientation (easy-axis versus easy-plane), but to understand easy-cone behavior, one needs $Q_4$~\cite{skomski2008book}.

The $R$ elements considered in this paper have $Q_4>0$ (Tb, Er) and $Q_4<0$ (Dy, Ho), as schematically shown in ~\rfig{fig:mae_4f_shapes}.
CF charges in both metals and nonmetals are usually negative~\cite{newman1989rpp,skomski1999book}, so that the Mn coordination of the $R$ atoms in $\rmnsn$ (about 50\degree) yields a negative $A_4^0$ and realizes the situation outlined in \rfig{fig:mae_4f_shapes}.
In a nutshell, for Dy and Ho, the combination of $Q_2$ and $Q_4$ creates a bone-like $4f$ charge distribution, and the electrostatic repulsion between the CF charges (Mn) and the negatively charged $4f$ electrons causes the magnetization direction to deviate from the $c$-axis.
This repulsion is exemplified, in \rfig{fig:mae_4f_shapes}, by dashed red lines near red-colored regions.
In contrast, $Q_4>0$ in Tb and Er results in an energy maximum near $\theta\approx45\degree$.
Moreover, Tb and Er have similar fourth-order Stevens coefficients and their opposite $Q_2$ (oblate versus prolate shape, respectively) produce easy-axis and easy-plane anisotropy, respectively.
Note that $Q_l/\avg{r^l}(\text{Ho})=-Q_l/\avg{r^l}(\text{Er})$, resulting in the roughly opposite $E(\theta)$ in $\ermnsn$ and $\homnsn$.
This can be understood considering that the total seven $4f$ electrons from Ho and Er will produce a nearly (or exactly, if we ignore element dependence of $\avg{r^l}$) spherical charge cloud with vanishing anisotropy.

Note that rare-earth anisotropy constants of order $n>2$ are normally much smaller than second-order anisotropy constants~\cite{skomski1999book}, which explains the relatively rare overall occurrence of easy-cone magnetism.
The high fourth-order anisotropy is a unique consequence of the Mn-coordination of the rare-earth atoms in the structure, which have 12 nearby Mn atoms in adjacent planes.
CFP are proportional to the number of neighbors, each contributing an intrinsic CF contribution $A'_n$, and these intrinsic contributions are multiplied by coordination factors~\cite{newman1989rpp,skomski1999book}.
For $A_4^0$, the coordination factor is $P_4(\cos\Theta)=(35\cos^4(\Theta)-30\cos^2(\Theta)+3)/8$, which has an extreme of $-0.429$ at $49.1\degree$ (see Fig. S6~\cite{supp}).
Moreover, it is worth comparing $\rmnsn$ and the well-studied $R$Co$_5$ system.
Despite the great structural similarity between $\rmnsn$ and $R$Co$_5$ systems~\cite{kirchmayr1978jmmm}, $A_2^0$ is smaller in $\rmnsn$ compared to $R$Co$_5$, because there are Sn near neighbors both axially and in the plane, while in $R$Co$_5$, without the axial Sn and the dissimilarity between transition metal atom and Sn, the large 2nd-order anisotropy ($K_1$) dictates the anisotropy.

The above phenomenological CF model provides an intuitive understanding of the easy directions in $\rmnsn$.
To better quantify the CF model of the anisotropy, in the following we present a more quantitative analytical model of anisotropy using the CF energies from DFT.

\subsection{Rare-earth anisotropy II: Analytical modeling using Crystal field levels}
\begin{figure}[th]
\centering
\begin{tabular}{c}
    \includegraphics[width=.99\linewidth,clip]{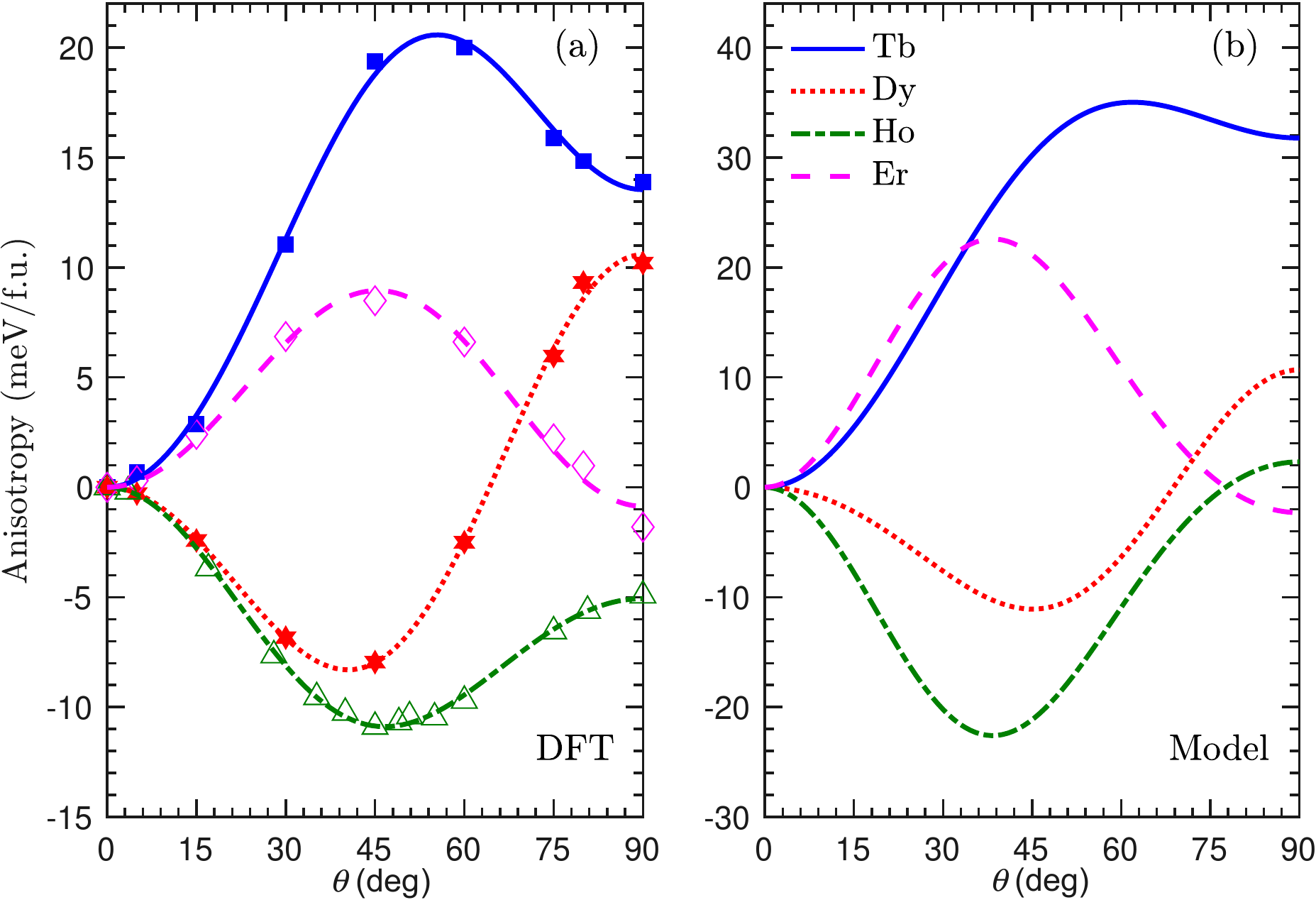} \\  
\end{tabular}%
\caption{$R$-$4f$ only single-ion anisotropy in $\rmnsn$ calculated in (a) DFT and (b) an analytic model \req{eq:kfn}} 
\label{fig:kR_model}
\end{figure}

In the case of SOC dominating the CF energy ($\xi\gg d$), one can assume that, in the first approximation, when
the spin rotates, the angular moment follows it; for example, if the spin is rotated by $\theta$, so is the angular moment, and the SOC energy remains the same during the rotation.
Then, for instance, in the case of Tb, the wave function of its one $f$-electron is described by the \emph{complex} spherical harmonic $\tilde{Y}^l_{m}=\tilde{Y}^3_{3}$ with the $\tilde{z}$ axis is rotated by $\theta$ from the crystallographic $c$ axis.
To calculate the CF energy of this rotated state, we need to re-expand this harmonic in terms of the original ones, namely, $\tilde{Y}^3_{3}=\sum_{m}D_{3m}^{3} (\theta)Y^3_{m}$, where $D$ are the reduced Wigner coefficients.

In the absence of SOC, CF splits the $4f$ states into five quenched levels characterized by \emph{real} spherical harmonics $\rY^l_m$, which are linear combinations of $Y^l_{\pm m}$.
Explicitly, $[\rY^l_m]^\intercal ={\bf U} [Y^l_m]^\intercal$, with $m=-3 \cdots 3$.

\begin{table}[H]
\def\arraystretch{1.3}
  \begin{tabular*}{\linewidth}{l@{\extracolsep{\fill}}ccccc}  
$a_{2u}$ & $e_{1u}$          & $e_{2u}$          & $b_{1u}$                           & $b_{2u}$\\
$z^{3}$  & $z^{2}(x\pm iy)$  & $z(x\pm iy)^{2}$  & $x(x^{2}-3y^{2})$                  & $y(3x^{2}-y^{2})$\\
$\rY^3_0$ & $\rY^3_{\pm1}$ & $\rY^3_{\pm2}$ & $\rY^3_{-3}$ & $\rY^3_3$    
  \end{tabular*}
\end{table}

Then the CF Hamiltonian becomes
\begin{align}
\langle \tilde{Y}_{m}|H_\text{CF}|\tilde{Y}_{m'}\rangle  
& = (\mathbf{D^\dagger U^\dagger E  U D})_{mm'}.
\label{eq:hcf-1}
\end{align}
Here, ${\bf E}$ is the diagonal matrix of CF levels $e_{m}$, and ${\bf D}={\bf D}(\theta)$ is the Wigner coefficient matrix corresponding to the Euler angles $(0,\theta,0)$.
Note that $e_0=E(a_{2u})$, $e_{\pm 1}=E(e_{1u})$, $e_{\pm 2}=E(e_{2u})$, $e_{-3}=E(b_{1u})$ and $e_{3}=E(b_{2u})$.
The contribution to $E(\theta)$ from orbital $m$ can be expanded in $\cos(i\theta)$ with $i=0,2,4,6$.

\begin{equation}
E_m(\theta) =\sum_{i=0,2,4,6}C_i^{m}\cos(i\theta).
\label{eq:kc0246}
\end{equation}
For the second half of the lanthanide series with configurations $f^{n_\downarrow}$, we have
\begin{equation}
  E(f^{n_\downarrow},\theta) = \sum_{m=-3}^{{n_\downarrow}-4}E_m(\theta)
  =\sum_{i=0,2,4,6}C_i^{n_\downarrow}\cos(i\theta).
\label{eq:kfn}
\end{equation}
Coefficients $C^m_{i}$ and $C^{f^{n_\downarrow}}_{i}$ (with $i=2,4,6$) are linear combinations of $e_m$ (see details in Table S1~\cite{supp}). 

We next extract CF levels $e_m$ in $\gdmnsn$ within DFT+$U$ and use them for all four $R$ elements for simplicity, although CF splitting should decrease in heavier $R$ compounds.
Most importantly, the unphysical self-interaction contribution to CF in DFT is mostly avoided in $\gdmnsn$, thanks to a half-filled $f$ shell.
Using the calculated $e_m$, the modeled $E(\theta)$ are calculated and compared to DFT results in \rfig{fig:kR_model}.
The modeled MA somewhat overestimates the calculated MA, partly due to using the larger CF splittings of $\gdmnsn$.
However, as crude as this approximation ($\xi\gg d)$ is, it captures the key features of first-principles calculations quantitatively: (i) the scale of the quartic term is comparable with the scale of the quadratic term, (ii) the sextic term is negligible in Tb, but becomes increasingly more important toward Ho and Er, and (iii) the magnetic anisotropy energy as a function of the angle is approximately opposite in Er and Ho.

\section{Band topology}
One of the most enticing features of the kagome lattice is the fact that, in the single-orbital nearest-neighbor tight-binding (TB) model, the electronic structures show a flat band and a DC at the K point in the Brillouin zone, where the latter is topologically protected while the former is not.
In Chern-gapped insulators, edge states may significantly contribute to the transport properties by avoiding backscattering when $\ef$ is located within the Chern gap.

$\tbmnsn$ is metallic. 
In the work of Yin $\etal$~\cite{yin2020n}, the anomalous Hall effects were observed and related to possible 2D-like (weak $k_z$-dependent) SOC-gapped DC, mainly consisting of Mn inplane orbitals, slightly above $\ef$ at the $K$ point.
However, Jones $\etal$~\cite{GMU} directly calculate the Berry curvatures and found that AHE actually comes from other parts of the BZ.
To understand this discrepancy, we should analyze the nature and characters of multiple Dirac bands in the systems.

Here, we systematically investigate how the band structures near the Fermi level in $\rmnsn$ evolve with $R$, electron correlations, and spin reorientation.
As we shown below, we found that the only quasi 2D DC is located about \SI{0.7}{\eV} above $\ef$, much higher than the value reported in the work of Yin $\etal$~\cite{yin2020n}, which explains why Jones $\etal$~\cite{GMU} do not find significant contributions to AHE at the $K$ point.

\subsection{Dirac crossings and gap openings}

It is instructive to expand the single-orbital kagome model Hamiltonian onto a more realistic five $d$-orbital model.
In a hexagonal CF, the $d$-orbitals split into three levels: $a_{1g}\propto \rY^2_0$, $e_{g}^{\prime}\propto\{\rY^2_{1},\rY^2_{-1}\}$, and $e_{g}^{\prime\prime}\propto\{\rY^2_2,\rY^2_{-2}\}$.
The $e_g'$ orbital is odd with respect to mirror reflection about the kagome plane, while the others are even.
At the $\Gamma$ point, they are orthogonal and protected by the six-fold rotation symmetry.
At a generic quasimomentum, the bands from different orbitals can hybridize due to the absence of mirror symmetry about the kagome plane.
Nevertheless, it is instructive to examine the energy bands of these orbitals on the kagome lattice.

We focus on the 2D momentum space with $k_z=0$, where the DCs appear and the system is invariant under mirror operations about the $ab$ plane.
Thus, we can simplify the model to a 2D kagome lattice without loss of generality.
Since the $a_{1g}$ state is rotation invariant about the $c$-axis, the hopping is the same along all three bonds characterized by vectors $\mathbf{a}_{1}$, $\mathbf{a}_{2}$, and $\mathbf{a}_{3}$.
Considering only the nearest hopping, the Hamiltonian can be written as:

\begin{equation}
\hat{H}_{0}=t_{0}\hat{\mathcal{H}}=t_{0}
\begin{pmatrix}
0 & \cos(\mathbf{k\cdot a}_{1}) & \cos(\mathbf{k\cdot a}_{2})\\
\cos(\mathbf{k\cdot a}_{1}) & 0 & \cos(\mathbf{k\cdot a}_{3})\\
\cos(\mathbf{k\cdot a}_{2}) & \cos(\mathbf{k\cdot a}_{3}) & 0
\end{pmatrix}
,
\end{equation}
which gives the well-known band structure with one flat band and one DC at the K point.

The complex $e_{g}^{\prime}$ orbitals can combine to form real $d_{yz}$ and $d_{zx}$ orbitals.
The TB energy bands of these orbitals on a kagome lattice, along with two Dirac cones (DCs) at the K point, can be found in the Supplementary Material~\cite{supp}.
Similarly, another two DCs at the K point can be attributed to the $e_{g}^{\prime\prime}$ orbitals, corresponding to the $d_{xy}$ and $d_{x^2-y^2}$ orbitals.
Consequently, without considering hybridization between them, we anticipate a total of five DCs per spin, per layer, resulting in five $k_{z}$-dependent Dirac lines.

The hybridizations between the five $d$ orbitals do not affect the existence of DCs.
It is noted that the high symmetric K point has $C_{3v}$ symmetry, which guarantees the decoupling among the $a_{1g}$, $e_g'$, and $e_g''$ orbitals in the absence of SOC.
Additionally, DCs are also robust against couplings within $e_{g}^{\prime}$ or within $e_{g}^{\prime\prime}$, although they can affect the position of DCs and the Dirac velocity.
Further details can be found in the supplementary information.
The $e_{g}^{\prime\prime}$ orbitals are less extended along the $z$ direction and are thus closer to a 2D electronic system.
All DCs in the same spin channel are spread over an energy range of the order of the Mn CF, that is, several eV.

Since two out of the five DCs are more two-dimensional, it becomes extremely important to identify them in the calculated band structure.
This can be achieved by plotting bands along the $K$-$H$ path or by plotting the band structures projected onto the surface BZ, where the dispersive (along $k_z$) band will be washed out and quasi-2D bands will be visible.

Near the Fermi energy, all five compounds share similar band structures, as the non-$4f$ electrons dominate in this energy range.
Multiple DCs occur at the $K$ point near $\ef$, both below and above $\ef$, as expected from the discussion above for the multiorbital kagome Mn lattice.
SOC splits the crossings and opens gaps of various sizes at the BZ corners if the spin is along the $z$ direction.
However, as expected, most of them strongly depend on $k_z$, reflecting the 3D nature of the corresponding bands.

To better illustrate the $k_z$ dependence of the band structures, we project all bands onto the surface BZ by integrating the ${\bf k}$-dependent spectral function over $k_z$, using the equation:
\begin{equation}
  I({\bf k}_{\parallel},\omega)
=\int_{0}^{1} \mathrm{d}k_z\ \sum_i \delta[\omega-E_i({\bf k}_{\parallel}, k_z)].
\label{eq:DOS_projected}
\end{equation}
Here, $k_z$ is integrated from 0 to 1 r.l.u., while ${\bf k}_\parallel$ is in the basal plane.

\begin{figure}[htb]
\centering
\begin{tabular}{c}
  \includegraphics[width=0.95\linewidth,clip]{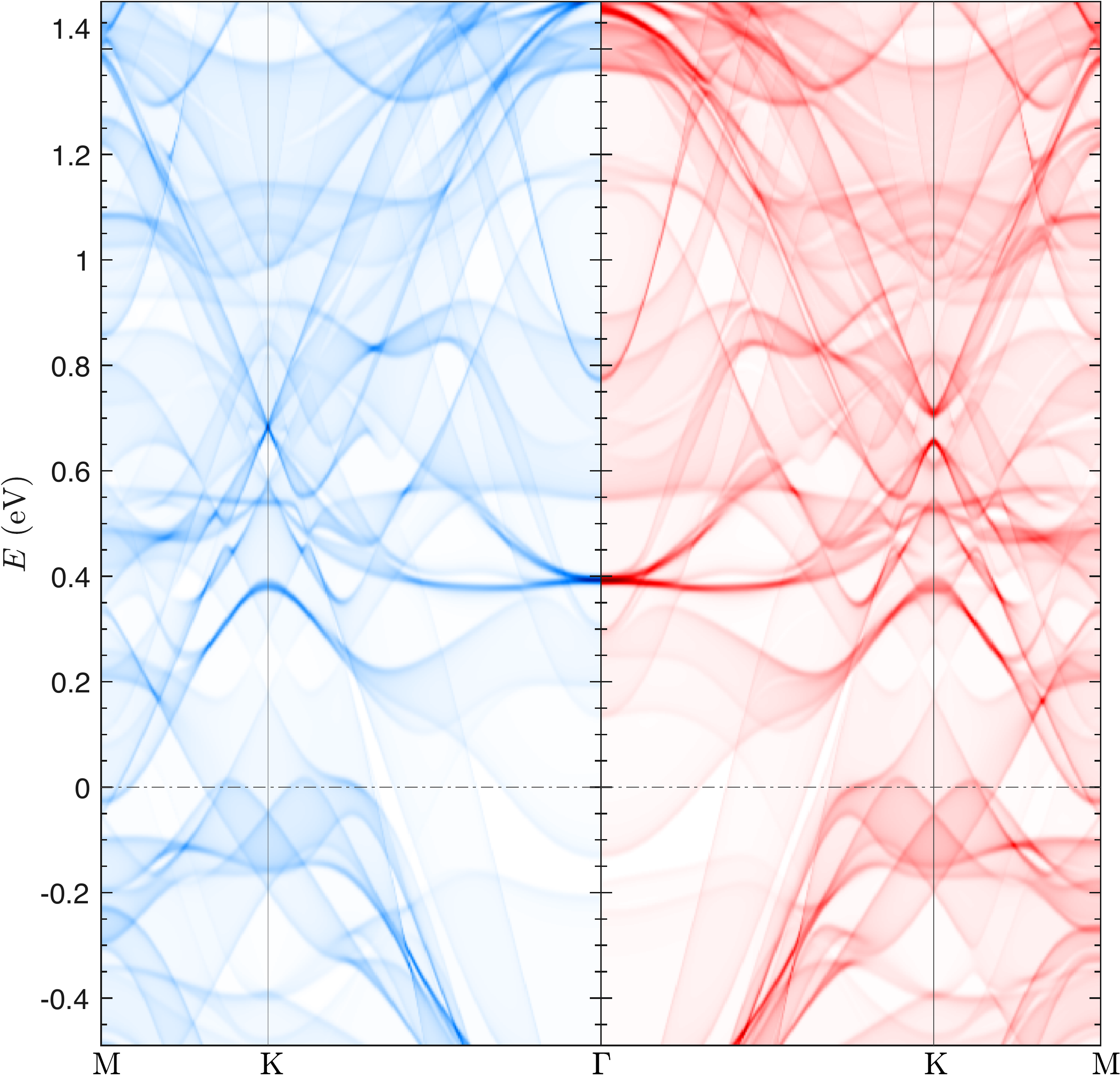} 
\end{tabular}%
\caption{Band structures projected on surface BZ calculated without (blue) and with (red) SOC in $\tbmnsn$.
  The $k$-dependent DOS are integrated along $k_z$ (see \req{eq:DOS_projected}) and are calculated in DFT.
} 
\label{fig:tb_bnd}
\end{figure}

\rFig{fig:tb_bnd} compares the projected $\tbmnsn$ bands along the 2D path $\Gamma$-$K$-$M$, calculated without and with SOC in DFT, shown as blue and red bands, respectively.
Two occupied DCs occur at approximately $0.05$ and $0.2$ eV below $\ef$, respectively, and their gaps are barely opened by SOC. The most prominent $k_z$-independent DC lies at around $0.7$ eV above $\ef$ and is dominated by Mn-$3d$ characters (see Table S2 in the Supplemental Material~\cite{supp}). In contrast to the two occupied DCs, a much larger gap is induced at this DC when SOC is included, which is consistent with the previous report~\cite{yin2020n}. It should be noted that the position of this DC is much higher than the previously reported value of $\sim0.13$ eV above $\ef$ (see the Extended Data Fig. 9 in Ref. \cite{yin2020n}), and it is unlikely to play a significant role in transport properties. The gap size depends on the band characters at these DCs and how effectively SOC can couple them.
Other $\rmnsn$ compounds show overall similar band structures (see Fig.~S3 in the Supplemental Material~\cite{supp} for comparison of the projected band structure of $\rmnsn$ with $R$= Gd, Tb, Dy, Ho, and Er.)

\subsection{Effects of non-$4f$ electron correlation}

TbMn$_6$Sn$_6$ is, as mentioned, a good metal, and Mn electrons are on the itinerant side.
Yet, these $d$ electrons are still considerably, albeit not strongly localized, so correlation effects may be important.
By analogy with such systems as Sr$_2$RuO$_4$ and Fe-based superconductors, one may expect a ``Hund's metal'' behavior.
This is rather hard to capture in static methods such as DFT+$U$ or hybrid functionals.
Even the dynamical mean-field theory (DMFT), the most common method to account for fluctuational correlations, faces serious problems in materials like ours, where long-range correlations are expected and hybridization with Sn is crucial.
In this subsection, to go beyond the standard DFT treatment of non-$4f$ electrons and better address the electron-correlation effects in a more unambiguous way, we employ the QSGW method based on a many-body perturbation approach~\cite{van-schilfgaarde2006prl, kotani2007prb, chantis2007prb, chantis2008prb}.

\begin{figure}[htb]
\centering
\begin{tabular}{c}
  \includegraphics[width=0.95\linewidth,clip]{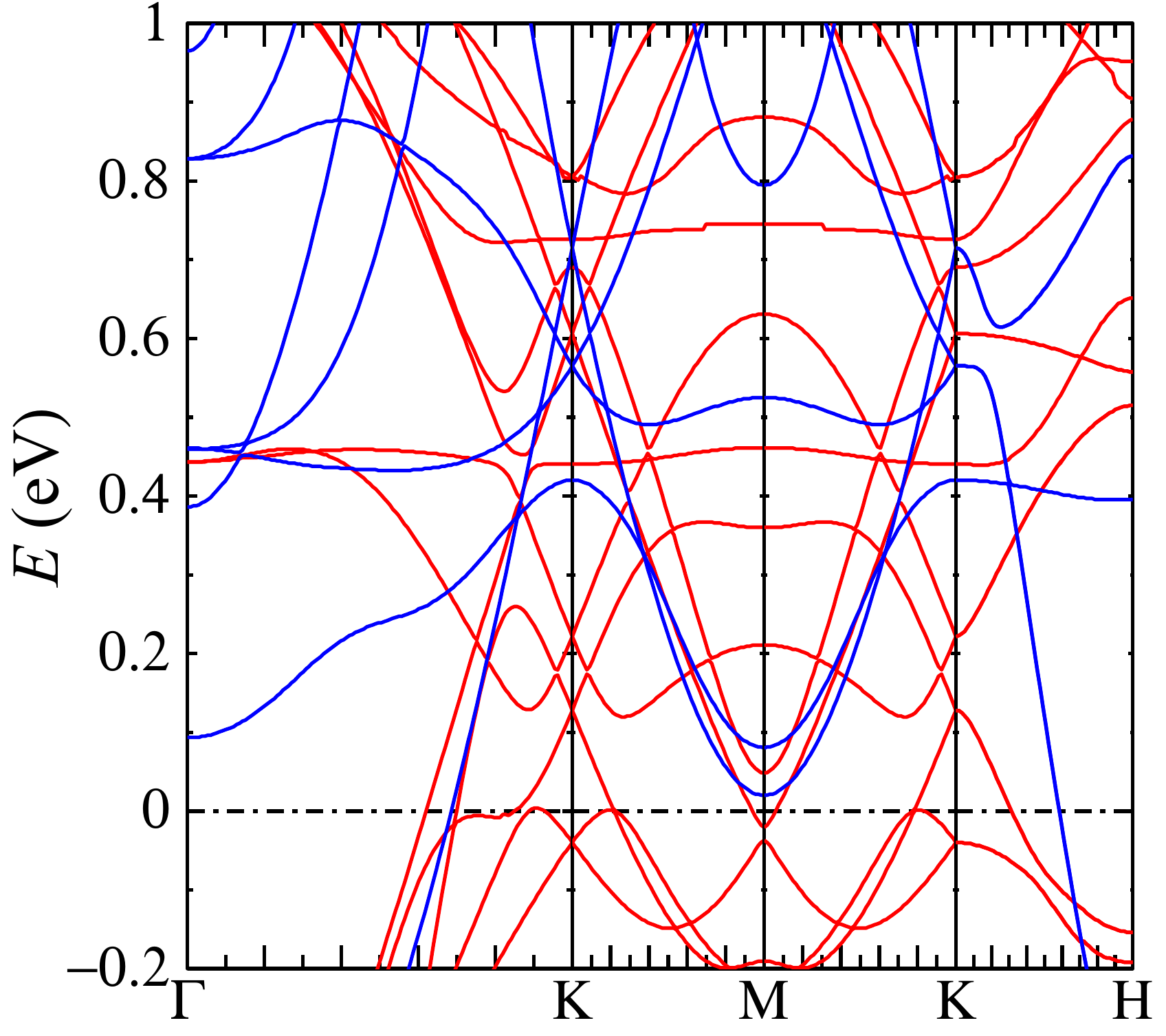} \\
\end{tabular}%
\caption{   
  The scalar-relativistic band structure near $\ef$ in $\tbmnsn$ calculated within QSGW.
  The majority-spin and minority-spin, referred to Mn site, are in blue and red, respectively.
} 
\label{fig:tb_bnd_qsgw_tb}
\end{figure}

\para{QSGW calculations.}
\rFig{fig:tb_bnd_qsgw_tb} shows the scalar-relativistic band structure near $\ef$ of $\tbmnsn$ calculated using QSGW.
The overall non-$4f$ band structure is similar to that obtained from DFT, although QSGW slightly lowers the quasi-2D DCs by approximately 0.1 eV.
This suggests that non-$4f$ electron correlations are not significant in these metallic compounds, and the quasi-2D DCs are still too far above $\ef$ to be related to the observed anomalous Hall conductivity.
It is worth noting that recent experiments on TbV$_6$Sn$_6$ have also shown that the plain DFT treatment of V-$3d$ states provides a reasonable description of the band structures near $\ef$ compared to ARPES measurements \cite{rosenberg2022prb}.

\subsection{Effects of spin orientation}
\begin{figure}[thb]    
\centering
\begin{tabular}{c}
   \includegraphics[width=.97\linewidth,clip]{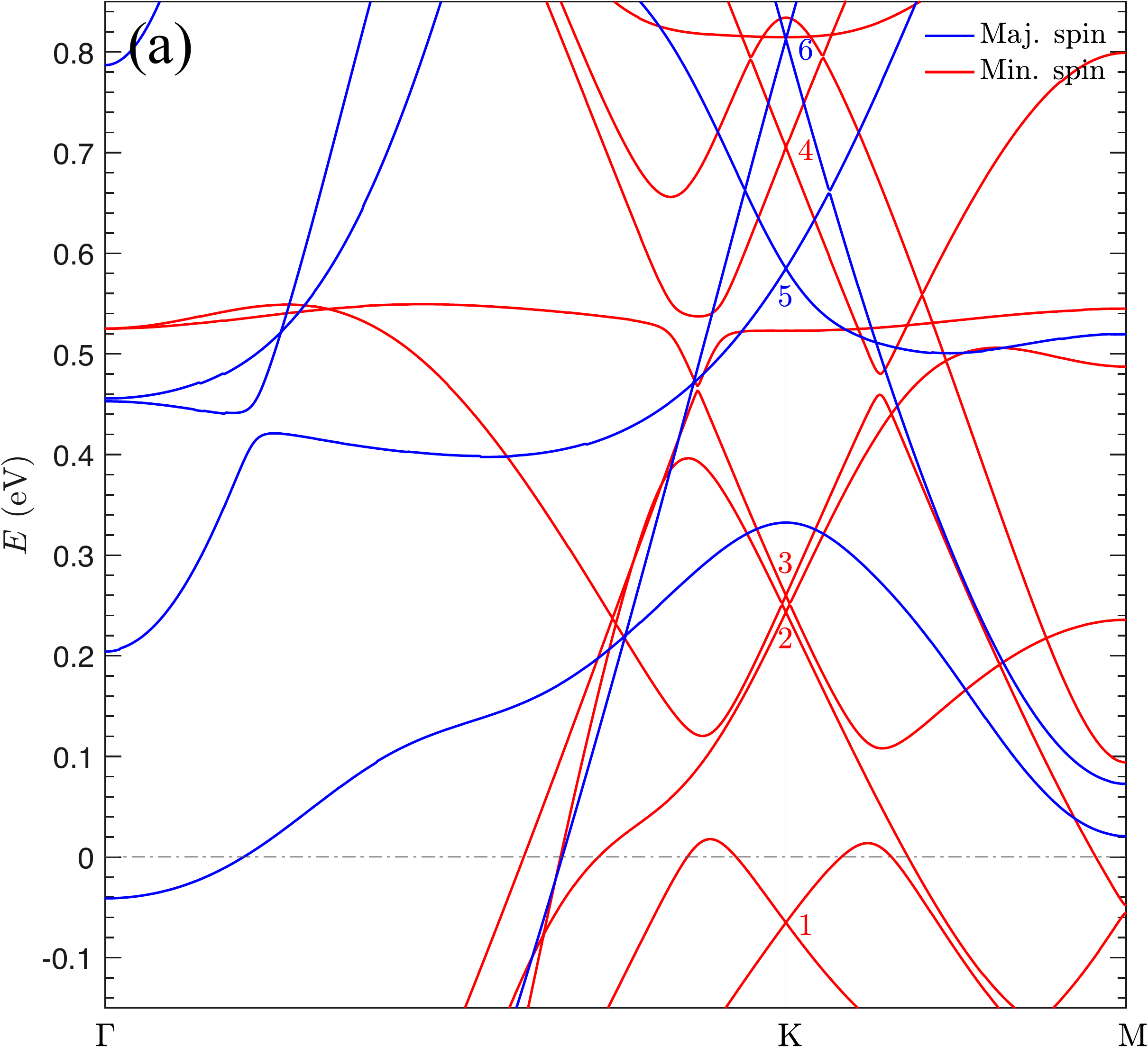} \\
   \includegraphics[width=.97\linewidth,clip]{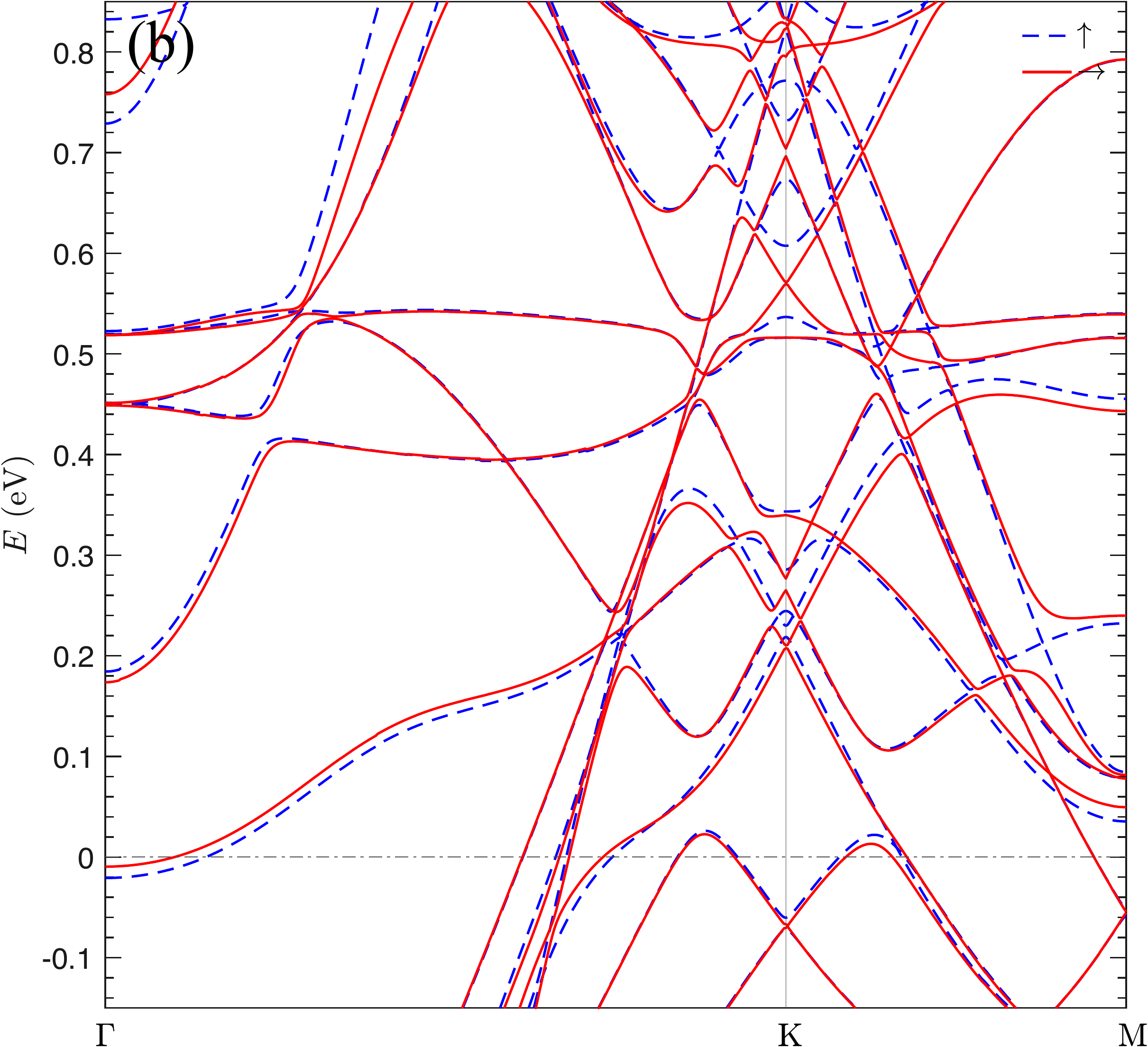}  
\end{tabular}%
\caption{ Band structure near $\ef$ in $\homnsn$ calculated (a) without SOC and (b) with SOC.
 In panel (a), the majority-spin and minority-spin, referred to Mn site, are in blue and red, respectively.
 In panel (b), the band structures are calculated with the spin-quantization axis along the out-of-plane (blue dashed line) and in-plane (red solid line) directions. Both magnetic sublattices are ordered. The gap sizes depend on spin orientations.}
\label{fig:ho_band_kz0}
\end{figure}

It is well known that kagome materials in the presence of SOC and out-of-plane magnetization effectively realize the Haldane model for a Chern insulator without Landau levels~\cite{haldane1988prl,tang2011prl,xu2015prl,yin2020n}.
This model describes spin-polarized electrons hopping in a background of staggered magnetic fluxes on a lattice that supports Dirac crossings in the absence of a magnetic field.
In $\rmnsn$, the bands that are mostly localized in the Mn kagome layer naturally exhibit DCs at the $K$ and $K'$ points near $\ef$, as shown in \rfig{fig:ho_band_kz0}.
Due to the FM order, these DCs occur within a single spin channel, which can be Chern-gapped by intrinsic SOC (see Eq. (S2)~\cite{supp}).
In addition to the itinerant band character, e.g., the $3d$-orbital characters of Mn atoms in the kagome lattice, the size of the SOC-induced gaps also depends on the spin orientations of the magnetic Mn atoms, which can evolve with the $R$ element type and with temperature~\cite{guo2007jac}.
Temperature- and substitution-induced spin reorientations thus have direct consequences on topological transport properties, such as the quantum anomalous Hall conductivity, if these (gapped) crossings occur close to the Fermi energy.

For example, the gap size is expected to vary when $\rmnsn$ goes from the easy-axis $\tbmnsn$ to the easy-cone $\homnsn$ or when $\rmnsn$ is heated above the spin-reorientation temperatures.
Figures~\ref{fig:ho_band_kz0}(a) and ~\ref{fig:ho_band_kz0}(b) show the band structures of $\homnsn$ calculated without and with SOC, respectively.
For the simplicity of illustration, here we focus on the large gap of the DC at \SI{0.7}{\eV}, labeled as DC4 in \rfig{fig:ho_band_kz0}(a).
In \rfig{fig:ho_band_kz0}(b), the gap almost vanishes when the spin-quantization axis rotates from the out-of-plane direction to the in-plane direction.
This can be understood by starting from the non-SOC band structures and treating SOC within perturbation theory. 

DC4 mainly consists of  $\rY^2_{\pm2}$ and $\rY^2_{0}$ Mn-$3d$ characters (see Table S2~\cite{supp}) in the minority spin channel.
Since the DCs occur within the same spin channel, the gap size $\Delta$ is proportional to the spin-parallel part of $H_\text{so}$, as shown in Eq.~(2), and can be written as 
\begin{equation}
  \Delta \propto L_z\cos(\theta) + f(L_{+},L_{-},\theta,\varphi).
\label{eq:gap_vs_theta}
\end{equation}
The second term in \req{eq:gap_vs_theta} vanishes because $L_{\pm}$ do not couple between $\rY^2_{\pm2}$ and $\rY^2_{0}$ states \cite{ke2015prb}.
Therefore, the gap size is solely determined by $L_z\cos(\theta)$, which vanishes at $\theta=90\degree$ with in-plane spin orientations.
If the DCs near $K$ are responsible for the observed AHE, one may expect a significant change in the measurement near $\tsr$.

The band characters of other DCs may consist of orbitals that can also be coupled by $L_\pm$.
The corresponding SOC-induced gap can remain open when the spin is in-plane.
Moreover, DCs containing a larger Sn component can have a larger gap, as Sn has a much larger SOC constant than Mn.
Finally, when DCs are next to each other, multiple DCs can be coupled by SOC, which complicates the analysis.

\subsection{Surface effects on magnetism and bandstructure}
Finally, we investigate the effects of surfaces on the magnetism and electronic structures in $\rmnsn$.
In experiments, purely Mn kagome lattices without detectable defects have been observed over a large field of view in $\tbmnsn$~\cite{yin2020n}.
Here, we calculate the electronic structures in monolayer and bilayer $\tbmnsn$ with a terminating Mn surface on one side and an $R$-Sn surface on the other side.
Each layer has a thickness of one formula unit, as depicted in \rfig{fig:xtal}, and consists of two Mn kagome planes.
To avoid interactions between neighboring slabs due to periodic boundary conditions, a sufficiently large vacuum space is included in the unit cell.
The structure is relaxed to ensure that the force on each atom is less than 1 mRy/a.u.

\begin{figure}[htb]
\centering
\begin{tabular}{c}
    \includegraphics[width=.99\linewidth,clip]{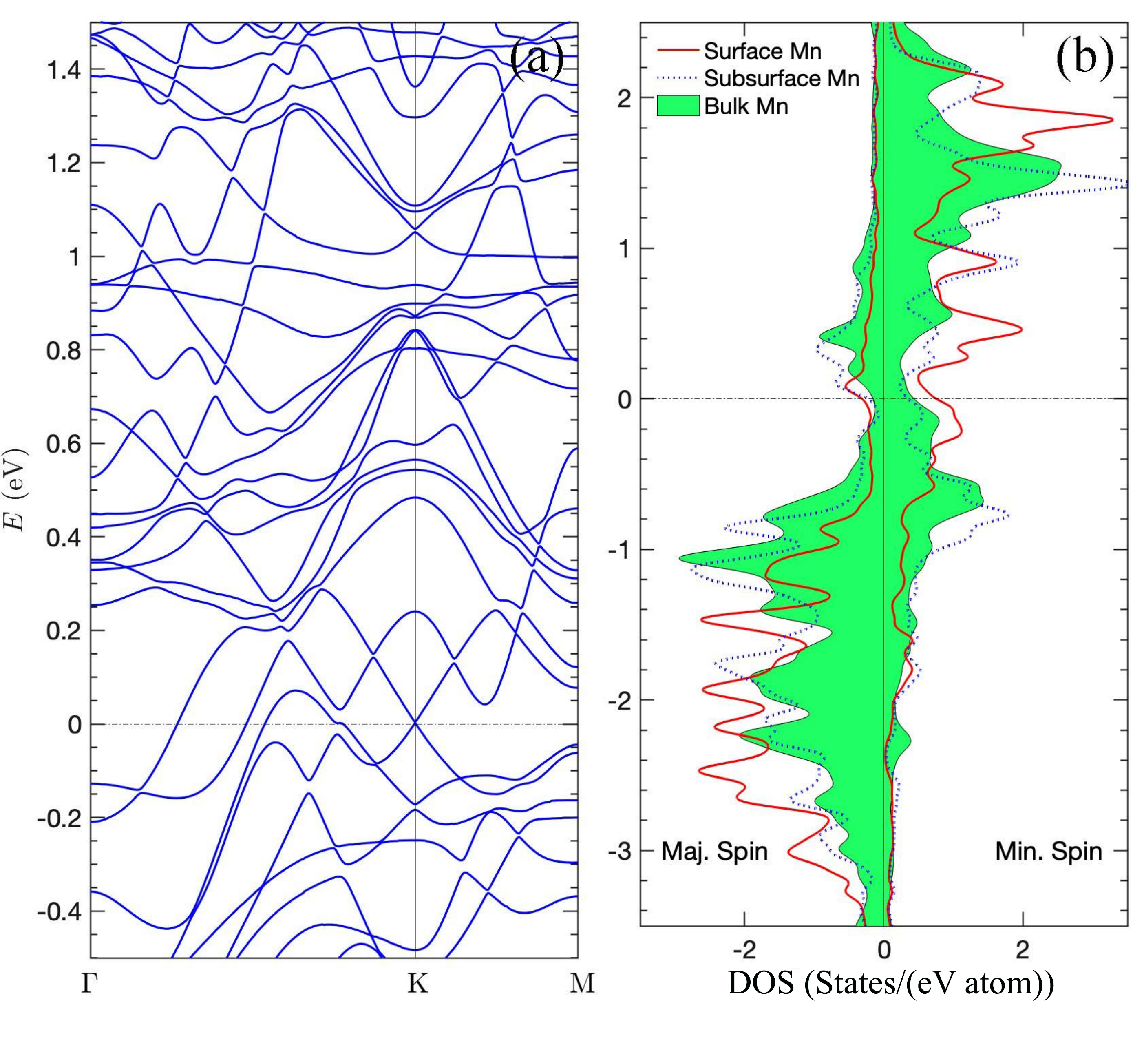}
\end{tabular}%
\caption{(a) Band structures and (b) partial density of states projected on the surface (red solid line) and subsurface (blue dashed line) Mn sites in monolayer $\tbmnsn$. In panel (b), bulk Mn (green filled area) DOS is also shown to compare.
The calculations were performed in plain DFT without SOC.
}
\label{fig:tb_mono_bandos}
\end{figure}

Both the monolayer and bilayer structures of $\tbmnsn$ maintain their metallic nature, similar to the bulk material.
In a bilayer $\tbmnsn$, the Tb atom on the surface exhibits a slightly larger magnetic spin moment compared to the subsurface Tb.
On the other hand, the surface Mn atoms exhibit significantly larger magnetic moments and stronger exchange splittings compared to the bulk.
In both monolayer and bilayer cases, the surface Mn atoms have a magnetic moment of $\sim\SI{3.3}{\mu_B/Mn}$, while the subsurface Mn layers maintain a similar moment of $\sim\SI{2.4}{\mu_B/Mn}$ as in the bulk.
Consequently, near the Fermi level, the spin splitting of the surface Mn states becomes more pronounced.
The band structures and partial density of states projected on the surface and subsurface Mn layers in monolayer $\tbmnsn$ calculated without SOC are shown in \rfig{fig:tb_mono_bandos}.
The band structures exhibit significant changes compared to the bulk bands, and notably, a DC appears at the $K$ point at $\ef$.
The larger spin splitting observed in the surface Mn states, resulting from their larger moments, is illustrated in \rfig{fig:tb_mono_bandos}(b).
In contrast to 2D van der Waals materials, where the calculated on-site moment and intralayer magnetic couplings remain similar between bulk and monolayer forms, $\rmnsn$ exhibits a distinct behavior.
The predicted enhancement of the surface Mn moment awaits experimental confirmation.

\section{Conclusions}
In summary, we have systematically investigate the electronic structures and intrinsic magnetic properties of $\rmnsn$ with $R=$ Gd, Tb, Dy, Ho, and Er.

We have demonstrated how the topological band structures near the $\ef$, including Dirac crossings and SOC-induced gaps, evolve with the choice of $R$ atom, electron correlations, spin reorientation, and surface effects.
The presence of multiple Dirac crossings can be qualitatively understood by solving a five-$d$-orbitals tight-binding model.
Our DFT calculations reveal a prominent SOC-gapped 2D-like Dirac crossing located approximately \SI{0.7}{\eV} above the $\ef$.
The inclusion of additional electron correlation effects using many-body Green's function-based methods only slightly affects the band structure near the $\ef$.
Thus, we have conclusively demonstrated that the observed anomalous Hall conductivity has a 3D character and is not related to the quasi-2D Dirac points.

Our calculations accurately reproduce the experimentally observed easy directions for all $\rmnsn$ compounds.
More importantly, by combining \textit{ab initio}, phenomenological, and analytical methods, we have gained a fundamental understanding of the microscopic origin of magnetism in these materials.
Specifically, we have discovered that the unique Mn coordination with the $R$ atoms leads to significant high-order crystal-field parameters and magnetic anisotropy constants, which are particularly relevant in the context of topological magnets.
The higher-order nature of the $R$ anisotropy, other than the previously believed competition between easy-axis $R$ MA and easy-plane Mn MA, is the true cause of easy-cone anisotropy in $\dymnsn$ and $\homnsn$.
This can be experimentally validated by measuring the easy directions of other $R$166 systems that feature nonmagnetic $3d$ sublattices, such as DyV$_6$Sn$_6$ and HoV$_6$Sn$_6$, where one would expect easy-cone anisotropy instead of easy-axis anisotropy in the ground state.
Additionally, future INS experiments can be employed to quantify the CF parameters in $\rmnsn$ and validate our predictions of the $E(\theta)$ profile, including the existence of a significant barrier between in-plane and out-of-plane spin orientations in ErMn$_6$Sn$_6$.

Methodologically, our work provides a comprehensive investigation of anisotropy in a series of rare-earth materials.
In particular, we have demonstrated that the seemingly irregular variation of the easy direction with different rare-earth elements can be accurately described analytically, without the need for adjustable parameters, based on the mathematical properties of Wigner matrices.
In the future, we can apply our analytical anisotropy modeling approach to other well-established rare-earth-based systems, such as 1-5, 2-17, and 2-14-1 rare-earth-transition-metal systems, to further demonstrate its effectiveness.

\acknowledgments

The authors gratefully acknowledge discussions with Q.~Niu and N.~Ghimire.
L.K. and Y.L. are supported by the U.S. Department of Energy, Office of Science Early Career Research Program through the Office of Basic Energy Sciences, Materials Sciences and Engineering Division.
R.J.M. and P.P.O. are supported by the U.S.~Department of Energy, Office of Basic Energy Sciences, Division of Materials Sciences and Engineering.
Ames Laboratory is operated for the U.S.~Department of Energy by Iowa State University under Contract No.~DE-AC02-07CH11358.
The work in Nebraska is supported by the National Science Foundation/EPSCoR RII Track-1: Emergent Quantum Materials and Technologies (EQUATE), Award OIA-2044049.
A.K.P. acknowledges the funding support from the National Science Foundation under Award No. DMR-2213412.
I.M. acknowledges support from DOE under Grant No. DE-SC0021089. 
This research used resources of the National Energy Research Scientific Computing Center (NERSC), a U.S.~Department of Energy Office of Science User Facility operated under Contract No.~DE-AC02-05CH11231.

\bibliography{aaa}

\begin{thebibliography}{67}%
\makeatletter
\providecommand \@ifxundefined [1]{%
 \@ifx{#1\undefined}
}%
\providecommand \@ifnum [1]{%
 \ifnum #1\expandafter \@firstoftwo
 \else \expandafter \@secondoftwo
 \fi
}%
\providecommand \@ifx [1]{%
 \ifx #1\expandafter \@firstoftwo
 \else \expandafter \@secondoftwo
 \fi
}%
\providecommand \natexlab [1]{#1}%
\providecommand \enquote  [1]{``#1''}%
\providecommand \bibnamefont  [1]{#1}%
\providecommand \bibfnamefont [1]{#1}%
\providecommand \citenamefont [1]{#1}%
\providecommand \href@noop [0]{\@secondoftwo}%
\providecommand \href [0]{\begingroup \@sanitize@url \@href}%
\providecommand \@href[1]{\@@startlink{#1}\@@href}%
\providecommand \@@href[1]{\endgroup#1\@@endlink}%
\providecommand \@sanitize@url [0]{\catcode `\\12\catcode `\$12\catcode
  `\&12\catcode `\#12\catcode `\^12\catcode `\_12\catcode `\%12\relax}%
\providecommand \@@startlink[1]{}%
\providecommand \@@endlink[0]{}%
\providecommand \url  [0]{\begingroup\@sanitize@url \@url }%
\providecommand \@url [1]{\endgroup\@href {#1}{\urlprefix }}%
\providecommand \urlprefix  [0]{URL }%
\providecommand \Eprint [0]{\href }%
\providecommand \doibase [0]{https://doi.org/}%
\providecommand \selectlanguage [0]{\@gobble}%
\providecommand \bibinfo  [0]{\@secondoftwo}%
\providecommand \bibfield  [0]{\@secondoftwo}%
\providecommand \translation [1]{[#1]}%
\providecommand \BibitemOpen [0]{}%
\providecommand \bibitemStop [0]{}%
\providecommand \bibitemNoStop [0]{.\EOS\space}%
\providecommand \EOS [0]{\spacefactor3000\relax}%
\providecommand \BibitemShut  [1]{\csname bibitem#1\endcsname}%
\let\auto@bib@innerbib\@empty
\bibitem [{\citenamefont {Norman}(2016)}]{Norman2016}%
  \BibitemOpen
  \bibfield  {author} {\bibinfo {author} {\bibfnamefont {M.~R.}\ \bibnamefont
  {Norman}},\ }\bibfield  {title} {\bibinfo {title} {{Colloquium:
  Herbertsmithite and the search for the quantum spin liquid}},\ }\href
  {https://doi.org/10.1103/RevModPhys.88.041002} {\bibfield  {journal}
  {\bibinfo  {journal} {Rev. Mod. Phys.}\ }\textbf {\bibinfo {volume} {88}},\
  \bibinfo {pages} {041002} (\bibinfo {year} {2016})}\BibitemShut {NoStop}%
\bibitem [{\citenamefont {Mazin}\ \emph {et~al.}(2014)\citenamefont {Mazin},
  \citenamefont {Jeschke}, \citenamefont {Lechermann}, \citenamefont {Lee},
  \citenamefont {Fink}, \citenamefont {Thomale},\ and\ \citenamefont
  {Valent{\'i}}}]{Mazin2014}%
  \BibitemOpen
  \bibfield  {author} {\bibinfo {author} {\bibfnamefont {I.~I.}\ \bibnamefont
  {Mazin}}, \bibinfo {author} {\bibfnamefont {H.~O.}\ \bibnamefont {Jeschke}},
  \bibinfo {author} {\bibfnamefont {F.}~\bibnamefont {Lechermann}}, \bibinfo
  {author} {\bibfnamefont {H.}~\bibnamefont {Lee}}, \bibinfo {author}
  {\bibfnamefont {M.}~\bibnamefont {Fink}}, \bibinfo {author} {\bibfnamefont
  {R.}~\bibnamefont {Thomale}},\ and\ \bibinfo {author} {\bibfnamefont
  {R.}~\bibnamefont {Valent{\'i}}},\ }\bibfield  {title} {\bibinfo {title}
  {Theoretical prediction of a strongly correlated dirac metal},\ }\href
  {https://doi.org/10.1038/ncomms5261} {\bibfield  {journal} {\bibinfo
  {journal} {Nature Communications}\ }\textbf {\bibinfo {volume} {5}},\
  \bibinfo {pages} {4261} (\bibinfo {year} {2014})}\BibitemShut {NoStop}%
\bibitem [{\citenamefont {Ghimire}\ and\ \citenamefont
  {Mazin}(2020)}]{Ghimire2020}%
  \BibitemOpen
  \bibfield  {author} {\bibinfo {author} {\bibfnamefont {N.~J.}\ \bibnamefont
  {Ghimire}}\ and\ \bibinfo {author} {\bibfnamefont {I.~I.}\ \bibnamefont
  {Mazin}},\ }\bibfield  {title} {\bibinfo {title} {Topology and correlations
  on the kagome lattice},\ }\href {https://doi.org/10.1038/s41563-019-0589-8}
  {\bibfield  {journal} {\bibinfo  {journal} {Nature Materials}\ }\textbf
  {\bibinfo {volume} {19}},\ \bibinfo {pages} {137} (\bibinfo {year}
  {2020})}\BibitemShut {NoStop}%
\bibitem [{\citenamefont {Xu}\ \emph {et~al.}(2015)\citenamefont {Xu},
  \citenamefont {Lian},\ and\ \citenamefont {Zhang}}]{xu2015prl}%
  \BibitemOpen
  \bibfield  {author} {\bibinfo {author} {\bibfnamefont {G.}~\bibnamefont
  {Xu}}, \bibinfo {author} {\bibfnamefont {B.}~\bibnamefont {Lian}},\ and\
  \bibinfo {author} {\bibfnamefont {S.-C.}\ \bibnamefont {Zhang}},\ }\bibfield
  {title} {\bibinfo {title} {{Intrinsic Quantum Anomalous Hall Effect in the
  Kagome Lattice Cs$_2$LiMn$_3$F$_{12}$}},\ }\href
  {https://doi.org/10.1103/PhysRevLett.115.186802} {\bibfield  {journal}
  {\bibinfo  {journal} {Phys. Rev. Lett.}\ }\textbf {\bibinfo {volume} {115}},\
  \bibinfo {pages} {186802} (\bibinfo {year} {2015})}\BibitemShut {NoStop}%
\bibitem [{\citenamefont {Tang}\ \emph {et~al.}(2011)\citenamefont {Tang},
  \citenamefont {Mei},\ and\ \citenamefont {Wen}}]{tang2011prl}%
  \BibitemOpen
  \bibfield  {author} {\bibinfo {author} {\bibfnamefont {E.}~\bibnamefont
  {Tang}}, \bibinfo {author} {\bibfnamefont {J.-W.}\ \bibnamefont {Mei}},\ and\
  \bibinfo {author} {\bibfnamefont {X.-G.}\ \bibnamefont {Wen}},\ }\bibfield
  {title} {\bibinfo {title} {{High-temperature fractional quantum Hall
  states}},\ }\href {https://doi.org/10.1103/PhysRevLett.106.236802} {\bibfield
   {journal} {\bibinfo  {journal} {Phys. Rev. Lett.}\ }\textbf {\bibinfo
  {volume} {106}},\ \bibinfo {pages} {236802} (\bibinfo {year}
  {2011})}\BibitemShut {NoStop}%
\bibitem [{\citenamefont {Neupert}\ \emph {et~al.}(2011)\citenamefont
  {Neupert}, \citenamefont {Santos}, \citenamefont {Chamon},\ and\
  \citenamefont {Mudry}}]{neupert2011prl}%
  \BibitemOpen
  \bibfield  {author} {\bibinfo {author} {\bibfnamefont {T.}~\bibnamefont
  {Neupert}}, \bibinfo {author} {\bibfnamefont {L.}~\bibnamefont {Santos}},
  \bibinfo {author} {\bibfnamefont {C.}~\bibnamefont {Chamon}},\ and\ \bibinfo
  {author} {\bibfnamefont {C.}~\bibnamefont {Mudry}},\ }\bibfield  {title}
  {\bibinfo {title} {Fractional quantum {Hall} states at zero magnetic field},\
  }\href {https://doi.org/10.1103/PhysRevLett.106.236804} {\bibfield  {journal}
  {\bibinfo  {journal} {Phys. Rev. Lett.}\ }\textbf {\bibinfo {volume} {106}},\
  \bibinfo {pages} {236804} (\bibinfo {year} {2011})}\BibitemShut {NoStop}%
\bibitem [{\citenamefont {{El Idrissi}}\ \emph {et~al.}(1991)\citenamefont {{El
  Idrissi}}, \citenamefont {Venturini}, \citenamefont {Malaman},\ and\
  \citenamefont {Fruchart}}]{idrissi1991jlcm}%
  \BibitemOpen
  \bibfield  {author} {\bibinfo {author} {\bibfnamefont {B.}~\bibnamefont {{El
  Idrissi}}}, \bibinfo {author} {\bibfnamefont {G.}~\bibnamefont {Venturini}},
  \bibinfo {author} {\bibfnamefont {B.}~\bibnamefont {Malaman}},\ and\ \bibinfo
  {author} {\bibfnamefont {D.}~\bibnamefont {Fruchart}},\ }\bibfield  {title}
  {\bibinfo {title} {Magnetic structures of {TbMn$_6$Sn$_6$} and
  {HoMn$_6$Sn$_6$} compounds from neutron diffraction study},\ }\href
  {https://doi.org/https://doi.org/10.1016/0022-5088(91)90359-C} {\bibfield
  {journal} {\bibinfo  {journal} {Journal of the Less Common Metals}\ }\textbf
  {\bibinfo {volume} {175}},\ \bibinfo {pages} {143} (\bibinfo {year}
  {1991})}\BibitemShut {NoStop}%
\bibitem [{\citenamefont {Venturini}\ \emph {et~al.}(1991)\citenamefont
  {Venturini}, \citenamefont {Idrissi},\ and\ \citenamefont
  {Malaman}}]{venturini1991jmmm}%
  \BibitemOpen
  \bibfield  {author} {\bibinfo {author} {\bibfnamefont {G.}~\bibnamefont
  {Venturini}}, \bibinfo {author} {\bibfnamefont {B.~E.}\ \bibnamefont
  {Idrissi}},\ and\ \bibinfo {author} {\bibfnamefont {B.}~\bibnamefont
  {Malaman}},\ }\bibfield  {title} {\bibinfo {title} {Magnetic properties of
  {$R$Mn$_6$Sn$_6$} {($R$ = Sc, Y, Gd--Tm, Lu) compounds with
  HfFe$_6$Ge$_6$-type structure}},\ }\href
  {https://doi.org/https://doi.org/10.1016/0304-8853(91)90108-M} {\bibfield
  {journal} {\bibinfo  {journal} {Journal of Magnetism and Magnetic Materials}\
  }\textbf {\bibinfo {volume} {94}},\ \bibinfo {pages} {35} (\bibinfo {year}
  {1991})}\BibitemShut {NoStop}%
\bibitem [{\citenamefont {Dirken}\ \emph {et~al.}(1991)\citenamefont {Dirken},
  \citenamefont {Thiel}, \citenamefont {Brabers}, \citenamefont {{de Boer}},\
  and\ \citenamefont {Buschow}}]{dirken1991jac}%
  \BibitemOpen
  \bibfield  {author} {\bibinfo {author} {\bibfnamefont {M.}~\bibnamefont
  {Dirken}}, \bibinfo {author} {\bibfnamefont {R.}~\bibnamefont {Thiel}},
  \bibinfo {author} {\bibfnamefont {J.}~\bibnamefont {Brabers}}, \bibinfo
  {author} {\bibfnamefont {F.}~\bibnamefont {{de Boer}}},\ and\ \bibinfo
  {author} {\bibfnamefont {K.}~\bibnamefont {Buschow}},\ }\bibfield  {title}
  {\bibinfo {title} {$^{155}${Gd M}\"ossbauer effect and magnetic properties of
  {GdMn$_6$Sn$_6$}},\ }\href
  {https://doi.org/https://doi.org/10.1016/0925-8388(91)90050-6} {\bibfield
  {journal} {\bibinfo  {journal} {Journal of Alloys and Compounds}\ }\textbf
  {\bibinfo {volume} {177}},\ \bibinfo {pages} {L11} (\bibinfo {year}
  {1991})}\BibitemShut {NoStop}%
\bibitem [{\citenamefont {Venturini}\ \emph {et~al.}(1993)\citenamefont
  {Venturini}, \citenamefont {Welter}, \citenamefont {Malaman},\ and\
  \citenamefont {Ressouche}}]{venturini1993jac}%
  \BibitemOpen
  \bibfield  {author} {\bibinfo {author} {\bibfnamefont {G.}~\bibnamefont
  {Venturini}}, \bibinfo {author} {\bibfnamefont {R.}~\bibnamefont {Welter}},
  \bibinfo {author} {\bibfnamefont {B.}~\bibnamefont {Malaman}},\ and\ \bibinfo
  {author} {\bibfnamefont {E.}~\bibnamefont {Ressouche}},\ }\bibfield  {title}
  {\bibinfo {title} {Magnetic structure of {YMn$_6$Ge$_6$} and room temperature
  magnetic structure of {LuMn$_6$Sn$_6$} obtained from neutron diffraction
  study},\ }\href
  {https://doi.org/https://doi.org/10.1016/0925-8388(93)90470-8} {\bibfield
  {journal} {\bibinfo  {journal} {Journal of Alloys and Compounds}\ }\textbf
  {\bibinfo {volume} {200}},\ \bibinfo {pages} {51} (\bibinfo {year}
  {1993})}\BibitemShut {NoStop}%
\bibitem [{\citenamefont {Venturini}\ \emph {et~al.}(1996)\citenamefont
  {Venturini}, \citenamefont {Fruchart},\ and\ \citenamefont
  {Malaman}}]{venturini1996jac}%
  \BibitemOpen
  \bibfield  {author} {\bibinfo {author} {\bibfnamefont {G.}~\bibnamefont
  {Venturini}}, \bibinfo {author} {\bibfnamefont {D.}~\bibnamefont
  {Fruchart}},\ and\ \bibinfo {author} {\bibfnamefont {B.}~\bibnamefont
  {Malaman}},\ }\bibfield  {title} {\bibinfo {title} {Incommensurate magnetic
  structures of {$R$Mn$_6$Sn$_6$ ($R$ = Sc, Y, Lu)} compounds from neutron
  diffraction study},\ }\href
  {https://doi.org/https://doi.org/10.1016/0925-8388(95)01998-7} {\bibfield
  {journal} {\bibinfo  {journal} {Journal of Alloys and Compounds}\ }\textbf
  {\bibinfo {volume} {236}},\ \bibinfo {pages} {102} (\bibinfo {year}
  {1996})}\BibitemShut {NoStop}%
\bibitem [{\citenamefont {Lefèvre}\ \emph {et~al.}(2003)\citenamefont
  {Lefèvre}, \citenamefont {Venturini},\ and\ \citenamefont
  {Malaman}}]{lefevre2003jac}%
  \BibitemOpen
  \bibfield  {author} {\bibinfo {author} {\bibfnamefont {C.}~\bibnamefont
  {Lefèvre}}, \bibinfo {author} {\bibfnamefont {G.}~\bibnamefont
  {Venturini}},\ and\ \bibinfo {author} {\bibfnamefont {B.}~\bibnamefont
  {Malaman}},\ }\bibfield  {title} {\bibinfo {title} {Neutron diffraction study
  of the magnetocrystalline anisotropy in {TbMn$_6$Sn$_{5.8}$Ga$_{0.2}$,
  TbMn$_{6}$Sn$_{5}$Ga, HoMn${_6}$Sn$_5$Ga and HoMn$_6$Sn$_5$In} compounds},\
  }\href {https://doi.org/https://doi.org/10.1016/S0925-8388(03)00135-X}
  {\bibfield  {journal} {\bibinfo  {journal} {Journal of Alloys and Compounds}\
  }\textbf {\bibinfo {volume} {358}},\ \bibinfo {pages} {29} (\bibinfo {year}
  {2003})}\BibitemShut {NoStop}%
\bibitem [{\citenamefont {Yin}\ \emph {et~al.}(2020)\citenamefont {Yin},
  \citenamefont {Ma}, \citenamefont {Cochran}, \citenamefont {Xu},
  \citenamefont {Zhang}, \citenamefont {Tien}, \citenamefont {Shumiya},
  \citenamefont {Cheng}, \citenamefont {Jiang}, \citenamefont {Lian} \emph
  {et~al.}}]{yin2020n}%
  \BibitemOpen
  \bibfield  {author} {\bibinfo {author} {\bibfnamefont {J.-X.}\ \bibnamefont
  {Yin}}, \bibinfo {author} {\bibfnamefont {W.}~\bibnamefont {Ma}}, \bibinfo
  {author} {\bibfnamefont {T.~A.}\ \bibnamefont {Cochran}}, \bibinfo {author}
  {\bibfnamefont {X.}~\bibnamefont {Xu}}, \bibinfo {author} {\bibfnamefont
  {S.~S.}\ \bibnamefont {Zhang}}, \bibinfo {author} {\bibfnamefont {H.-J.}\
  \bibnamefont {Tien}}, \bibinfo {author} {\bibfnamefont {N.}~\bibnamefont
  {Shumiya}}, \bibinfo {author} {\bibfnamefont {G.}~\bibnamefont {Cheng}},
  \bibinfo {author} {\bibfnamefont {K.}~\bibnamefont {Jiang}}, \bibinfo
  {author} {\bibfnamefont {B.}~\bibnamefont {Lian}}, \emph {et~al.},\
  }\bibfield  {title} {\bibinfo {title} {{Quantum-limit Chern topological
  magnetism in TbMn$_6$Sn$_6$}},\ }\href@noop {} {\bibfield  {journal}
  {\bibinfo  {journal} {Nature}\ }\textbf {\bibinfo {volume} {583}},\ \bibinfo
  {pages} {533} (\bibinfo {year} {2020})}\BibitemShut {NoStop}%
\bibitem [{\citenamefont {Xu}\ \emph {et~al.}(2022)\citenamefont {Xu},
  \citenamefont {Yin}, \citenamefont {Ma}, \citenamefont {Tien}, \citenamefont
  {Qiang}, \citenamefont {Reddy}, \citenamefont {Zhou}, \citenamefont {Shen},
  \citenamefont {Lu}, \citenamefont {Chang}, \citenamefont {Qu},\ and\
  \citenamefont {Jia}}]{xu2022topological}%
  \BibitemOpen
  \bibfield  {author} {\bibinfo {author} {\bibfnamefont {X.}~\bibnamefont
  {Xu}}, \bibinfo {author} {\bibfnamefont {J.-X.}\ \bibnamefont {Yin}},
  \bibinfo {author} {\bibfnamefont {W.}~\bibnamefont {Ma}}, \bibinfo {author}
  {\bibfnamefont {H.-J.}\ \bibnamefont {Tien}}, \bibinfo {author}
  {\bibfnamefont {X.-B.}\ \bibnamefont {Qiang}}, \bibinfo {author}
  {\bibfnamefont {P.~V.~S.}\ \bibnamefont {Reddy}}, \bibinfo {author}
  {\bibfnamefont {H.}~\bibnamefont {Zhou}}, \bibinfo {author} {\bibfnamefont
  {J.}~\bibnamefont {Shen}}, \bibinfo {author} {\bibfnamefont {H.-Z.}\
  \bibnamefont {Lu}}, \bibinfo {author} {\bibfnamefont {T.-R.}\ \bibnamefont
  {Chang}}, \bibinfo {author} {\bibfnamefont {Z.}~\bibnamefont {Qu}},\ and\
  \bibinfo {author} {\bibfnamefont {S.}~\bibnamefont {Jia}},\ }\bibfield
  {title} {\bibinfo {title} {{Topological charge-entropy scaling in kagome
  Chern magnet TbMn$_6$Sn$_6$}},\ }\href
  {https://doi.org/10.1038/s41467-022-28796-6} {\bibfield  {journal} {\bibinfo
  {journal} {Nature Communications}\ }\textbf {\bibinfo {volume} {13}},\
  \bibinfo {pages} {1197} (\bibinfo {year} {2022})}\BibitemShut {NoStop}%
\bibitem [{\citenamefont {Gao}\ \emph {et~al.}(2021)\citenamefont {Gao},
  \citenamefont {Shen}, \citenamefont {Wang}, \citenamefont {Shi},
  \citenamefont {Zhao}, \citenamefont {Li}, \citenamefont {Cao}, \citenamefont
  {Pei}, \citenamefont {Ge}, \citenamefont {Li}, \citenamefont {Li},
  \citenamefont {Chen}, \citenamefont {Yan},\ and\ \citenamefont
  {Qi}}]{gao2021apl}%
  \BibitemOpen
  \bibfield  {author} {\bibinfo {author} {\bibfnamefont {L.}~\bibnamefont
  {Gao}}, \bibinfo {author} {\bibfnamefont {S.}~\bibnamefont {Shen}}, \bibinfo
  {author} {\bibfnamefont {Q.}~\bibnamefont {Wang}}, \bibinfo {author}
  {\bibfnamefont {W.}~\bibnamefont {Shi}}, \bibinfo {author} {\bibfnamefont
  {Y.}~\bibnamefont {Zhao}}, \bibinfo {author} {\bibfnamefont {C.}~\bibnamefont
  {Li}}, \bibinfo {author} {\bibfnamefont {W.}~\bibnamefont {Cao}}, \bibinfo
  {author} {\bibfnamefont {C.}~\bibnamefont {Pei}}, \bibinfo {author}
  {\bibfnamefont {J.-Y.}\ \bibnamefont {Ge}}, \bibinfo {author} {\bibfnamefont
  {G.}~\bibnamefont {Li}}, \bibinfo {author} {\bibfnamefont {J.}~\bibnamefont
  {Li}}, \bibinfo {author} {\bibfnamefont {Y.}~\bibnamefont {Chen}}, \bibinfo
  {author} {\bibfnamefont {S.}~\bibnamefont {Yan}},\ and\ \bibinfo {author}
  {\bibfnamefont {Y.}~\bibnamefont {Qi}},\ }\bibfield  {title} {\bibinfo
  {title} {{Anomalous Hall effect in ferrimagnetic metal $R$Mn$_6$Sn$_6$ ($R$ =
  Tb, Dy, Ho) with clean Mn kagome lattice}},\ }\href
  {https://doi.org/10.1063/5.0061260} {\bibfield  {journal} {\bibinfo
  {journal} {Applied Physics Letters}\ }\textbf {\bibinfo {volume} {119}},\
  \bibinfo {pages} {092405} (\bibinfo {year} {2021})}\BibitemShut {NoStop}%
\bibitem [{\citenamefont {Riberolles}\ \emph {et~al.}(2022)\citenamefont
  {Riberolles}, \citenamefont {Slade}, \citenamefont {Abernathy}, \citenamefont
  {Granroth}, \citenamefont {Li}, \citenamefont {Lee}, \citenamefont
  {Canfield}, \citenamefont {Ueland}, \citenamefont {Ke},\ and\ \citenamefont
  {McQueeney}}]{riberolles2022prx}%
  \BibitemOpen
  \bibfield  {author} {\bibinfo {author} {\bibfnamefont {S.~X.~M.}\
  \bibnamefont {Riberolles}}, \bibinfo {author} {\bibfnamefont {T.~J.}\
  \bibnamefont {Slade}}, \bibinfo {author} {\bibfnamefont {D.~L.}\ \bibnamefont
  {Abernathy}}, \bibinfo {author} {\bibfnamefont {G.~E.}\ \bibnamefont
  {Granroth}}, \bibinfo {author} {\bibfnamefont {B.}~\bibnamefont {Li}},
  \bibinfo {author} {\bibfnamefont {Y.}~\bibnamefont {Lee}}, \bibinfo {author}
  {\bibfnamefont {P.~C.}\ \bibnamefont {Canfield}}, \bibinfo {author}
  {\bibfnamefont {B.~G.}\ \bibnamefont {Ueland}}, \bibinfo {author}
  {\bibfnamefont {L.}~\bibnamefont {Ke}},\ and\ \bibinfo {author}
  {\bibfnamefont {R.~J.}\ \bibnamefont {McQueeney}},\ }\bibfield  {title}
  {\bibinfo {title} {{Low-Temperature Competing Magnetic Energy Scales in the
  Topological Ferrimagnet TbMn$_6$Sn$_6$}},\ }\href
  {https://doi.org/10.1103/PhysRevX.12.021043} {\bibfield  {journal} {\bibinfo
  {journal} {Phys. Rev. X}\ }\textbf {\bibinfo {volume} {12}},\ \bibinfo
  {pages} {021043} (\bibinfo {year} {2022})}\BibitemShut {NoStop}%
\bibitem [{\citenamefont {Mielke~III}\ \emph {et~al.}(2022)\citenamefont
  {Mielke~III}, \citenamefont {Ma}, \citenamefont {Pomjakushin}, \citenamefont
  {Zaharko}, \citenamefont {Sturniolo}, \citenamefont {Liu}, \citenamefont
  {Ukleev}, \citenamefont {White}, \citenamefont {Yin}, \citenamefont
  {Tsirkin}, \citenamefont {Larsen}, \citenamefont {Cochran}, \citenamefont
  {Medarde}, \citenamefont {Por{\'e}e}, \citenamefont {Das}, \citenamefont
  {Gupta}, \citenamefont {Wang}, \citenamefont {Chang}, \citenamefont {Wang},
  \citenamefont {Khasanov}, \citenamefont {Neupert}, \citenamefont {Amato},
  \citenamefont {Liborio}, \citenamefont {Jia}, \citenamefont {Hasan},
  \citenamefont {Luetkens},\ and\ \citenamefont {Guguchia}}]{mielkeiii2022cp}%
  \BibitemOpen
  \bibfield  {author} {\bibinfo {author} {\bibfnamefont {C.}~\bibnamefont
  {Mielke~III}}, \bibinfo {author} {\bibfnamefont {W.~L.}\ \bibnamefont {Ma}},
  \bibinfo {author} {\bibfnamefont {V.}~\bibnamefont {Pomjakushin}}, \bibinfo
  {author} {\bibfnamefont {O.}~\bibnamefont {Zaharko}}, \bibinfo {author}
  {\bibfnamefont {S.}~\bibnamefont {Sturniolo}}, \bibinfo {author}
  {\bibfnamefont {X.}~\bibnamefont {Liu}}, \bibinfo {author} {\bibfnamefont
  {V.}~\bibnamefont {Ukleev}}, \bibinfo {author} {\bibfnamefont {J.~S.}\
  \bibnamefont {White}}, \bibinfo {author} {\bibfnamefont {J.-X.}\ \bibnamefont
  {Yin}}, \bibinfo {author} {\bibfnamefont {S.~S.}\ \bibnamefont {Tsirkin}},
  \bibinfo {author} {\bibfnamefont {C.~B.}\ \bibnamefont {Larsen}}, \bibinfo
  {author} {\bibfnamefont {T.~A.}\ \bibnamefont {Cochran}}, \bibinfo {author}
  {\bibfnamefont {M.}~\bibnamefont {Medarde}}, \bibinfo {author} {\bibfnamefont
  {V.}~\bibnamefont {Por{\'e}e}}, \bibinfo {author} {\bibfnamefont
  {D.}~\bibnamefont {Das}}, \bibinfo {author} {\bibfnamefont {R.}~\bibnamefont
  {Gupta}}, \bibinfo {author} {\bibfnamefont {C.~N.}\ \bibnamefont {Wang}},
  \bibinfo {author} {\bibfnamefont {J.}~\bibnamefont {Chang}}, \bibinfo
  {author} {\bibfnamefont {Z.~Q.}\ \bibnamefont {Wang}}, \bibinfo {author}
  {\bibfnamefont {R.}~\bibnamefont {Khasanov}}, \bibinfo {author}
  {\bibfnamefont {T.}~\bibnamefont {Neupert}}, \bibinfo {author} {\bibfnamefont
  {A.}~\bibnamefont {Amato}}, \bibinfo {author} {\bibfnamefont
  {L.}~\bibnamefont {Liborio}}, \bibinfo {author} {\bibfnamefont
  {S.}~\bibnamefont {Jia}}, \bibinfo {author} {\bibfnamefont {M.~Z.}\
  \bibnamefont {Hasan}}, \bibinfo {author} {\bibfnamefont {H.}~\bibnamefont
  {Luetkens}},\ and\ \bibinfo {author} {\bibfnamefont {Z.}~\bibnamefont
  {Guguchia}},\ }\bibfield  {title} {\bibinfo {title} {Low-temperature magnetic
  crossover in the topological kagome magnet tbmn$_6$sn$_6$},\ }\href
  {https://doi.org/10.1038/s42005-022-00885-4} {\bibfield  {journal} {\bibinfo
  {journal} {Communications Physics}\ }\textbf {\bibinfo {volume} {5}},\
  \bibinfo {pages} {107} (\bibinfo {year} {2022})}\BibitemShut {NoStop}%
\bibitem [{\citenamefont {Jones}\ \emph {et~al.}(2022)\citenamefont {Jones},
  \citenamefont {Das}, \citenamefont {Bhandari}, \citenamefont {Siegfried},
  \citenamefont {Ghimire}, \citenamefont {Tsirkin}, \citenamefont {Mazin},\
  and\ \citenamefont {Ghimire}}]{GMU}%
  \BibitemOpen
  \bibfield  {author} {\bibinfo {author} {\bibfnamefont {D.~C.}\ \bibnamefont
  {Jones}}, \bibinfo {author} {\bibfnamefont {S.}~\bibnamefont {Das}}, \bibinfo
  {author} {\bibfnamefont {H.}~\bibnamefont {Bhandari}}, \bibinfo {author}
  {\bibfnamefont {P.}~\bibnamefont {Siegfried}}, \bibinfo {author}
  {\bibfnamefont {M.~P.}\ \bibnamefont {Ghimire}}, \bibinfo {author}
  {\bibfnamefont {S.}~\bibnamefont {Tsirkin}}, \bibinfo {author} {\bibfnamefont
  {I.~I.}\ \bibnamefont {Mazin}},\ and\ \bibinfo {author} {\bibfnamefont
  {N.~J.}\ \bibnamefont {Ghimire}},\ }\href@noop {} {\bibinfo {title} {{Origin
  of spin reorientation and intrinsic anomalous Hall effect in the kagome
  ferrimagnet TbMn$_6$Sn$_6$}}} (\bibinfo {year} {2022}),\ \Eprint
  {https://arxiv.org/abs/2203.17246} {arXiv:2203.17246} \BibitemShut {NoStop}%
\bibitem [{\citenamefont {Blaha}\ \emph {et~al.}(2018)\citenamefont {Blaha},
  \citenamefont {Schwarz}, \citenamefont {Madsen}, \citenamefont {Kvasnicka},
  \citenamefont {Luitz}, \citenamefont {Laskowski}, \citenamefont {Tran},\ and\
  \citenamefont {Marks}}]{WIEN2k}%
  \BibitemOpen
  \bibfield  {author} {\bibinfo {author} {\bibfnamefont {P.}~\bibnamefont
  {Blaha}}, \bibinfo {author} {\bibfnamefont {K.}~\bibnamefont {Schwarz}},
  \bibinfo {author} {\bibfnamefont {G.~K.~H.}\ \bibnamefont {Madsen}}, \bibinfo
  {author} {\bibfnamefont {D.}~\bibnamefont {Kvasnicka}}, \bibinfo {author}
  {\bibfnamefont {J.}~\bibnamefont {Luitz}}, \bibinfo {author} {\bibfnamefont
  {R.}~\bibnamefont {Laskowski}}, \bibinfo {author} {\bibfnamefont
  {F.}~\bibnamefont {Tran}},\ and\ \bibinfo {author} {\bibfnamefont {L.~D.}\
  \bibnamefont {Marks}},\ }\href@noop {} {\emph {\bibinfo {title} {{WIEN2k: An
  Augmented Plane Wave plus Local Orbitals Program for Calculating Crystal
  Properties}}}}\ (\bibinfo  {publisher} {Vienna University of Technology},\
  \bibinfo {address} {Austria},\ \bibinfo {year} {2018})\BibitemShut {NoStop}%
\bibitem [{\citenamefont {Perdew}\ \emph {et~al.}(1996)\citenamefont {Perdew},
  \citenamefont {Burke},\ and\ \citenamefont {Ernzerhof}}]{perdew1996}%
  \BibitemOpen
  \bibfield  {author} {\bibinfo {author} {\bibfnamefont {J.~P.}\ \bibnamefont
  {Perdew}}, \bibinfo {author} {\bibfnamefont {K.}~\bibnamefont {Burke}},\ and\
  \bibinfo {author} {\bibfnamefont {M.}~\bibnamefont {Ernzerhof}},\ }\bibfield
  {title} {\bibinfo {title} {Generalized gradient approximation made simple},\
  }\href {https://doi.org/10.1103/PhysRevLett.77.3865} {\bibfield  {journal}
  {\bibinfo  {journal} {Phys. Rev. Lett.}\ }\textbf {\bibinfo {volume} {77}},\
  \bibinfo {pages} {3865} (\bibinfo {year} {1996})}\BibitemShut {NoStop}%
\bibitem [{\citenamefont {{van Schilfgaarde}}\ \emph
  {et~al.}(2006)\citenamefont {{van Schilfgaarde}}, \citenamefont {Kotani},\
  and\ \citenamefont {Faleev}}]{van-schilfgaarde2006prl}%
  \BibitemOpen
  \bibfield  {author} {\bibinfo {author} {\bibfnamefont {M.}~\bibnamefont {{van
  Schilfgaarde}}}, \bibinfo {author} {\bibfnamefont {T.}~\bibnamefont
  {Kotani}},\ and\ \bibinfo {author} {\bibfnamefont {S.}~\bibnamefont
  {Faleev}},\ }\bibfield  {title} {\bibinfo {title} {Quasiparticle
  self-consistent {$GW$} theory},\ }\href
  {https://doi.org/10.1103/PhysRevLett.96.226402} {\bibfield  {journal}
  {\bibinfo  {journal} {Phys. Rev. Lett.}\ }\textbf {\bibinfo {volume} {96}},\
  \bibinfo {pages} {226402} (\bibinfo {year} {2006})}\BibitemShut {NoStop}%
\bibitem [{\citenamefont {Kotani}\ \emph {et~al.}(2007)\citenamefont {Kotani},
  \citenamefont {van Schilfgaarde},\ and\ \citenamefont
  {Faleev}}]{kotani2007prb}%
  \BibitemOpen
  \bibfield  {author} {\bibinfo {author} {\bibfnamefont {T.}~\bibnamefont
  {Kotani}}, \bibinfo {author} {\bibfnamefont {M.}~\bibnamefont {van
  Schilfgaarde}},\ and\ \bibinfo {author} {\bibfnamefont {S.~V.}\ \bibnamefont
  {Faleev}},\ }\bibfield  {title} {\bibinfo {title} {Quasiparticle
  self-consistent {$GW$} method: A basis for the independent-particle
  approximation},\ }\href {https://doi.org/10.1103/PhysRevB.76.165106}
  {\bibfield  {journal} {\bibinfo  {journal} {Phys. Rev. B}\ }\textbf {\bibinfo
  {volume} {76}},\ \bibinfo {pages} {165106} (\bibinfo {year}
  {2007})}\BibitemShut {NoStop}%
\bibitem [{\citenamefont {Hedin}(1965)}]{hedin1965pr}%
  \BibitemOpen
  \bibfield  {author} {\bibinfo {author} {\bibfnamefont {L.}~\bibnamefont
  {Hedin}},\ }\bibfield  {title} {\bibinfo {title} {{New Method for Calculating
  the One-Particle Green's Function with Application to the Electron-Gas
  Problem}},\ }\href {https://doi.org/10.1103/PhysRev.139.A796} {\bibfield
  {journal} {\bibinfo  {journal} {Phys. Rev.}\ }\textbf {\bibinfo {volume}
  {139}},\ \bibinfo {pages} {A796} (\bibinfo {year} {1965})}\BibitemShut
  {NoStop}%
\bibitem [{\citenamefont {Kutepov}(2017)}]{kutepov2017prb}%
  \BibitemOpen
  \bibfield  {author} {\bibinfo {author} {\bibfnamefont {A.~L.}\ \bibnamefont
  {Kutepov}},\ }\bibfield  {title} {\bibinfo {title} {Self-consistent solution
  of hedin's equations: Semiconductors and insulators},\ }\href
  {https://doi.org/10.1103/PhysRevB.95.195120} {\bibfield  {journal} {\bibinfo
  {journal} {Phys. Rev. B}\ }\textbf {\bibinfo {volume} {95}},\ \bibinfo
  {pages} {195120} (\bibinfo {year} {2017})}\BibitemShut {NoStop}%
\bibitem [{\citenamefont {Grumet}\ \emph {et~al.}(2018)\citenamefont {Grumet},
  \citenamefont {Liu}, \citenamefont {Kaltak}, \citenamefont {Klime\v{s}},\
  and\ \citenamefont {Kresse}}]{grumet2018prb}%
  \BibitemOpen
  \bibfield  {author} {\bibinfo {author} {\bibfnamefont {M.}~\bibnamefont
  {Grumet}}, \bibinfo {author} {\bibfnamefont {P.}~\bibnamefont {Liu}},
  \bibinfo {author} {\bibfnamefont {M.}~\bibnamefont {Kaltak}}, \bibinfo
  {author} {\bibfnamefont {J.}~\bibnamefont {Klime\v{s}}},\ and\ \bibinfo
  {author} {\bibfnamefont {G.}~\bibnamefont {Kresse}},\ }\bibfield  {title}
  {\bibinfo {title} {Beyond the quasiparticle approximation: Fully
  self-consistent {$GW$} calculations},\ }\href
  {https://doi.org/10.1103/PhysRevB.98.155143} {\bibfield  {journal} {\bibinfo
  {journal} {Phys. Rev. B}\ }\textbf {\bibinfo {volume} {98}},\ \bibinfo
  {pages} {155143} (\bibinfo {year} {2018})}\BibitemShut {NoStop}%
\bibitem [{\citenamefont {Deguchi}\ \emph {et~al.}(2016)\citenamefont
  {Deguchi}, \citenamefont {Sato}, \citenamefont {Kino},\ and\ \citenamefont
  {Kotani}}]{deguchi2016jsap}%
  \BibitemOpen
  \bibfield  {author} {\bibinfo {author} {\bibfnamefont {D.}~\bibnamefont
  {Deguchi}}, \bibinfo {author} {\bibfnamefont {K.}~\bibnamefont {Sato}},
  \bibinfo {author} {\bibfnamefont {H.}~\bibnamefont {Kino}},\ and\ \bibinfo
  {author} {\bibfnamefont {T.}~\bibnamefont {Kotani}},\ }\bibfield  {title}
  {\bibinfo {title} {Accurate energy bands calculated by the hybrid
  quasiparticle self-consistent {$GW$} method implemented in the ecalj
  package},\ }\href {https://doi.org/10.7567/JJAP.55.051201} {\bibfield
  {journal} {\bibinfo  {journal} {Japanese Journal of Applied Physics}\
  }\textbf {\bibinfo {volume} {55}},\ \bibinfo {pages} {051201} (\bibinfo
  {year} {2016})}\BibitemShut {NoStop}%
\bibitem [{\citenamefont {Chantis}\ \emph {et~al.}(2007)\citenamefont
  {Chantis}, \citenamefont {van Schilfgaarde},\ and\ \citenamefont
  {Kotani}}]{chantis2007prb}%
  \BibitemOpen
  \bibfield  {author} {\bibinfo {author} {\bibfnamefont {A.~N.}\ \bibnamefont
  {Chantis}}, \bibinfo {author} {\bibfnamefont {M.}~\bibnamefont {van
  Schilfgaarde}},\ and\ \bibinfo {author} {\bibfnamefont {T.}~\bibnamefont
  {Kotani}},\ }\bibfield  {title} {\bibinfo {title} {Quasiparticle
  self-consistent {$GW$} method applied to localized $4f$ electron systems},\
  }\href {https://doi.org/10.1103/PhysRevB.76.165126} {\bibfield  {journal}
  {\bibinfo  {journal} {Phys. Rev. B}\ }\textbf {\bibinfo {volume} {76}},\
  \bibinfo {pages} {165126} (\bibinfo {year} {2007})}\BibitemShut {NoStop}%
\bibitem [{\citenamefont {Chantis}\ \emph {et~al.}(2008)\citenamefont
  {Chantis}, \citenamefont {Albers}, \citenamefont {Jones}, \citenamefont {van
  Schilfgaarde},\ and\ \citenamefont {Kotani}}]{chantis2008prb}%
  \BibitemOpen
  \bibfield  {author} {\bibinfo {author} {\bibfnamefont {A.~N.}\ \bibnamefont
  {Chantis}}, \bibinfo {author} {\bibfnamefont {R.~C.}\ \bibnamefont {Albers}},
  \bibinfo {author} {\bibfnamefont {M.~D.}\ \bibnamefont {Jones}}, \bibinfo
  {author} {\bibfnamefont {M.}~\bibnamefont {van Schilfgaarde}},\ and\ \bibinfo
  {author} {\bibfnamefont {T.}~\bibnamefont {Kotani}},\ }\bibfield  {title}
  {\bibinfo {title} {Many-body electronic structure of metallic
  $\ensuremath{\alpha}$-uranium},\ }\href
  {https://doi.org/10.1103/PhysRevB.78.081101} {\bibfield  {journal} {\bibinfo
  {journal} {Phys. Rev. B}\ }\textbf {\bibinfo {volume} {78}},\ \bibinfo
  {pages} {081101} (\bibinfo {year} {2008})}\BibitemShut {NoStop}%
\bibitem [{\citenamefont {Lee}\ \emph {et~al.}(2020)\citenamefont {Lee},
  \citenamefont {Kotani},\ and\ \citenamefont {Ke}}]{lee2020prbr}%
  \BibitemOpen
  \bibfield  {author} {\bibinfo {author} {\bibfnamefont {Y.}~\bibnamefont
  {Lee}}, \bibinfo {author} {\bibfnamefont {T.}~\bibnamefont {Kotani}},\ and\
  \bibinfo {author} {\bibfnamefont {L.}~\bibnamefont {Ke}},\ }\bibfield
  {title} {\bibinfo {title} {{Role of nonlocality in exchange correlation for
  magnetic two-dimensional van der Waals materials}},\ }\href
  {https://doi.org/10.1103/PhysRevB.101.241409} {\bibfield  {journal} {\bibinfo
   {journal} {Phys. Rev. B}\ }\textbf {\bibinfo {volume} {101}},\ \bibinfo
  {pages} {241409} (\bibinfo {year} {2020})}\BibitemShut {NoStop}%
\bibitem [{\citenamefont {Ke}\ and\ \citenamefont
  {Katsnelson}(2021)}]{ke2021ncm}%
  \BibitemOpen
  \bibfield  {author} {\bibinfo {author} {\bibfnamefont {L.}~\bibnamefont
  {Ke}}\ and\ \bibinfo {author} {\bibfnamefont {M.~I.}\ \bibnamefont
  {Katsnelson}},\ }\bibfield  {title} {\bibinfo {title} {{Electron correlation
  effects on exchange interactions and spin excitations in 2D van der Waals
  materials}},\ }\href {https://doi.org/10.1038/s41524-020-00469-2} {\bibfield
  {journal} {\bibinfo  {journal} {npj Computational Materials}\ }\textbf
  {\bibinfo {volume} {7}},\ \bibinfo {pages} {1} (\bibinfo {year}
  {2021})}\BibitemShut {NoStop}%
\bibitem [{\citenamefont {Ke}\ \emph {et~al.}(2013)\citenamefont {Ke},
  \citenamefont {Belashchenko}, \citenamefont {van Schilfgaarde}, \citenamefont
  {Kotani},\ and\ \citenamefont {Antropov}}]{ke2013prb}%
  \BibitemOpen
  \bibfield  {author} {\bibinfo {author} {\bibfnamefont {L.}~\bibnamefont
  {Ke}}, \bibinfo {author} {\bibfnamefont {K.~D.}\ \bibnamefont
  {Belashchenko}}, \bibinfo {author} {\bibfnamefont {M.}~\bibnamefont {van
  Schilfgaarde}}, \bibinfo {author} {\bibfnamefont {T.}~\bibnamefont
  {Kotani}},\ and\ \bibinfo {author} {\bibfnamefont {V.~P.}\ \bibnamefont
  {Antropov}},\ }\bibfield  {title} {\bibinfo {title} {Effects of alloying and
  strain on the magnetic properties of {Fe$_{16}$N$_2$}},\ }\href
  {https://doi.org/10.1103/PhysRevB.88.024404} {\bibfield  {journal} {\bibinfo
  {journal} {Phys. Rev. B}\ }\textbf {\bibinfo {volume} {88}},\ \bibinfo
  {pages} {024404} (\bibinfo {year} {2013})}\BibitemShut {NoStop}%
\bibitem [{\citenamefont {Kotani}(2014)}]{kotani2014jpsj}%
  \BibitemOpen
  \bibfield  {author} {\bibinfo {author} {\bibfnamefont {T.}~\bibnamefont
  {Kotani}},\ }\bibfield  {title} {\bibinfo {title} {Quasiparticle
  self-consistent {$GW$} method based on the augmented plane-wave and
  muffin-tin orbital method},\ }\href@noop {} {\bibfield  {journal} {\bibinfo
  {journal} {Journal of the Physical Society of Japan}\ }\textbf {\bibinfo
  {volume} {83}},\ \bibinfo {pages} {094711} (\bibinfo {year}
  {2014})}\BibitemShut {NoStop}%
\bibitem [{\citenamefont {Malaman}\ \emph {et~al.}(1999)\citenamefont
  {Malaman}, \citenamefont {Venturini}, \citenamefont {Welter}, \citenamefont
  {Sanchez}, \citenamefont {Vulliet},\ and\ \citenamefont
  {Ressouche}}]{malaman1999jmmm}%
  \BibitemOpen
  \bibfield  {author} {\bibinfo {author} {\bibfnamefont {B.}~\bibnamefont
  {Malaman}}, \bibinfo {author} {\bibfnamefont {G.}~\bibnamefont {Venturini}},
  \bibinfo {author} {\bibfnamefont {R.}~\bibnamefont {Welter}}, \bibinfo
  {author} {\bibfnamefont {J.}~\bibnamefont {Sanchez}}, \bibinfo {author}
  {\bibfnamefont {P.}~\bibnamefont {Vulliet}},\ and\ \bibinfo {author}
  {\bibfnamefont {E.}~\bibnamefont {Ressouche}},\ }\bibfield  {title} {\bibinfo
  {title} {Magnetic properties of {$R$Mn$_6$Sn$_6$} {($R=$Gd--Er) compounds
  from neutron diffraction and M\"ossbauer measurements}},\ }\href
  {https://doi.org/https://doi.org/10.1016/S0304-8853(99)00300-5} {\bibfield
  {journal} {\bibinfo  {journal} {Journal of Magnetism and Magnetic Materials}\
  }\textbf {\bibinfo {volume} {202}},\ \bibinfo {pages} {519} (\bibinfo {year}
  {1999})}\BibitemShut {NoStop}%
\bibitem [{\citenamefont {Clatterbuck}\ and\ \citenamefont
  {Gschneidner}(1999)}]{clatterbuck1999jmmm}%
  \BibitemOpen
  \bibfield  {author} {\bibinfo {author} {\bibfnamefont {D.}~\bibnamefont
  {Clatterbuck}}\ and\ \bibinfo {author} {\bibfnamefont {K.}~\bibnamefont
  {Gschneidner}},\ }\bibfield  {title} {\bibinfo {title} {Magnetic properties
  of {$R$Mn$_6$Sn$_6$ ($R=$Tb, Ho, Er, Tm, Lu)} single crystals},\ }\href
  {https://doi.org/https://doi.org/10.1016/S0304-8853(99)00571-5} {\bibfield
  {journal} {\bibinfo  {journal} {Journal of Magnetism and Magnetic Materials}\
  }\textbf {\bibinfo {volume} {207}},\ \bibinfo {pages} {78} (\bibinfo {year}
  {1999})}\BibitemShut {NoStop}%
\bibitem [{\citenamefont {Kimura}\ \emph {et~al.}(2006)\citenamefont {Kimura},
  \citenamefont {Matsuo}, \citenamefont {Yoshii}, \citenamefont {Kindo},
  \citenamefont {Zhang}, \citenamefont {Brück}, \citenamefont {Buschow},
  \citenamefont {{de Boer}}, \citenamefont {Lefèvre},\ and\ \citenamefont
  {Venturini}}]{kimura2006jac}%
  \BibitemOpen
  \bibfield  {author} {\bibinfo {author} {\bibfnamefont {S.}~\bibnamefont
  {Kimura}}, \bibinfo {author} {\bibfnamefont {A.}~\bibnamefont {Matsuo}},
  \bibinfo {author} {\bibfnamefont {S.}~\bibnamefont {Yoshii}}, \bibinfo
  {author} {\bibfnamefont {K.}~\bibnamefont {Kindo}}, \bibinfo {author}
  {\bibfnamefont {L.}~\bibnamefont {Zhang}}, \bibinfo {author} {\bibfnamefont
  {E.}~\bibnamefont {Brück}}, \bibinfo {author} {\bibfnamefont
  {K.}~\bibnamefont {Buschow}}, \bibinfo {author} {\bibfnamefont
  {F.}~\bibnamefont {{de Boer}}}, \bibinfo {author} {\bibfnamefont
  {C.}~\bibnamefont {Lefèvre}},\ and\ \bibinfo {author} {\bibfnamefont
  {G.}~\bibnamefont {Venturini}},\ }\bibfield  {title} {\bibinfo {title}
  {High-field magnetization of {$R$Mn$_6$Sn$_6$} compounds with {$R$ = Gd, Tb,
  Dy and Ho}},\ }\href
  {https://doi.org/https://doi.org/10.1016/j.jallcom.2005.04.087} {\bibfield
  {journal} {\bibinfo  {journal} {Journal of Alloys and Compounds}\ }\textbf
  {\bibinfo {volume} {408-412}},\ \bibinfo {pages} {169} (\bibinfo {year}
  {2006})},\ \bibinfo {note} {proceedings of Rare Earths'04 in Nara,
  Japan}\BibitemShut {NoStop}%
\bibitem [{\citenamefont {{Tabatabai Yazdi}}\ \emph {et~al.}(2012)\citenamefont
  {{Tabatabai Yazdi}}, \citenamefont {Tajabor}, \citenamefont {{Rezaee
  Roknabadi}}, \citenamefont {Behdani},\ and\ \citenamefont
  {Pourarian}}]{yazdi2012jmmm}%
  \BibitemOpen
  \bibfield  {author} {\bibinfo {author} {\bibfnamefont {S.}~\bibnamefont
  {{Tabatabai Yazdi}}}, \bibinfo {author} {\bibfnamefont {N.}~\bibnamefont
  {Tajabor}}, \bibinfo {author} {\bibfnamefont {M.}~\bibnamefont {{Rezaee
  Roknabadi}}}, \bibinfo {author} {\bibfnamefont {M.}~\bibnamefont {Behdani}},\
  and\ \bibinfo {author} {\bibfnamefont {F.}~\bibnamefont {Pourarian}},\
  }\bibfield  {title} {\bibinfo {title} {Magnetoelastic properties of
  {ErMn$_6$Sn$_6$} intermetallic compound},\ }\href
  {https://doi.org/https://doi.org/10.1016/j.jmmm.2011.09.004} {\bibfield
  {journal} {\bibinfo  {journal} {Journal of Magnetism and Magnetic Materials}\
  }\textbf {\bibinfo {volume} {324}},\ \bibinfo {pages} {723} (\bibinfo {year}
  {2012})}\BibitemShut {NoStop}%
\bibitem [{\citenamefont {Verhoef}\ \emph {et~al.}(1990)\citenamefont
  {Verhoef}, \citenamefont {Quang}, \citenamefont {Franse},\ and\ \citenamefont
  {Radwański}}]{verhoef1990jmmm}%
  \BibitemOpen
  \bibfield  {author} {\bibinfo {author} {\bibfnamefont {R.}~\bibnamefont
  {Verhoef}}, \bibinfo {author} {\bibfnamefont {P.}~\bibnamefont {Quang}},
  \bibinfo {author} {\bibfnamefont {J.}~\bibnamefont {Franse}},\ and\ \bibinfo
  {author} {\bibfnamefont {R.}~\bibnamefont {Radwański}},\ }\bibfield  {title}
  {\bibinfo {title} {{The strength of the $R$-T exchange coupling in
  $R_2$Fe$_{14}$B compounds}},\ }\href
  {https://doi.org/https://doi.org/10.1016/0304-8853(90)90460-8} {\bibfield
  {journal} {\bibinfo  {journal} {Journal of Magnetism and Magnetic Materials}\
  }\textbf {\bibinfo {volume} {83}},\ \bibinfo {pages} {139} (\bibinfo {year}
  {1990})}\BibitemShut {NoStop}%
\bibitem [{\citenamefont {Liu}\ \emph {et~al.}(1994)\citenamefont {Liu},
  \citenamefont {{de Boer}}, \citenamefont {{de Châtel}}, \citenamefont
  {Coehoorn},\ and\ \citenamefont {Buschow}}]{liu1994jmmm}%
  \BibitemOpen
  \bibfield  {author} {\bibinfo {author} {\bibfnamefont {J.}~\bibnamefont
  {Liu}}, \bibinfo {author} {\bibfnamefont {F.}~\bibnamefont {{de Boer}}},
  \bibinfo {author} {\bibfnamefont {P.}~\bibnamefont {{de Châtel}}}, \bibinfo
  {author} {\bibfnamefont {R.}~\bibnamefont {Coehoorn}},\ and\ \bibinfo
  {author} {\bibfnamefont {K.}~\bibnamefont {Buschow}},\ }\bibfield  {title}
  {\bibinfo {title} {On the $4f$-$3d$ exchange interaction in intermetallic
  compounds},\ }\href
  {https://doi.org/https://doi.org/10.1016/0304-8853(94)90310-7} {\bibfield
  {journal} {\bibinfo  {journal} {Journal of Magnetism and Magnetic Materials}\
  }\textbf {\bibinfo {volume} {132}},\ \bibinfo {pages} {159} (\bibinfo {year}
  {1994})}\BibitemShut {NoStop}%
\bibitem [{\citenamefont {Duc}\ \emph {et~al.}(1993)\citenamefont {Duc},
  \citenamefont {Hien}, \citenamefont {Givord}, \citenamefont {Franse},\ and\
  \citenamefont {{de Boer}}}]{duc1993jmmm}%
  \BibitemOpen
  \bibfield  {author} {\bibinfo {author} {\bibfnamefont {N.}~\bibnamefont
  {Duc}}, \bibinfo {author} {\bibfnamefont {T.}~\bibnamefont {Hien}}, \bibinfo
  {author} {\bibfnamefont {D.}~\bibnamefont {Givord}}, \bibinfo {author}
  {\bibfnamefont {J.}~\bibnamefont {Franse}},\ and\ \bibinfo {author}
  {\bibfnamefont {F.}~\bibnamefont {{de Boer}}},\ }\bibfield  {title} {\bibinfo
  {title} {{Exchange interactions in rare earth-transition metal compounds}},\
  }\href {https://doi.org/https://doi.org/10.1016/0304-8853(93)90131-K}
  {\bibfield  {journal} {\bibinfo  {journal} {Journal of Magnetism and Magnetic
  Materials}\ }\textbf {\bibinfo {volume} {124}},\ \bibinfo {pages} {305}
  (\bibinfo {year} {1993})}\BibitemShut {NoStop}%
\bibitem [{\citenamefont {Radwa\'{n}ski}(1986)}]{radwanski1986pssb}%
  \BibitemOpen
  \bibfield  {author} {\bibinfo {author} {\bibfnamefont {R.~J.}\ \bibnamefont
  {Radwa\'{n}ski}},\ }\bibfield  {title} {\bibinfo {title} {{The
  Intersublattice Molecular Fields in the Rare Earth-Cobalt Intermetallics}},\
  }\href {https://doi.org/https://doi.org/10.1002/pssb.2221370210} {\bibfield
  {journal} {\bibinfo  {journal} {physica status solidi (b)}\ }\textbf
  {\bibinfo {volume} {137}},\ \bibinfo {pages} {487} (\bibinfo {year}
  {1986})}\BibitemShut {NoStop}%
\bibitem [{\citenamefont {Belorizky}\ \emph {et~al.}(1987)\citenamefont
  {Belorizky}, \citenamefont {Fremy}, \citenamefont {Gavigan}, \citenamefont
  {Givord},\ and\ \citenamefont {Li}}]{belorizky1987jap}%
  \BibitemOpen
  \bibfield  {author} {\bibinfo {author} {\bibfnamefont {E.}~\bibnamefont
  {Belorizky}}, \bibinfo {author} {\bibfnamefont {M.~A.}\ \bibnamefont
  {Fremy}}, \bibinfo {author} {\bibfnamefont {J.~P.}\ \bibnamefont {Gavigan}},
  \bibinfo {author} {\bibfnamefont {D.}~\bibnamefont {Givord}},\ and\ \bibinfo
  {author} {\bibfnamefont {H.~S.}\ \bibnamefont {Li}},\ }\bibfield  {title}
  {\bibinfo {title} {{Evidence in rare-earth (R)-transition metal (M)
  intermetallics for a systematic dependence of R-M exchange interactions on
  the nature of the R atom}},\ }\href {https://doi.org/10.1063/1.338573}
  {\bibfield  {journal} {\bibinfo  {journal} {Journal of Applied Physics}\
  }\textbf {\bibinfo {volume} {61}},\ \bibinfo {pages} {3971} (\bibinfo {year}
  {1987})}\BibitemShut {NoStop}%
\bibitem [{\citenamefont {Mermin}\ and\ \citenamefont
  {Wagner}(1966)}]{mermin1966prl}%
  \BibitemOpen
  \bibfield  {author} {\bibinfo {author} {\bibfnamefont {N.~D.}\ \bibnamefont
  {Mermin}}\ and\ \bibinfo {author} {\bibfnamefont {H.}~\bibnamefont
  {Wagner}},\ }\bibfield  {title} {\bibinfo {title} {{Absence of Ferromagnetism
  or Antiferromagnetism in One- or Two-Dimensional Isotropic Heisenberg
  Models}},\ }\href {https://doi.org/10.1103/PhysRevLett.17.1133} {\bibfield
  {journal} {\bibinfo  {journal} {Phys. Rev. Lett.}\ }\textbf {\bibinfo
  {volume} {17}},\ \bibinfo {pages} {1133} (\bibinfo {year}
  {1966})}\BibitemShut {NoStop}%
\bibitem [{\citenamefont {Mkhitaryan}\ and\ \citenamefont
  {Ke}(2021)}]{mkhitaryan2021prb}%
  \BibitemOpen
  \bibfield  {author} {\bibinfo {author} {\bibfnamefont {V.~V.}\ \bibnamefont
  {Mkhitaryan}}\ and\ \bibinfo {author} {\bibfnamefont {L.}~\bibnamefont
  {Ke}},\ }\bibfield  {title} {\bibinfo {title} {Self-consistently renormalized
  spin-wave theory of layered ferromagnets on the honeycomb lattice},\ }\href
  {https://doi.org/10.1103/PhysRevB.104.064435} {\bibfield  {journal} {\bibinfo
   {journal} {Phys. Rev. B}\ }\textbf {\bibinfo {volume} {104}},\ \bibinfo
  {pages} {064435} (\bibinfo {year} {2021})}\BibitemShut {NoStop}%
\bibitem [{\citenamefont {Kirchmayr}\ and\ \citenamefont
  {Poldy}(1978)}]{kirchmayr1978jmmm}%
  \BibitemOpen
  \bibfield  {author} {\bibinfo {author} {\bibfnamefont {H.}~\bibnamefont
  {Kirchmayr}}\ and\ \bibinfo {author} {\bibfnamefont {C.}~\bibnamefont
  {Poldy}},\ }\bibfield  {title} {\bibinfo {title} {Magnetism in rare earth-3d
  intermetallics},\ }\href
  {https://doi.org/https://doi.org/10.1016/0304-8853(78)90073-2} {\bibfield
  {journal} {\bibinfo  {journal} {Journal of Magnetism and Magnetic Materials}\
  }\textbf {\bibinfo {volume} {8}},\ \bibinfo {pages} {1} (\bibinfo {year}
  {1978})}\BibitemShut {NoStop}%
\bibitem [{\citenamefont {Ke}(2019)}]{ke2019prb}%
  \BibitemOpen
  \bibfield  {author} {\bibinfo {author} {\bibfnamefont {L.}~\bibnamefont
  {Ke}},\ }\bibfield  {title} {\bibinfo {title} {Intersublattice
  magnetocrystalline anisotropy using a realistic tight-binding method based on
  maximally localized {Wannier} functions},\ }\href
  {https://doi.org/10.1103/PhysRevB.99.054418} {\bibfield  {journal} {\bibinfo
  {journal} {Phys. Rev. B}\ }\textbf {\bibinfo {volume} {99}},\ \bibinfo
  {pages} {054418} (\bibinfo {year} {2019})}\BibitemShut {NoStop}%
\bibitem [{\citenamefont {Li}\ \emph {et~al.}(2020)\citenamefont {Li},
  \citenamefont {Yan}, \citenamefont {Pajerowski}, \citenamefont {Gordon},
  \citenamefont {Nedi\'{c}}, \citenamefont {Sizyuk}, \citenamefont {Ke},
  \citenamefont {Orth}, \citenamefont {Vaknin},\ and\ \citenamefont
  {McQueeney}}]{li2020prl}%
  \BibitemOpen
  \bibfield  {author} {\bibinfo {author} {\bibfnamefont {B.}~\bibnamefont
  {Li}}, \bibinfo {author} {\bibfnamefont {J.-Q.}\ \bibnamefont {Yan}},
  \bibinfo {author} {\bibfnamefont {D.~M.}\ \bibnamefont {Pajerowski}},
  \bibinfo {author} {\bibfnamefont {E.}~\bibnamefont {Gordon}}, \bibinfo
  {author} {\bibfnamefont {A.-M.}\ \bibnamefont {Nedi\'{c}}}, \bibinfo {author}
  {\bibfnamefont {Y.}~\bibnamefont {Sizyuk}}, \bibinfo {author} {\bibfnamefont
  {L.}~\bibnamefont {Ke}}, \bibinfo {author} {\bibfnamefont {P.~P.}\
  \bibnamefont {Orth}}, \bibinfo {author} {\bibfnamefont {D.}~\bibnamefont
  {Vaknin}},\ and\ \bibinfo {author} {\bibfnamefont {R.~J.}\ \bibnamefont
  {McQueeney}},\ }\bibfield  {title} {\bibinfo {title} {{Competing Magnetic
  Interactions in the Antiferromagnetic Topological Insulator
  MnBi$_2$Te$_4$}},\ }\href {https://doi.org/10.1103/PhysRevLett.124.167204}
  {\bibfield  {journal} {\bibinfo  {journal} {Phys. Rev. Lett.}\ }\textbf
  {\bibinfo {volume} {124}},\ \bibinfo {pages} {167204} (\bibinfo {year}
  {2020})}\BibitemShut {NoStop}%
\bibitem [{\citenamefont {Gordon}\ \emph {et~al.}(2021)\citenamefont {Gordon},
  \citenamefont {Mkhitaryan}, \citenamefont {Zhao}, \citenamefont {Lee},\ and\
  \citenamefont {Ke}}]{gordon2021jpda}%
  \BibitemOpen
  \bibfield  {author} {\bibinfo {author} {\bibfnamefont {E.}~\bibnamefont
  {Gordon}}, \bibinfo {author} {\bibfnamefont {V.}~\bibnamefont {Mkhitaryan}},
  \bibinfo {author} {\bibfnamefont {H.}~\bibnamefont {Zhao}}, \bibinfo {author}
  {\bibfnamefont {Y.}~\bibnamefont {Lee}},\ and\ \bibinfo {author}
  {\bibfnamefont {L.}~\bibnamefont {Ke}},\ }\bibfield  {title} {\bibinfo
  {title} {Magnetic interactions and spin excitations in van der {Waals
  ferromagnet VI$_3$}},\ }\href
  {http://iopscience.iop.org/article/10.1088/1361-6463/ac1bd3} {\bibfield
  {journal} {\bibinfo  {journal} {Journal of Physics D: Applied Physics}\
  }\textbf {\bibinfo {volume} {54}},\ \bibinfo {pages} {464001} (\bibinfo
  {year} {2021})}\BibitemShut {NoStop}%
\bibitem [{\citenamefont {Ghimire}\ \emph {et~al.}(2020)\citenamefont
  {Ghimire}, \citenamefont {Dally}, \citenamefont {Poudel}, \citenamefont
  {Jones}, \citenamefont {Michel}, \citenamefont {Magar}, \citenamefont
  {Bleuel}, \citenamefont {McGuire}, \citenamefont {Jiang}, \citenamefont
  {Mitchell}, \citenamefont {Lynn},\ and\ \citenamefont
  {Mazin}}]{nirmal2020sciadv}%
  \BibitemOpen
  \bibfield  {author} {\bibinfo {author} {\bibfnamefont {N.~J.}\ \bibnamefont
  {Ghimire}}, \bibinfo {author} {\bibfnamefont {R.~L.}\ \bibnamefont {Dally}},
  \bibinfo {author} {\bibfnamefont {L.}~\bibnamefont {Poudel}}, \bibinfo
  {author} {\bibfnamefont {D.~C.}\ \bibnamefont {Jones}}, \bibinfo {author}
  {\bibfnamefont {D.}~\bibnamefont {Michel}}, \bibinfo {author} {\bibfnamefont
  {N.~T.}\ \bibnamefont {Magar}}, \bibinfo {author} {\bibfnamefont
  {M.}~\bibnamefont {Bleuel}}, \bibinfo {author} {\bibfnamefont {M.~A.}\
  \bibnamefont {McGuire}}, \bibinfo {author} {\bibfnamefont {J.~S.}\
  \bibnamefont {Jiang}}, \bibinfo {author} {\bibfnamefont {J.~F.}\ \bibnamefont
  {Mitchell}}, \bibinfo {author} {\bibfnamefont {J.~W.}\ \bibnamefont {Lynn}},\
  and\ \bibinfo {author} {\bibfnamefont {I.~I.}\ \bibnamefont {Mazin}},\
  }\bibfield  {title} {\bibinfo {title} {{Competing magnetic phases and
  fluctuation-driven scalar spin chirality in the kagome metal
  YMn$_6$Sn$_6$}},\ }\href {https://doi.org/10.1126/sciadv.abe2680} {\bibfield
  {journal} {\bibinfo  {journal} {Science Advances}\ }\textbf {\bibinfo
  {volume} {6}},\ \bibinfo {pages} {eabe2680} (\bibinfo {year}
  {2020})}\BibitemShut {NoStop}%
\bibitem [{\citenamefont {Bethe}(1929)}]{bethe1929adp}%
  \BibitemOpen
  \bibfield  {author} {\bibinfo {author} {\bibfnamefont {H.}~\bibnamefont
  {Bethe}},\ }\bibfield  {title} {\bibinfo {title} {Termaufspaltung in
  kristallen},\ }\href@noop {} {\bibfield  {journal} {\bibinfo  {journal}
  {Annalen der Physik}\ }\textbf {\bibinfo {volume} {395}},\ \bibinfo {pages}
  {133} (\bibinfo {year} {1929})}\BibitemShut {NoStop}%
\bibitem [{\citenamefont {Ballhausen}(1962)}]{ballhausen1962book}%
  \BibitemOpen
  \bibfield  {author} {\bibinfo {author} {\bibfnamefont {C.~J.}\ \bibnamefont
  {Ballhausen}},\ }\href@noop {} {\emph {\bibinfo {title} {{Introduction to
  Ligand Field Theory}}}}\ (\bibinfo  {publisher} {McGraw-Hill, New York, NY},\
  \bibinfo {year} {1962})\BibitemShut {NoStop}%
\bibitem [{\citenamefont {Newman}\ and\ \citenamefont
  {Ng}(1989)}]{newman1989rpp}%
  \BibitemOpen
  \bibfield  {author} {\bibinfo {author} {\bibfnamefont {D.~J.}\ \bibnamefont
  {Newman}}\ and\ \bibinfo {author} {\bibfnamefont {B.}~\bibnamefont {Ng}},\
  }\bibfield  {title} {\bibinfo {title} {The superposition model of crystal
  fields},\ }\href {https://doi.org/10.1088/0034-4885/52/6/002} {\bibfield
  {journal} {\bibinfo  {journal} {Reports on Progress in Physics}\ }\textbf
  {\bibinfo {volume} {52}},\ \bibinfo {pages} {699} (\bibinfo {year}
  {1989})}\BibitemShut {NoStop}%
\bibitem [{\citenamefont {Hutchings}(1964)}]{hutchings1964ssp}%
  \BibitemOpen
  \bibfield  {author} {\bibinfo {author} {\bibfnamefont {M.}~\bibnamefont
  {Hutchings}},\ }\bibfield  {title} {\bibinfo {title} {{Point-Charge
  Calculations of Energy Levels of Magnetic Ions in Crystalline Electric
  Fields}}\ }(\bibinfo  {publisher} {Academic Press, San Diego, CA},\ \bibinfo
  {year} {1964})\ pp.\ \bibinfo {pages} {227--273}\BibitemShut {NoStop}%
\bibitem [{\citenamefont {Skomski}\ and\ \citenamefont
  {Coey}(1999)}]{skomski1999book}%
  \BibitemOpen
  \bibfield  {author} {\bibinfo {author} {\bibfnamefont {R.}~\bibnamefont
  {Skomski}}\ and\ \bibinfo {author} {\bibfnamefont {J.~M.~D.}\ \bibnamefont
  {Coey}},\ }\href@noop {} {\emph {\bibinfo {title} {Permanent magnetism}}}\
  (\bibinfo  {publisher} {Institute of Physics Publishing, Bristol, UK},\
  \bibinfo {year} {1999})\BibitemShut {NoStop}%
\bibitem [{\citenamefont {Skomski}(2008)}]{skomski2008book}%
  \BibitemOpen
  \bibfield  {author} {\bibinfo {author} {\bibfnamefont {R.}~\bibnamefont
  {Skomski}},\ }\href@noop {} {\emph {\bibinfo {title} {{Simple Models of
  Magnetism}}}},\ Oxford Graduate Texts\ (\bibinfo  {publisher} {University
  Press, Oxford},\ \bibinfo {year} {2008})\BibitemShut {NoStop}%
\bibitem [{\citenamefont {Taylor}\ and\ \citenamefont
  {Darby}(1972)}]{taylor1972book}%
  \BibitemOpen
  \bibfield  {author} {\bibinfo {author} {\bibfnamefont {K.~N.~R.}\
  \bibnamefont {Taylor}}\ and\ \bibinfo {author} {\bibfnamefont {M.~I.}\
  \bibnamefont {Darby}},\ }\href@noop {} {\emph {\bibinfo {title} {{Physics of
  Rare Earth Solids}}}}\ (\bibinfo  {publisher} {Chapman and Hall, London,
  UK},\ \bibinfo {year} {1972})\BibitemShut {NoStop}%
\bibitem [{sup()}]{supp}%
  \BibitemOpen
  \href@noop {} {}\bibinfo {note} {See Supplemental Material at [http link]
  for: the crystal structure of $\rmnsn$ that better illustrates the Mn kagome
  lattice; the atomic coordination origin of large high-order crystal-field
  parameters and anisotropy contributions in $\rmnsn$; the band structures
  projected on the surface BZ in $\rmnsn$ with $R=$ Gd, Tb, Dy, Ho, and Er; the
  band structures in $\tbmnsn$ calculated without and with spin-orbit coupling;
  the scalar-relativistic band structures in $\yymnsn$ calculated within QSGW;
  the effects of Sn SOC on anisotropy; the orbital characters of Dirac
  crossings near $\ef$; the detailed MAE analytical modeling using calculated
  CF energies; and the detailed tight-binding model of band structures of $d$
  orbitals on a kagome lattice. The Supplemental Material also contains
  Refs.~\cite{ke2016prbA, ke2016prb, marzari1997prb, mostofi2014cpc,
  souza2001prb, marzari2012rmp}.}\BibitemShut {Stop}%
\bibitem [{\citenamefont {Rosenberg}\ \emph {et~al.}(2022)\citenamefont
  {Rosenberg}, \citenamefont {DeStefano}, \citenamefont {Guo}, \citenamefont
  {Oh}, \citenamefont {Hashimoto}, \citenamefont {Lu}, \citenamefont
  {Birgeneau}, \citenamefont {Lee}, \citenamefont {Ke}, \citenamefont {Yi},\
  and\ \citenamefont {Chu}}]{rosenberg2022prb}%
  \BibitemOpen
  \bibfield  {author} {\bibinfo {author} {\bibfnamefont {E.}~\bibnamefont
  {Rosenberg}}, \bibinfo {author} {\bibfnamefont {J.~M.}\ \bibnamefont
  {DeStefano}}, \bibinfo {author} {\bibfnamefont {Y.}~\bibnamefont {Guo}},
  \bibinfo {author} {\bibfnamefont {J.~S.}\ \bibnamefont {Oh}}, \bibinfo
  {author} {\bibfnamefont {M.}~\bibnamefont {Hashimoto}}, \bibinfo {author}
  {\bibfnamefont {D.}~\bibnamefont {Lu}}, \bibinfo {author} {\bibfnamefont
  {R.~J.}\ \bibnamefont {Birgeneau}}, \bibinfo {author} {\bibfnamefont
  {Y.}~\bibnamefont {Lee}}, \bibinfo {author} {\bibfnamefont {L.}~\bibnamefont
  {Ke}}, \bibinfo {author} {\bibfnamefont {M.}~\bibnamefont {Yi}},\ and\
  \bibinfo {author} {\bibfnamefont {J.-H.}\ \bibnamefont {Chu}},\ }\bibfield
  {title} {\bibinfo {title} {Uniaxial ferromagnetism in the kagome metal
  {TbV$_{6}$Sn$_6$}},\ }\href {https://doi.org/10.1103/PhysRevB.106.115139}
  {\bibfield  {journal} {\bibinfo  {journal} {Phys. Rev. B}\ }\textbf {\bibinfo
  {volume} {106}},\ \bibinfo {pages} {115139} (\bibinfo {year}
  {2022})}\BibitemShut {NoStop}%
\bibitem [{\citenamefont {Haldane}(1988)}]{haldane1988prl}%
  \BibitemOpen
  \bibfield  {author} {\bibinfo {author} {\bibfnamefont {F.~D.~M.}\
  \bibnamefont {Haldane}},\ }\bibfield  {title} {\bibinfo {title} {Model for a
  quantum hall effect without {Landau} levels: Condensed-matter realization of
  the "parity anomaly"},\ }\href {https://doi.org/10.1103/PhysRevLett.61.2015}
  {\bibfield  {journal} {\bibinfo  {journal} {Phys. Rev. Lett.}\ }\textbf
  {\bibinfo {volume} {61}},\ \bibinfo {pages} {2015} (\bibinfo {year}
  {1988})}\BibitemShut {NoStop}%
\bibitem [{\citenamefont {hua Guo}\ and\ \citenamefont {bei
  Zhang}(2007)}]{guo2007jac}%
  \BibitemOpen
  \bibfield  {author} {\bibinfo {author} {\bibfnamefont {G.}~\bibnamefont {hua
  Guo}}\ and\ \bibinfo {author} {\bibfnamefont {H.}~\bibnamefont {bei Zhang}},\
  }\bibfield  {title} {\bibinfo {title} {Magnetocrystalline anisotropy and spin
  reorientation transition of {HoMn$_6$Sn$_6$} compound},\ }\href
  {https://doi.org/https://doi.org/10.1016/j.jallcom.2006.04.008} {\bibfield
  {journal} {\bibinfo  {journal} {Journal of Alloys and Compounds}\ }\textbf
  {\bibinfo {volume} {429}},\ \bibinfo {pages} {46} (\bibinfo {year}
  {2007})}\BibitemShut {NoStop}%
\bibitem [{\citenamefont {Ke}\ and\ \citenamefont {van
  Schilfgaarde}(2015)}]{ke2015prb}%
  \BibitemOpen
  \bibfield  {author} {\bibinfo {author} {\bibfnamefont {L.}~\bibnamefont
  {Ke}}\ and\ \bibinfo {author} {\bibfnamefont {M.}~\bibnamefont {van
  Schilfgaarde}},\ }\bibfield  {title} {\bibinfo {title} {Band-filling effect
  on magnetic anisotropy using a {G}reen's function method},\ }\href
  {https://doi.org/10.1103/PhysRevB.92.014423} {\bibfield  {journal} {\bibinfo
  {journal} {Phys. Rev. B}\ }\textbf {\bibinfo {volume} {92}},\ \bibinfo
  {pages} {014423} (\bibinfo {year} {2015})}\BibitemShut {NoStop}%
\bibitem [{\citenamefont {Ke}\ \emph {et~al.}(2016)\citenamefont {Ke},
  \citenamefont {Kukusta},\ and\ \citenamefont {Johnson}}]{ke2016prbA}%
  \BibitemOpen
  \bibfield  {author} {\bibinfo {author} {\bibfnamefont {L.}~\bibnamefont
  {Ke}}, \bibinfo {author} {\bibfnamefont {D.~A.}\ \bibnamefont {Kukusta}},\
  and\ \bibinfo {author} {\bibfnamefont {D.~D.}\ \bibnamefont {Johnson}},\
  }\bibfield  {title} {\bibinfo {title} {Origin of magnetic anisotropy in doped
  {Ce$_2$Co$_{17}$} alloys},\ }\href
  {https://doi.org/10.1103/PhysRevB.94.144429} {\bibfield  {journal} {\bibinfo
  {journal} {Phys. Rev. B}\ }\textbf {\bibinfo {volume} {94}},\ \bibinfo
  {pages} {144429} (\bibinfo {year} {2016})}\BibitemShut {NoStop}%
\bibitem [{\citenamefont {Ke}\ and\ \citenamefont {Johnson}(2016)}]{ke2016prb}%
  \BibitemOpen
  \bibfield  {author} {\bibinfo {author} {\bibfnamefont {L.}~\bibnamefont
  {Ke}}\ and\ \bibinfo {author} {\bibfnamefont {D.~D.}\ \bibnamefont
  {Johnson}},\ }\bibfield  {title} {\bibinfo {title} {Intrinsic magnetic
  properties in {$R$(Fe$_{1-x}$Co$_x$)$_{11}$Ti$Z$ ($R$=Y and Ce; $Z$=H, C, and
  N)}},\ }\href {https://doi.org/10.1103/PhysRevB.94.024423} {\bibfield
  {journal} {\bibinfo  {journal} {Phys. Rev. B}\ }\textbf {\bibinfo {volume}
  {94}},\ \bibinfo {pages} {024423} (\bibinfo {year} {2016})}\BibitemShut
  {NoStop}%
\bibitem [{\citenamefont {Marzari}\ and\ \citenamefont
  {Vanderbilt}(1997)}]{marzari1997prb}%
  \BibitemOpen
  \bibfield  {author} {\bibinfo {author} {\bibfnamefont {N.}~\bibnamefont
  {Marzari}}\ and\ \bibinfo {author} {\bibfnamefont {D.}~\bibnamefont
  {Vanderbilt}},\ }\bibfield  {title} {\bibinfo {title} {Maximally localized
  generalized {Wannier} functions for composite energy bands},\ }\href@noop {}
  {\bibfield  {journal} {\bibinfo  {journal} {Phys. Rev. B}\ }\textbf {\bibinfo
  {volume} {56}},\ \bibinfo {pages} {12847} (\bibinfo {year}
  {1997})}\BibitemShut {NoStop}%
\bibitem [{\citenamefont {Mostofi}\ \emph {et~al.}(2014)\citenamefont
  {Mostofi}, \citenamefont {Yates}, \citenamefont {Pizzi}, \citenamefont {Lee},
  \citenamefont {Souza}, \citenamefont {Vanderbilt},\ and\ \citenamefont
  {Marzari}}]{mostofi2014cpc}%
  \BibitemOpen
  \bibfield  {author} {\bibinfo {author} {\bibfnamefont {A.~A.}\ \bibnamefont
  {Mostofi}}, \bibinfo {author} {\bibfnamefont {J.~R.}\ \bibnamefont {Yates}},
  \bibinfo {author} {\bibfnamefont {G.}~\bibnamefont {Pizzi}}, \bibinfo
  {author} {\bibfnamefont {Y.-S.}\ \bibnamefont {Lee}}, \bibinfo {author}
  {\bibfnamefont {I.}~\bibnamefont {Souza}}, \bibinfo {author} {\bibfnamefont
  {D.}~\bibnamefont {Vanderbilt}},\ and\ \bibinfo {author} {\bibfnamefont
  {N.}~\bibnamefont {Marzari}},\ }\bibfield  {title} {\bibinfo {title} {An
  updated version of wannier90: {A tool for obtaining maximally-localised
  Wannier} functions},\ }\href@noop {} {\bibfield  {journal} {\bibinfo
  {journal} {Computer Physics Communications}\ }\textbf {\bibinfo {volume}
  {185}},\ \bibinfo {pages} {2309} (\bibinfo {year} {2014})}\BibitemShut
  {NoStop}%
\bibitem [{\citenamefont {Souza}\ \emph {et~al.}(2001)\citenamefont {Souza},
  \citenamefont {Marzari},\ and\ \citenamefont {Vanderbilt}}]{souza2001prb}%
  \BibitemOpen
  \bibfield  {author} {\bibinfo {author} {\bibfnamefont {I.}~\bibnamefont
  {Souza}}, \bibinfo {author} {\bibfnamefont {N.}~\bibnamefont {Marzari}},\
  and\ \bibinfo {author} {\bibfnamefont {D.}~\bibnamefont {Vanderbilt}},\
  }\bibfield  {title} {\bibinfo {title} {Maximally localized {Wannier}
  functions for entangled energy bands},\ }\href@noop {} {\bibfield  {journal}
  {\bibinfo  {journal} {Phys. Rev. B}\ }\textbf {\bibinfo {volume} {65}},\
  \bibinfo {pages} {035109} (\bibinfo {year} {2001})}\BibitemShut {NoStop}%
\bibitem [{\citenamefont {Marzari}\ \emph {et~al.}(2012)\citenamefont
  {Marzari}, \citenamefont {Mostofi}, \citenamefont {Yates}, \citenamefont
  {Souza},\ and\ \citenamefont {Vanderbilt}}]{marzari2012rmp}%
  \BibitemOpen
  \bibfield  {author} {\bibinfo {author} {\bibfnamefont {N.}~\bibnamefont
  {Marzari}}, \bibinfo {author} {\bibfnamefont {A.~A.}\ \bibnamefont
  {Mostofi}}, \bibinfo {author} {\bibfnamefont {J.~R.}\ \bibnamefont {Yates}},
  \bibinfo {author} {\bibfnamefont {I.}~\bibnamefont {Souza}},\ and\ \bibinfo
  {author} {\bibfnamefont {D.}~\bibnamefont {Vanderbilt}},\ }\bibfield  {title}
  {\bibinfo {title} {{Maximally localized Wannier functions: Theory and
  applications}},\ }\href@noop {} {\bibfield  {journal} {\bibinfo  {journal}
  {Rev. Mod. Phys.}\ }\textbf {\bibinfo {volume} {84}},\ \bibinfo {pages}
  {1419} (\bibinfo {year} {2012})}\BibitemShut {NoStop}%
\bibitem [{\citenamefont {Harrison}(2012)}]{harrison2012electronic}%
  \BibitemOpen
  \bibfield  {author} {\bibinfo {author} {\bibfnamefont {W.~A.}\ \bibnamefont
  {Harrison}},\ }\href@noop {} {\emph {\bibinfo {title} {Electronic structure
  and the properties of solids: the physics of the chemical bond}}}\ (\bibinfo
  {publisher} {Courier Corporation},\ \bibinfo {year} {2012})\BibitemShut
  {NoStop}%
\end{thebibliography}%

\widetext

\clearpage
\textbf{\Large Supplementary information for:}
\newline
\begin{center}
\textbf{\large Interplay between magnetism and band topology in Kagome magnets $R$Mn$_6$Sn$_6$}
\end{center}
\setcounter{equation}{0}
\setcounter{figure}{0}
\setcounter{table}{0}
\setcounter{section}{0}
\makeatletter
\renewcommand{\theequation}{S\arabic{equation}}
\renewcommand{\thefigure}{S\arabic{figure}}
\renewcommand{\thetable}{S\arabic{table}}
\renewcommand{\bibnumfmt}[1]{[S#1]}

\vspace{20pt}

\textbf{
\begin{enumerate}[label=\Roman*]
\item Supplementary Figures
\item Supplementary Tables
\item Supplementary Theories and Methods
\item Supplementary Discussions
\end{enumerate}
}


\clearpage

\section{Supplementary Figures}

\begin{figure}[htb]
  \includegraphics[width=0.32\linewidth,clip]{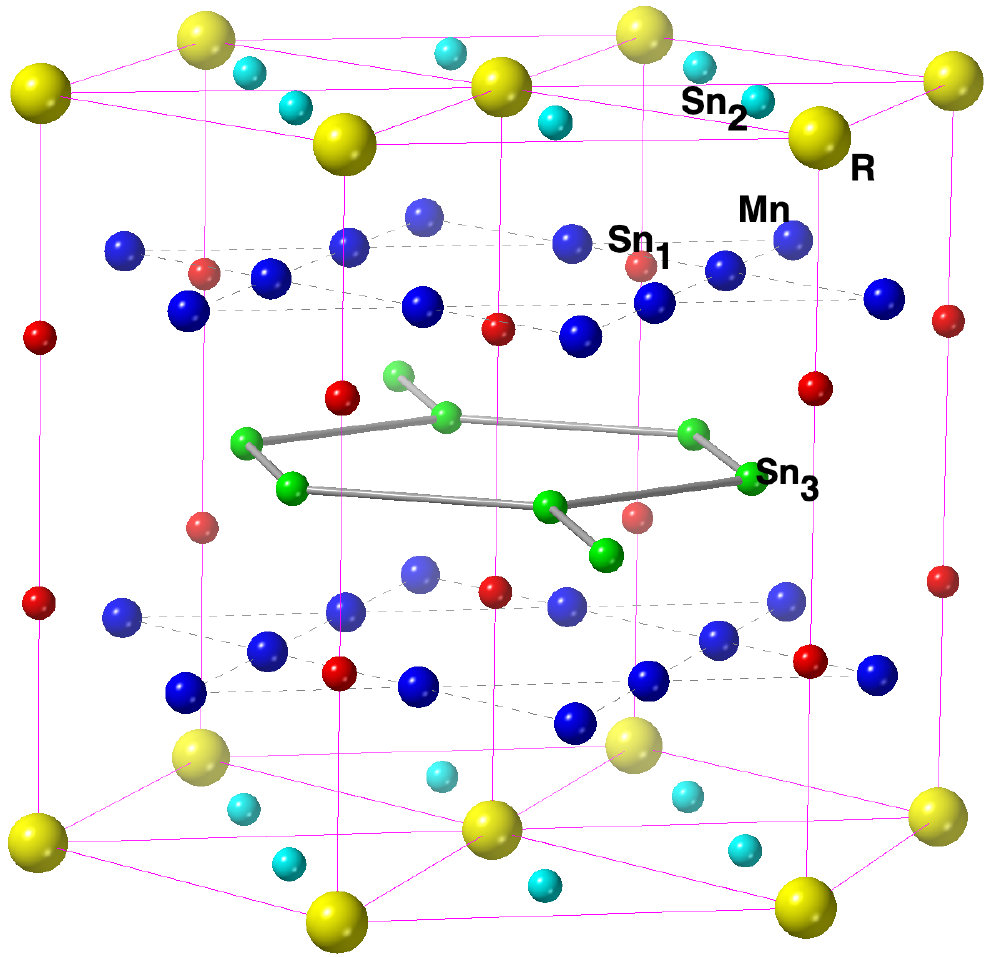}
\caption{Crystal structure of $\rmnsn$.
  The primitive unit cell is tripled to better illustrate the Mn Kagome lattice.
$\rmnsn$ with heavy $R$ elements crystallizes in the hexagonal HfFe$_6$Ge$_6$-type (or equivalently MgFe$_6$Ge$_6$-type; $P6/mmm$, space group no.~191) structure.
The primitive cell contains one formula unit (f.u.).
$R$ atoms occupy the  $1a\ (D_{6h}, \text{or } 6/mmm)$ site, forming a triangular lattice in the basal plane.
Sn$_2$ atoms occupy the $2d\ (\overline{6}m2)$ site located at the center of the $R$ triangles, forming a honeycomb lattice; vice versa, $R$ atoms are located in the centers of Sn$_2$ hexagons. 
Mn atoms occupy the $6i(2mm)$ site, forming two layers of the Kagome lattice in the unit cell.
Sn$_1$ atoms occupy the $2e(6mm)$ site, forming two layers of triangular lattice, adjacent to the Mn layers.
Sn$_1$ atoms form -Sn$_1$-Sn$_1$-$R$- chains along the $c$ axis with $R$ atoms and are pushed slightly off the Mn Kagome plane by $R$ atoms.
Sn$_3$ atoms occupy the $2c(\overline{6}m2)$ site, similar to Sn$_2$, forming a honeycomb lattice by itself and sandwiched between two Sn$_1$ layers.
These layers are stacked in the order of [$R$-Sn$_2$]-Mn-Sn$_1$-Sn$_3$-Sn$_1$-Mn-[$R$-Sn$_2$] along the $c$ axis.
}
\label{fig:xtal_0}
\end{figure}

\begin{figure}[htb]
\centering
\begin{tabular}{c}
  \includegraphics[width=.32\linewidth,clip]{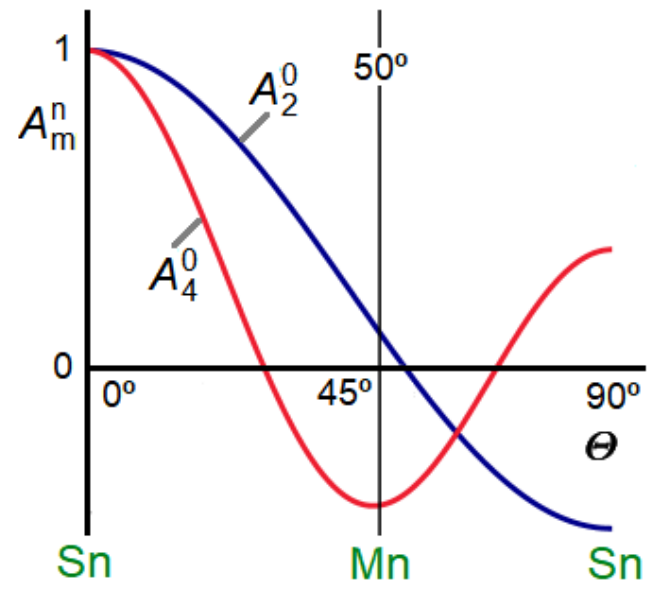}  
\end{tabular}%
\caption{Atomic coordination origin of large high-order crystal-field parameters and anisotropy contributions in $\rmnsn$.
The relation between the crystal-field parameters and the atomic structure is given by the intrinsic crystal-field parameters $A'_2$ and $A'_4$.
Simplifying somewhat, these parameters describe the interaction strength with the Mn and Sn ligands as far as the leading anisotropy contribution by the rare-earth atoms is concerned.
For example, $A_2^0 = \frac12 A'_2 (3\cos2\Theta - 1)$, where $\Theta$ is the coordination angle, means that axial coordination ($\Theta = 0\degree$) and in-plane coordination ($\Theta = 90\degree$) yield opposite anisotropy contributions.
The corresponding 4th-order expression is $A_4^0 = 1⁄8 A'_4 (35 \cos4\Theta - 30 \cos2\Theta + 3)$.
The presence of energy minimum or maximum near $45\degree$ reflects the competition between $K_1$ and $K_2$, that is, between $A_2^0$ and $A_4^0$.
The figure assumes normalized parameters $A'_2 = A'_4 = 1$, but in reality, $A'_2 \gg A'_4$.
To make $A'_4$ competitive, it is necessary to have Mn and Sn coordinations that minimize $A_2^0$ but maximize $A_4^0$.
Indeed, $A_2^0$ is very small, because the coordination of the Mn (about $50\degree$) is close to the point where $A_2^0 = 0$, whereas the contributions of the axially ($0\degree$) and in-plane ($90\degree$) coordinated Sn atoms largely cancel each other.
By contrast, the magnitudes of the $A_4^0$ contributions of both Mn and Sn are maximized.
We also estimated these phenomenological CFP using CF levels of  $\gdmnsn$ that calculated in DFT and obtained $A_2^0\avg{r^2}=5.75$ meV and $A_4^0\avg{r^4}=-9.45$ meV.
}
\label{fig:A_intrinsic}
\end{figure}

\begin{figure*}[htb]
\centering
\begin{tabular}{c}
\includegraphics[width=0.99\linewidth,clip]{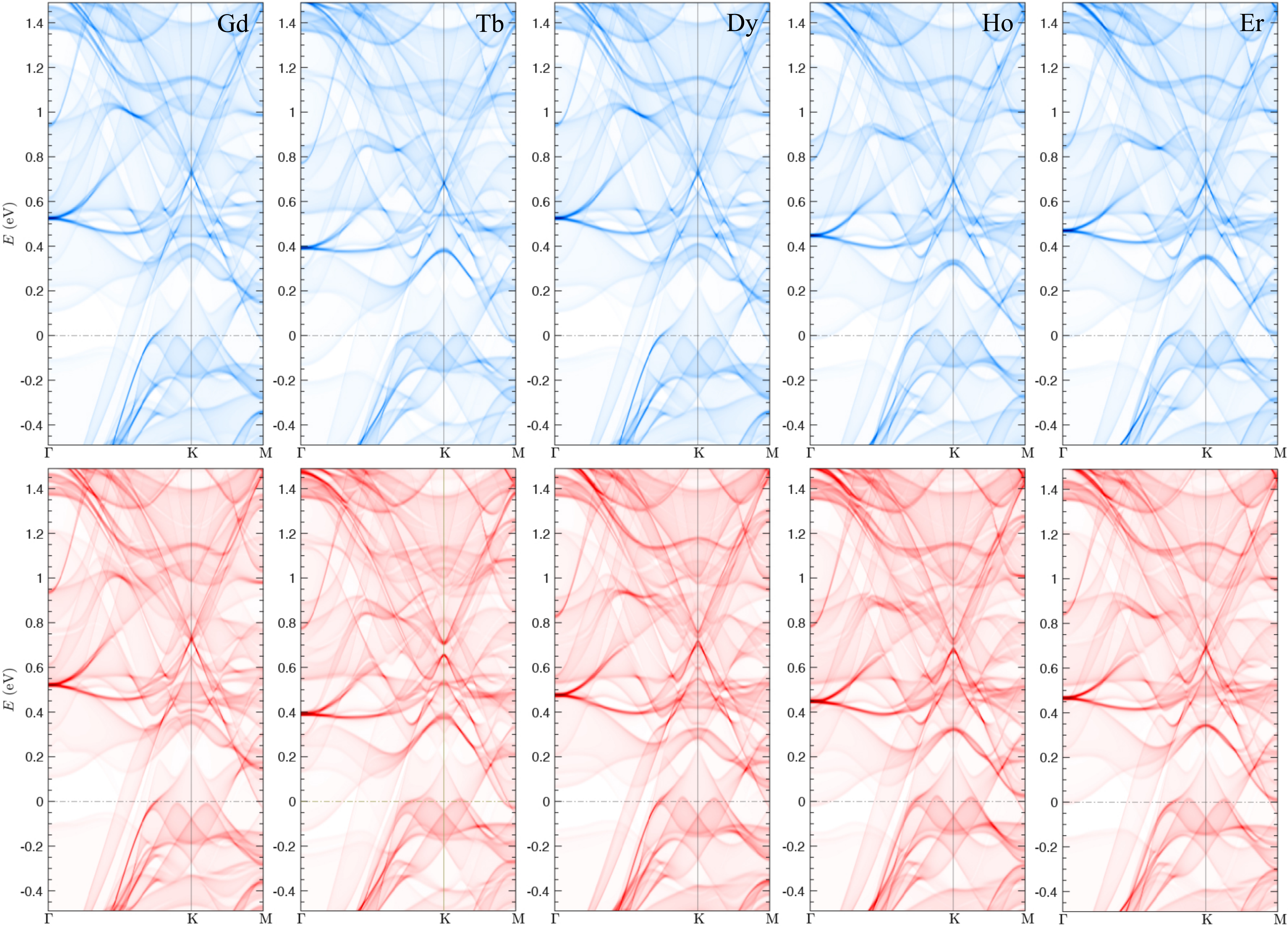}    
\end{tabular}%
\caption{The band structures projected on surface BZ near the Fermi level in $\rmnsn$ with $R=$ Gd, Tb, Dy, Ho, and Er, calculated without (in blue) and with (in red) SOC.
  $R$-$4f$ electrons are treated in the open-core approach in DFT and $R$-$4f$ electrons are configured  to satisfy Hund's first rule.
A Gaussian smearing of $\SI{5}{\meV}$ is used for the $k_z$-integration of spectral functions.  
For the SOC cases, the spin quantization axis directions are set to be along the easy direction of each compound. More specifically, $\theta_\text{Gd}=90\degree$, $\theta_\text{Tb}=0\degree$, $\theta_\text{Dy}=45\degree$, $\theta_\text{Ho}=49\degree$, and $\theta_\text{Er}=90\degree$.
Overall, the band structures near $\ef$, consisting of non-$4f$ states, share great similarities in all compounds.
Multiple DCs occur at $K$, below and above $\ef$.
All of them show an well expressed DC at $\sim\SI{0.7}{\eV}$, which can be gapped by SOC, as we discussed above for $R=$ Tb and Ho.
The difference between these band structures can be attributed to the variations of lattice parameters and the strength of $R$-$5d$ moment and exchange splittings enhanced by various sizes of $R$-$4f$ spin.
}
\label{fig:bnds_10in1}
\end{figure*}

\begin{figure*}[ht]
\centering
\begin{tabular}{cc}
  \includegraphics[width=.49\linewidth,clip]{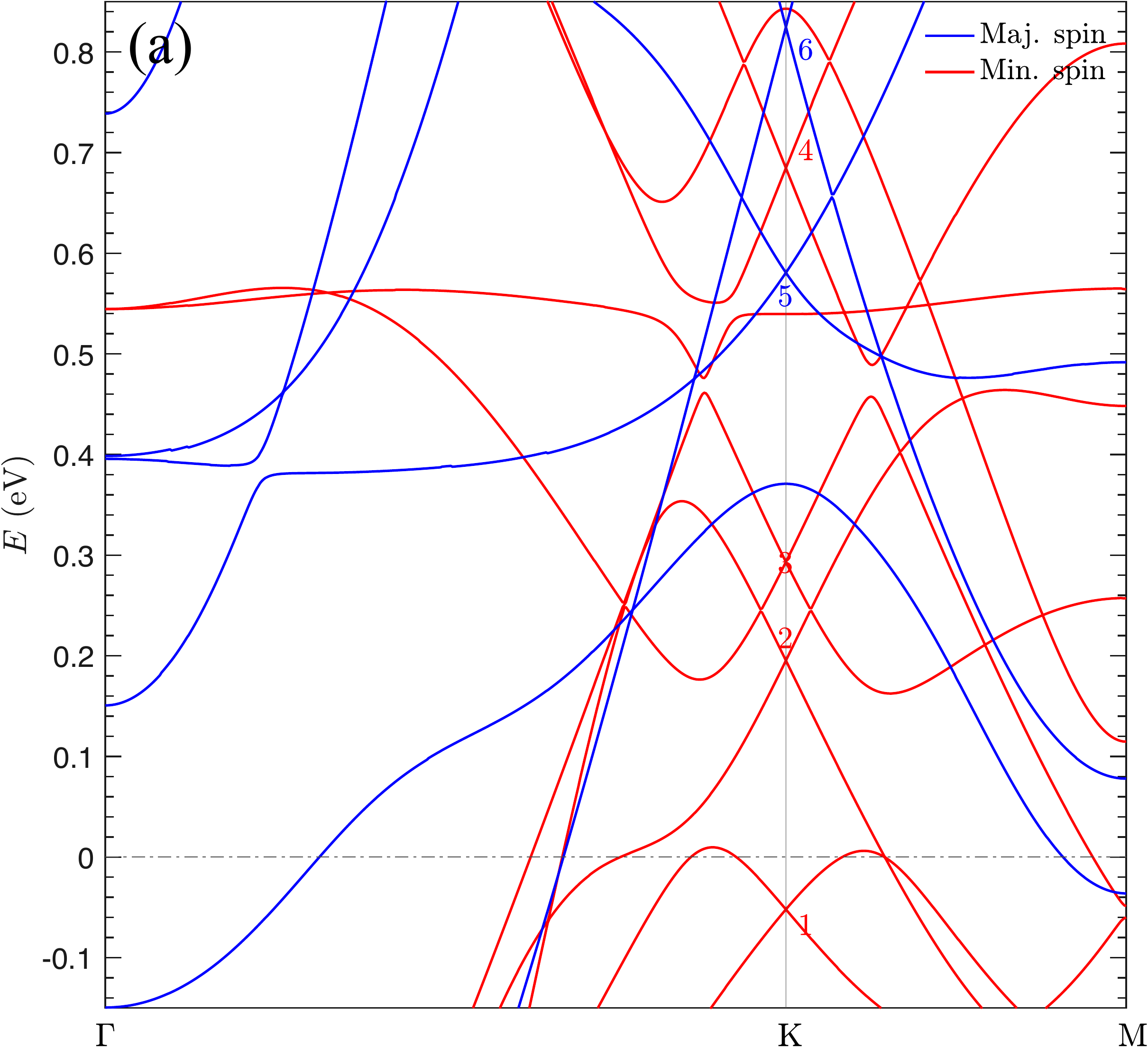} 
    \includegraphics[width=.49\linewidth,clip]{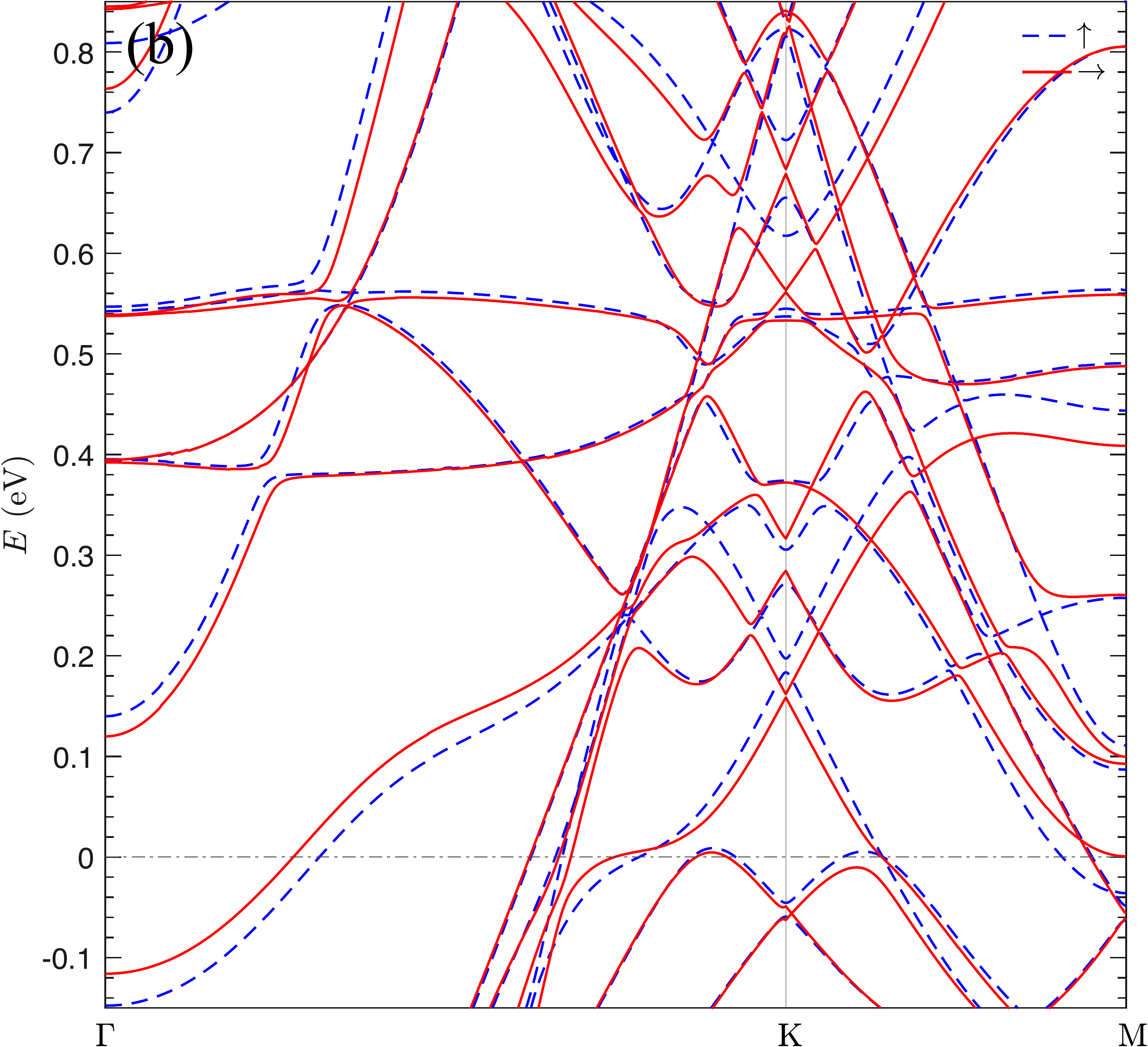}  
\end{tabular}%
\caption{ The band structures near $\ef$ in $\tbmnsn$ calculated (a) without SOC and (b) with SOC.
 In panel (a), the majority-spin and minority-spin, referred to Mn site, are in blue and red, respectively.
 In panel (b), the band structures are calculated with the spin-quantization axis along the out-of-plane (blue dashed line) and in-plane (red solid line) directions. Both magnetic sublattices are ordered. The gap sizes depend on spin orientations.}
\label{fig:tb_band_kz0}
\end{figure*}

\begin{figure}[htb]
\centering
\begin{tabular}{c}
  \includegraphics[width=0.49\linewidth,clip]{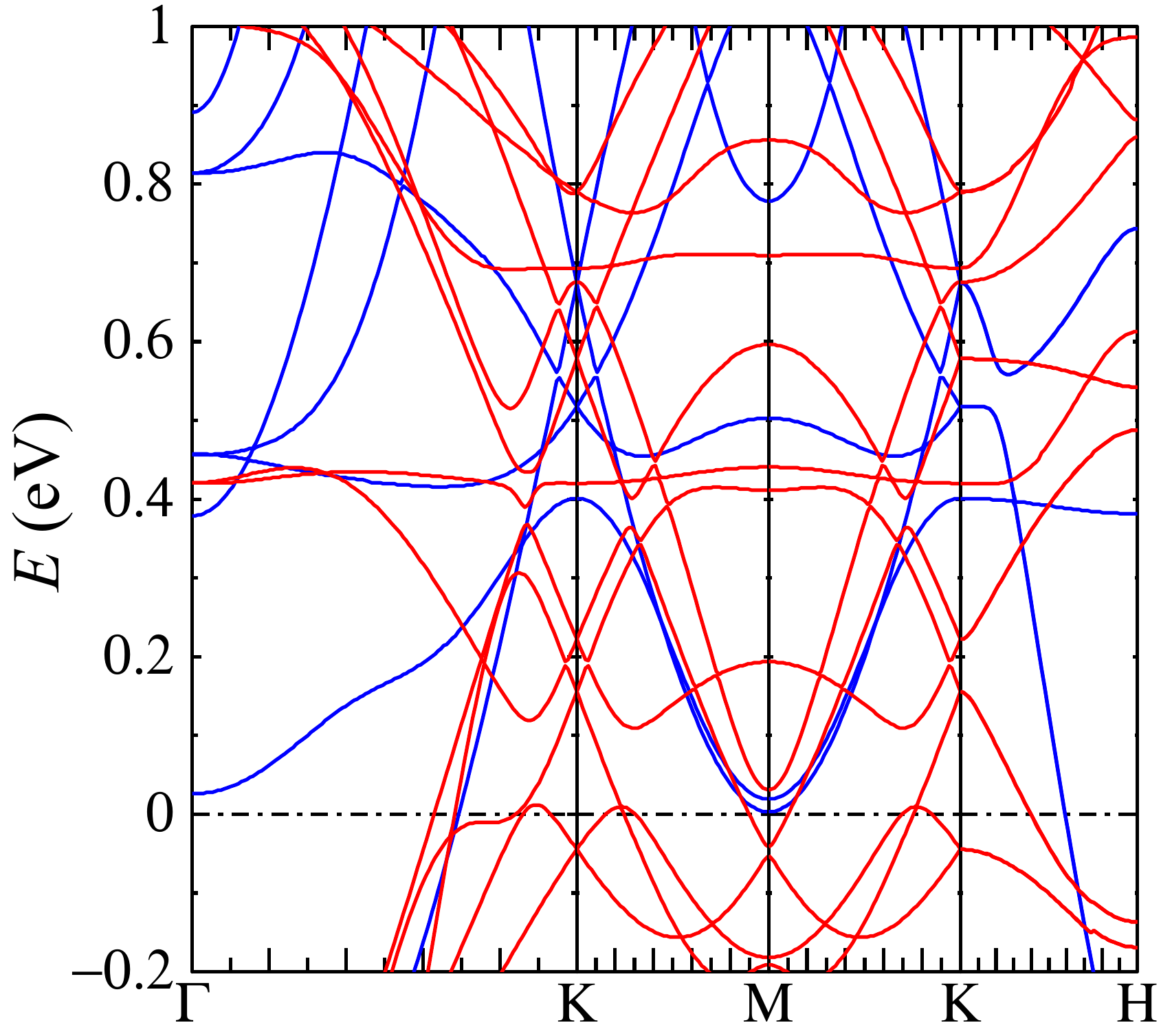} 
\end{tabular}%
\caption{   
The scalar-relativistic band structure near $\ef$ in $\yymnsn$ calculated within QSGW.
The majority-spin and minority-spin, referred to Mn site, are in blue and red, respectively.
The ferrogmagnetic Mn sublattice ordering is utilized in calculation for $\yymnsn$.   
} 
\label{fig:tb_bnd_qsgw}
\end{figure}

\begin{figure}[htb]
\centering
\begin{tabular}{c}
  \includegraphics[width=.52\linewidth,clip]{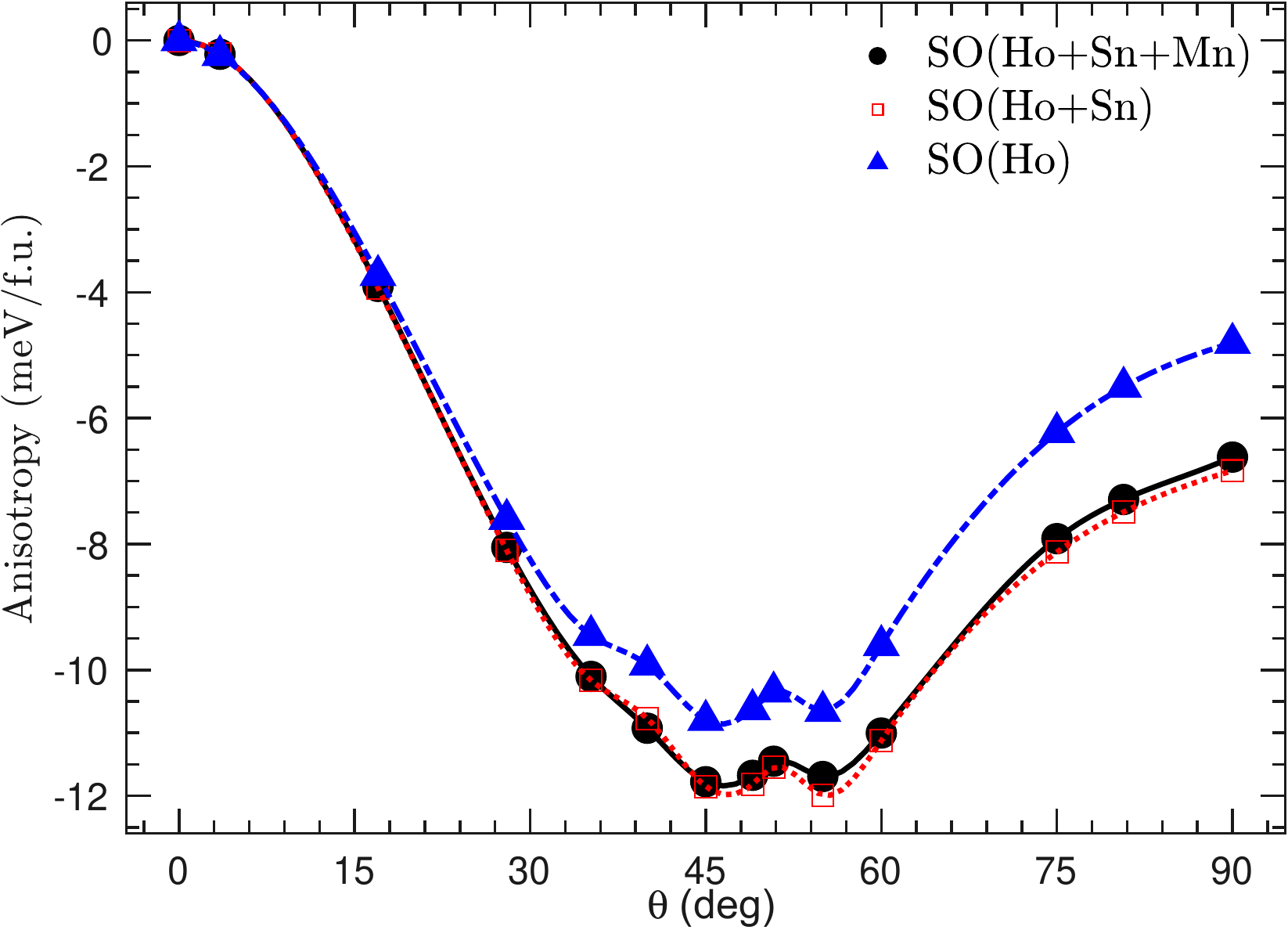}  
\end{tabular}%
\caption{Variation of magnetic energy (in meV/f.u.) as a function of spin-quantization-axis rotation in $\homnsn$, calculated with and without turning on SOC on Sn and Mn sites. $\theta$ is the angle between the spin direction and the out-of-plane direction.
  The black circles denote the regular full-SOC calculations with SOC turning on all sublattices.
  The red squares are calculated with SOC on Ho and Sn sublattices only, barely deviating from the full-SOC calculations (black circles).
  The blue triangles denote the calculations with Ho SOC only, showing the Ho sublattice, by itself, prefers the $\sim\theta=49\degree$ orientation.
  Turning off Sn SOC also increases $E(\theta=90\degree)$ by $\sim$\SI{2}{\meV/f.u.}, which is the value of the above discussed easy-plane Mn anisotropy.
  Thus, Mn MA originates from the interplay between the Mn-$3d$ spin polarization and the large Sn-$4p$ SOC.}
\label{fig:mae_atom}
\end{figure}

\clearpage
\newpage

\section{Supplementary Tables}

\begin{table}[h]
  \caption{
    The coefficients (multiply by 1024 for readability) of $e_0, e_{\pm1}, e_{\pm2}, e_{-3}$ in anisotropy parameters $C^m_i$ and $C_i^{n_\downarrow}$, with $i=0$, 2, 4, 6.
    Note that $C^{f^1}_i=C^{\pm3}_i$. We also have $C^{f^4}_i=-C^{f^3}_i$,  $C^{f^5}_i=-C^{f^2}_i$,  $C^{f^6}_i=-C^{f^1}_i$, and $C^{f^7}_i=0$.}
\label{tbl:c0246}
\bgroup
\def\arraystretch{1.3}
\begin{tabular*}{\linewidth}{l@{\extracolsep{\fill}}lrrrr rrrrr rrrrr rrrrr| rrrrr rrrrr}
  \hline\hline
       & &  \multicolumn{4}{c}{$m=\pm3$} &  &   \multicolumn{4}{c}{$m=\pm2$}  & &  \multicolumn{4}{c}{$m=\pm1$}  &   &  \multicolumn{3}{c}{$m=0$}   &  &  \multicolumn{4}{c}{$f^2$}  &   & \multicolumn{3}{c}{$f^3$}  \\   
   \cline{1-1}  \cline{3-6} \cline{8-11}  \cline{13-16}   \cline{18-20}   \cline{22-25}   \cline{27-29}
$C$    & &  $e_0$  &  $e_1$    &  $e_2$   & $e_{-3}$ &  &  $e_0$   &  $e_1$     &  $e_2$   & $e_{-3}$ & & $e_0$    &  $e_1$    &  $e_2$   & $e_{-3}$ &   & $e_0$    &  $e_1$    &  $e_2$    & & $e_0$    &  $e_1$    &  $e_2$   & $e_{-3}$ &   & $e_0$    &  $e_1$    &  $e_2$  \\ 
\hline
$C_0$  & &  100  &   210  &  252  & -276  &    &   120  &   260  &   392  &   96   &   &   156  &   398  &   260  &   180   & &  272 &  312 &  240    &    &   220  &   470  &  -380  & -180   &   &   376  &  -156  &  -120\\
$C_2$  & & -150  &  -255  &  -90  &  240  &    &   -60  &   -70  &   220  &        &   &    54  &   271  &   -70  &  -240   & &  312 &  108 & -120    &    &  -210  &  -325  &   130  &  240   &   &  -156  &   -54  &    60\\
$C_4$  & &   60  &    30  & -156  &   36  &    &  -120  &  -100  &   376  &  -96   &   &   -60  &   130  &  -100  &    60   & &  240 & -120 & -240    &    &   -60  &   -70  &   220  &  -60   &   &  -120  &    60  &   120\\
$C_6$  & &  -10  &    15  &   -6  &       &    &    60  &   -90  &    36  &        &   &  -150  &   225  &   -90  &         & &  200 & -300 &  120    &    &    50  &   -75  &    30  &        &   &  -100  &   150  &   -60\\
\hline
\hline
\end{tabular*}
\egroup
\end{table}

\begin{table}[h]
  \caption{$\homnsn$ band characters of six Dirac crossings near $\ef$ at BZ corners $K$,
    as indicated in Fig.~9(a) in the main text,
    resolved into sublattices (per sublattice) and Mn-$3d$ orbitals (per Mn atom).
    The majority and minority spin channels are referred to  Mn sites.
  }
	\label{tbl:band_character}
	\bgroup
	\def\arraystretch{1.3}
	\begin{tabular*}{\linewidth}{l@{\extracolsep{\fill}}cccccccc}
	  \hline\hline
          & &  \multicolumn{4}{c}{Minority Spin} & &\multicolumn{2}{c}{Majority Spin} \\    \cline{3-6} \cline{8-9}
          Index  & &  1 & 2 & 3 & 4 & & 5 & 6 \\ \hline
		  $E-\ef$ (eV) & & -0.065 & 0.243 & 0.260 & 0.705 & & 0.585 & 0.812 \\ \hline
		  Ho           & &  0.012 & 0.034 & 0.059 & 0.003 & & 0.053 & 0.102  \\
		  Mn           & &  0.831 & 0.697 & 0.690 & 0.856 & & 0.554 & 0.477  \\
		  Sn           & &  0.026 & 0.080 & 0.040 & 0.021 & & 0.190 & 0.138  \\
		  Interstitial & &  0.131 & 0.189 & 0.211 & 0.121 & & 0.203 & 0.283  \\
                  \hline
                  $d_{xy}$      & & 0.009 & 0.026 & 0.001 & 0.091 & & 0.075 & 0.000  \\
		 $d_{yz}$      & & 0.000 & 0.010 & 0.016 & 0.000 & & 0.000 & 0.052  \\		 
		  $d_{z^2}$     & & 0.110 & 0.001 & 0.003 & 0.021 & & 0.000 & 0.000  \\
		  $d_{xz}$      & & 0.002 & 0.000 & 0.093 & 0.000 & & 0.000 & 0.011  \\
		  $d_{x^2-y^2}$ & & 0.015 & 0.076 & 0.000 & 0.025 & & 0.012 & 0.000  \\
        	  \hline\hline
	\end{tabular*}
	\egroup
 \end{table}

\begin{table}[h]
  \caption{$\tbmnsn$ band characters of six Dirac crossings near $\ef$ at BZ corners $K$, as indicated in \rfig{fig:tb_band_kz0}, resolved into sublattices (per sublattice) and Mn-$3d$ orbitals (per Mn atom).
Majority and minority spin chanels are refered to Mn sites.
  }
	\label{tbl:band_character_tb}
	\bgroup
	\def\arraystretch{1.3}
	\begin{tabular*}{\linewidth}{l@{\extracolsep{\fill}}cccccccc}
	  \hline\hline
          & &  \multicolumn{4}{c}{Minority Spin} & &\multicolumn{2}{c}{Majority Spin} \\    \cline{3-6} \cline{8-9}
          Index  & &  1 & 2 & 3 & 4 & & 5 & 6 \\ \hline
		  $E-\ef$ (eV) & &  -0.052 & 0.195 & 0.293 & 0.684 & & 0.580 & 0.826 \\ \hline
		  Tb           & &   0.018 & 0.038 & 0.054 & 0.002 & & 0.057 & 0.062  \\
		  Mn           & &   0.817 & 0.690 & 0.699 & 0.854 & & 0.546 & 0.492  \\
		  Sn           & &   0.025 & 0.080 & 0.038 & 0.022 & & 0.189 & 0.144  \\
		  Interstitial & &   0.139 & 0.191 & 0.209 & 0.122 & & 0.210 & 0.301  \\
                  \hline
                  $d_{xy}$      & &  0.009 & 0.025 & 0.001 & 0.092 & & 0.073 & 0.000   \\
		  $d_{yz}$      & &  0.002 & 0.011 & 0.015 & 0.000 & & 0.000 & 0.054   \\
		  $d_{z^2}$     & &  0.105 & 0.001 & 0.007 & 0.021 & & 0.000 & 0.002   \\
		  $d_{xz}$      & &  0.004 & 0.002 & 0.090 & 0.000 & & 0.000 & 0.011   \\
      		  $d_{x^2-y^2}$ & &  0.015 & 0.076 & 0.002 & 0.024 & & 0.012 & 0.000   \\
	  \hline\hline
	\end{tabular*}
	\egroup
 \end{table}


\clearpage
\section{Supplementary Theories and Methods}

\subsection{Crystal structure}

\rFig{fig:xtal_0} shows the crystal structure, where the primitive cell is tripled to illustrate the Mn kagome lattice better.
Lattice parameters of $\rmnsn$ change with $R$.
Experiments found that $\tbmnsn$ has the largest volume, while $\gdmnsn$ has the smallest volume.
For all $\rmnsn$ except $\tbmnsn$, the distances between neighboring Mn layers are slightly larger across the Sn$_1$ and Sn$_3$ layers than the one across the [$R$-Sn$_2$] layer.
These two inter-Mn-plane distances are accidentally the same in  $\tbmnsn$.  
The crystal structure can also be described as a filled derivative of the CoSn $B35$-type structure and closely related to the CaCu$_5$-type~\cite{ke2016prbA} and ThMn$_{12}$-type~\cite{ke2016prb} structures~\cite{idrissi1991jlcm}.
While the $\rmnsn$ structure shares great similarity to that of the well-studied SmCo$_5$ magnet~\cite{idrissi1991jlcm,ke2016prbA}, in contrast to $R$Co$_5$, $R$ atoms in $\rmnsn$ have two different kinds of ligand atoms and also have the nearest neighbor Sn$_1$ along the axial direction, resulting in a drastically different $R$ anisotropy than in $R$Co$_5$.
$\rmnsn$ with light $R$ elements have a different crystal structure.

\subsection{\textit{Ab initio} methods: additional details}

The DFT calculations are performed in \textsc{Wien2k}~\cite{WIEN2k}.
To generate the self-consistent potential and charge, we employed $R_\text{MT}\cdot K_\text{max}=8.0$ with muffin-tin (MT) radii $R_\text{MT}=$ 2.7, 2.4, and \SI{2.5}{\atomicunit}, for $R$, Mn, and Sn, respectively.
The calculations are performed with 264 $k$-points in the irreducible Brillouin zone (IBZ).
They are iterated until charge differences between consecutive iterations are smaller than \num{e-3}~$e$ and the total energy differences lower than \SI{0.01}{mRy}.

\para{QSGW calculations.} The QSGW method is used to investigate the non-$4f$ bandstructures near $\ef$.
For $\yymnsn$, we apply QSGW on top of DFT, while for $\tbmnsn$, we apply QSGW on top of DFT+$U$, with $U$ = 0.52 Ry applied to Tb-$4f$ electrons.
The calculated band structures near $\ef$ exhibit great similarity, as shown in \rfig{fig:tb_bnd_qsgw}.
All QSGW calculations were performed within the scalar-relativistic approximation.
For the reciprocal space integration, we utilized an $8\times 8\times 4$ mesh in the Brillouin zone.
Our basis set in the muffin-tin spheres was defined by using an expansion in spherical harmonics up to $L_\text{max}=6$, whereas in the interstitial region, we used a cutoff of \SI{2.8}{\atomicunit} for the wave functions and a cutoff of \SI{2.8}{\atomicunit} for the product basis set.

\para{Realistic tight-binding calculations.} The band structures near $\ef$ are further analyzed using an in-house \textit{ab initio} tight-binding (TB) framework~\cite{ke2019prb}.
Realistic TB Hamiltonians are constructed via the maximally localized Wannier functions (MLWFs) method~\cite{marzari1997prb} as implemented in \textsc{Wannier90}~\cite{mostofi2014cpc} through a post-processing procedure~\cite{marzari1997prb,souza2001prb,marzari2012rmp} using the output of the self-consistent DFT calculations.
We construct the TB Hamiltonian using 118 MLWFs, which correspond to $d$-type orbitals for $R$ and Mn, and $s$- and $p$-type orbitals for Sn in the unit cell.
Note that the 59 orbitals are doubled to account for SOC, which mixes the two spin channels.
A real-space Hamiltonian $H({\bf R})$ with dimensions 118$\times$118 is constructed to accurately represent the band structures in the energy window of interest.

\subsection{Magnetocrystalline anisotropy in Rare-earth compounds}
In general, the Mn spins in $\rmnsn$ prefer the in-plane direction while the easy-axis of the $R$ atoms varies with the type of $R$ atom and may be incompatible with Mn.
For the $R$ sublattice, due to the strong SOC, electronic configurations of $4f$ shell obey the Hund's rules~\cite{skomski2008book}.
The magnetic moment of $4f$ electrons is strongly coupled with the anisotropic-shaped charge cloud, which is determined by superposing the occupied spherical harmonics of $|l=3,m\rangle$ with various $m$ channels.
The charge cloud orients accordingly with respect to the CF of surrounding lattices to minimize the Coulomb energy, giving the strong $4f$-electron MA.

We use various approaches to resolve the MAE contribution into different sublattices.
For example, to explore the non-$4f$ contribution to MAE dominated at high temperatures, we include $R$-$4f$ in the open core and constrain their moments to be zero to mimic a disordered $4f$ moment.
On the other hand, we also investigate the $R$-only contribution by turning off the SOC on Mn and Sn sites in MAE calculations.

\subsection{Spin-orientation-dependence of electronic structure}
The magnetic structures, more specifically, the spin orientations of $R$ and Mn atoms, are known to evolve with $R$ and temperature.
The moment directions and sizes directly impact the topology of the electronic band structure, and we calculate how band structures near $\ef$, especially the position of DCs and SOC-induced gap openings, depend on spin orientations.

The Dirac bands are mainly characterized by non-$4f$, mostly the Mn-$3d$ orbitals.
The sizes of SOC-induced gaps and their dependence on the spin-quantization-axis direction can be understood by including the SOC term in the single-particle Hamiltonian within a perturbation theory~\cite{ke2015prb}.
For an arbitrary spin-quantization direction $\hat{\bf n}=(\theta,\varphi)$, the SOC Hamiltonian can be written as
\begin{equation}
H_\text{so}(\hat{\bf n})=\frac{\xi}{2}U(\theta,\varphi)(\mathbf{L}\cdot \mathbf{S}) U^\dagger(\theta,\varphi) \,.
\end{equation}
Here, $\xi$ is the SOC constant depending on orbital $l$ and site $i$, and $U(\theta,\varphi)$ is the unitary transformation Wigner matrix (See details in Appendix $A$ in Ref.~[\onlinecite{ke2019prb}]). Thus, we obtain
\begin{equation}
H_\text{so}\left(\theta,\varphi\right)=\frac{\xi}{2}
\left(
\begin{array}{cc}
A & B \\
B^\dagger & -A \\
\end{array}
\right),
\label{eq:hso}
\end{equation}
where the spin-parallel component $A$ and the spin-flip component $B$ are written as
\begin{eqnarray}
  A(\theta,\varphi) &=&  \cos(\theta)L_{z} + \frac{1}{2} \sin (\theta ) \left(e^{i \varphi }L_{-} + e^{-i\varphi}L_{+}\right) \\
  B(\theta,\varphi) &=& -\sin(\theta)L_{z} +  \frac{1}{2} \Big( \left(\cos(\theta)+1\right) e^{i\varphi} L_{-}
   + \left( \cos(\theta)-1 \right) e^{-i \varphi} L_{+} \Big).
 \label{eq:hso_ab}
\end{eqnarray}  


\subsection{MAE analytical modeling using calculated CF energies}

The full Hamiltonian matrix of atom $R$ can be written as
\begin{equation}
  H_{mm^{\prime}}=H_\text{SO}+H_\text{CF}=\left\langle \tilde{Y}_{lm}|\xi \mathbf{L\cdot S}|\tilde{Y}_{lm^{\prime}}\right\rangle
  +\left\langle \tilde{Y}_{lm}|H_\text{CF}|\tilde{Y}_{lm^{\prime}}\right\rangle,
\end{equation}

The anisotropy energy of $4f$ states, the angular dependence of CF energy, can be expressed in terms of CF levels $\epsilon=[e_{-3}\, e_{-2}\, e_{-1}\, e_{0}\,  e_{1}\,  e_{2}\,  e_{3}]^\intercal$.
In the absence of SOC, orbitals are fully quenched and the eigenstates are characterized by the \emph{real} spherical harmonics $\rY^l_m$.
In the limit of $\xi\gg d$, instead, the eigenstates are characterized by rotated \emph{complex} spherical harmonics $\tilde{Y}_{m}(\theta)$, where $\tilde{Y}_{m}(\theta=0)={Y}_{m}$.
Expressing $\langle \tilde{Y}_{m}|H_\text{CF}|\tilde{Y}_{m'}\rangle$ in terms of $\left\langle \rY_{m}|H_\text{CF}|\rY_{m'}\right\rangle$ is achieved by a Wigner rotation followed by an unitary transformation $\tilde{Y} \to Y \to \rY$.

\begin{align}
  \tilde{Y}^l_{m_2}(\theta)  & = \sum_{m_1} Y^l_{m_1} D_{m_1 m_2}^{l}(\theta) \\
  Y^l_{m_2}  & = \sum_{m_1} \rY^l_{m_1} U_{m_1 m_2}
\end{align}

Note that, the corresponding Wigner rotation matrix $D_{mm'}^{l}(\theta)=D_{mm'}^{l}(\alpha=0,\beta=\theta,\gamma=0)$ is a real matrix and $D^\dagger=D^\intercal$. In DFT calculation, we rotate the spin axis from $z$ axis to $y$ axis, which corresponds to
rotations that characterized by the Euler angles $(\alpha=0,\beta=\theta,\gamma=0)$.
For the $f$ block, the unitary transformation matrix ${\bf U}$ is
\begin{equation}
 \mathbf{U}=\mathbf{U}_{\mathbb{R}\leftarrow\mathbb{C}}
=
\frac{1}{\sqrt{2}}\left(
\begin{array}
[c]{ccccccc}%
-i &  0 &  0 & 0 &  0 & 0 &-i\\ 
 0 & -i &  0 & 0 &  0 & i & 0\\
 0 &  0 & -i & 0 & -i & 0 & 0\\
 0 &  0 &  0 & \sqrt{2} & 0 & 0 & 0\\
 0 &  0 &  1 & 0 & -1 & 0 & 0\\
 0 &  1 &  0 & 0 &  0 & 1 & 0 \\
 1 &  0 &  0 & 0 &  0 & 0 &-1\\ 
\end{array}
\right)
\end{equation}

For simplicity, in the following, we absorb $(\theta)$ and $l=3$ in $\tilde{Y}^l_{m}(\theta)$ and $D_{mm'}^{l}(\theta)$.
The matrix element of $H_\text{CF}$ in the basis of $|\tilde{Y}_{m}\rangle$ becomes:

\begin{align}
\langle \tilde{Y}_{m}|H_\text{CF}|\tilde{Y}_{m'}\rangle  
& = \sum_{m_1 m_2 m_3 m_4} \langle \tilde{Y}_{m} | Y_{m_1}\rangle \langle Y_{m_1}|
 \rY_{m_2}\rangle \langle \rY_{m_2} | V | \rY_{m_3}\rangle \langle \rY_{m_3} | Y_{m_4}\rangle  \langle Y_{m_4}| \tilde{Y}_{m'} \rangle \\ \nonumber
& = \sum_{m_1 m_2 m_3 m_4} D^\dagger_{m,m_1} U^\dagger_{m,m_2} V_{m_2,m_3} U_{m_3,m_4} D_{m_4m'} \\ 
& = (\mathbf{D^\dagger U^\dagger E  U D})_{mm'} \nonumber
\label{eq:hcf-2}
\end{align}
Here, $\mathbf{E}$ is the diagonal matrix with element of $\langle \rY_{m_1}| V | \rY_{m_2}\rangle=e_{m_1}\delta_{m_1,m_2}$.
In $\rmnsn$, we have $e_0=E(a_{2u})$, $e_{\pm 1}=E(e_{1u})$, $e_{\pm 2}=E(e_{2u})$, $e_{-3}=E(b_{1u})$ and $e_{3}=E(b_{2u})$.
Note that, matrix $(\mathbf{U^\dagger E  U})$ have diagonal elements $(\mathbf{U^\dagger E  U})_{\pm3,\pm3}=(e_{-3}+e_{3})/2$ and off-diagonal elements $(\mathbf{U^\dagger E  U})_{\pm3,\mp3}=(e_{-3}-e_{3})/2$,
which differs from the diagonal matrix $\mathbf{E}$.
In the following, we set $e_3$ as the energy reference zero.

\begin{align}
  K_m &= \langle \tilde{Y}_{m}|H_\text{CF}|\tilde{Y}_{m}\rangle =  D_{-3,m}D_{3,m}e_{-3} + \sum_{m'} (D_{m'm})^2e_{m'} \\
  K_n &= \sum_{m=3}^{4-n} K_m
\end{align}

For Tb case, 
\begin{align}
K_\text{Tb}=K(f^1)=K_{m=3}=\langle \tilde{Y}_{3}|H_\text{CF}|\tilde{Y}_{3}\rangle = \sum_m (D_{m3})^2e_{m} + D_{3,-3}D_{33}e_{-3}
\end{align}

The $m$-orbital contribution to anisotropy, $K_m(\theta)$, can be written as
\begin{equation}
K_m(\theta) = C^m_0 + C^m_2\cos(2\theta) + C^m_4\cos(4\theta) + C^m_6\cos(6\theta)
\label{eq:kc0246b}
\end{equation}

Here, the parameters $C^m_i$ are linear combinations of $e_j$, and the corresponding coefficients are tabulated in \rtbl{tbl:c0246}.
For more than one $4f$ electrons, we have
\begin{equation}
K_{f^{n_\downarrow}}=\sum_{m=-3}^{n-4} K_m
\end{equation}
  
As expected, in this simple model, we have 
\begin{eqnarray}
K_m(\theta)  &=& K_{-m}(\theta) \\  \nonumber
\sum_{m=-3}^3 K_m(\theta) &=& 0
\label{eq:krule}
\end{eqnarray}
From these equations, we also have $K(f^n)=-K(f^{7-n})$.

For $f^{1}$,
\begin{align}
  K_\text{Tb}  &=\frac{1}{1024}\big( ( 100 e_{0}   +210  e_{1}    +252  e_{2}   -276 e_{-3})  \\ \nonumber
                                   &+ (-150 e_{0}   -255  e_{1}    -90  e_{2}    +240 e_{-3})\cos2\theta  \\ \nonumber
                                   &+ (  60 e_{0}   +30   e_{1}   -156  e_{2}    + 36 e_{-3})\cos4\theta  \\ \nonumber
                                   &+ ( -10 e_{0}   +15   e_{1}     -6  e_{2}              )\cos6\theta
  \big),
  \label{eq:ktb}
\end{align}

for $f^{2}$%
\begin{align}
  K_\text{Dy} &=\frac{1}{1024}\big( ( 220 e_{0}   +470  e_{1}   -380  e_{2}   -180 e_{-3})  \\ \nonumber
                                   & + (-210 e_{0}   -325  e_{1}   +130  e_{2}   +240 e_{-3})\cos2\theta  \\ \nonumber
                                   &+ ( -60 e_{0}    -70  e_{1}   +220  e_{2}    -60 e_{-3})\cos4\theta  \\ \nonumber
                                   & + (  50 e_{0}    -75  e_{1}    +30  e_{2}              )\cos6\theta
  \big),
\end{align}
for $f^{3}$%
\begin{align}
  K_\text{Ho}  &=\frac{1}{1024}\big( ( 376 e_{0}   -156  e_{1}   -120  e_{2}   )  \\ \nonumber
                                   &+ (-156 e_{0}    -54  e_{1}    +60  e_{2}   )\cos2\theta  \\ \nonumber
                                   &+ (-120 e_{0}    +60  e_{1}   +120  e_{2}   )\cos4\theta  \\ \nonumber
                                   &+ (-100 e_{0}   +150  e_{1}    -60  e_{2}   )\cos6\theta
  \big),
\end{align}

We use the CF energy levels calculated in GdMn$_6$Sn$_6$, $e_0=33.44, e_{\pm 1}=76.84, e_{\pm 2}=90.2, e_{-3}=67.42$ (meV).

\clearpage

\section{Supplementary Discussions}

\subsection{Orbital characters of Dirac crossings near $\ef$}
To understand the $k_z$ dependence of surface bands and how the Dirac crossings and SOC-induced gaps evolve with spin orientation, we analyze the corresponding orbital characters at these band crossings. 
Figure 9(a)
shows the band structure in $\homnsn$ calculated without SOC along the high symmetry path $\Gamma$-$K$-$M$ ($k_z=0$).
Six DCs, indexed as 1--6 in
Fig.~9(a),
occur within the energy window; four (DC1--DC4) in the minority Mn-spin channel, and two (DC5, DC6) in the majority Mn-spin channel.
The band characters of these DCs are resolved into atoms and Mn-$3d$ orbitals, as listed in \rtbl{tbl:band_character}.

Band characters are dominated by Mn-$3d$ orbitals while showing hybridization with Ho and Sn sites.
In comparison to other crossings, two DCs in the majority spin channel, DC5 and DC6, have more substantial amounts of contributions from Ho and Sn, especially the latter, resulting in a stronger $k_z$ dependence, as shown in
Fig.~7.
In contrast, DC4 and DC1 in the minority spin channel, consisting of the least amount of Sn and Ho characters, show strong intensity in \rfig{fig:bnds_10in1}.

The size of the SOC-induced gap and its dependence on spin quantization direction can be understood by further resolving the band characters into Mn-$3d$ orbitals.
DC4 consists $d_{xy}$ ($|m=-2\rangle$), $d_{z^2}$ ($|m=0\rangle$), and $d_{x^2-y^2}$ ($|m=2\rangle$) characters.
The SOC Hamiltonian, more specifically, the $L_z$ operator, couples $|m=\pm2\rangle$ states and effectively opens up a gap.
On the other hand, DC1 consists of more $|m=0\rangle$ states and less $|m=\pm2\rangle$ states, resulting in a much smaller SOC-induced gap.
DC5 contains Mn-$|\pm2\rangle$ states and also a substantial amount of Sn-$p$ states, which have a large SOC constant, giving a large gap, as shown in
Fig.~9(b).

$\tbmnsn$ shows the similar orbital characters, as shown in \rtbl{tbl:band_character} and \rfig{fig:tb_band_kz0}.

\subsection{Energy bands of $d$ orbitals on a Kagome lattice}
The $d$ orbital wavefunctions are highly anisotropic and thus the overlapping between two wavefunctions at different sites depends the relative orientation between these two sites.
These hoppings can be captured by the Slater-Koster two center approximation that decomposes the hoppings into three specific configurations with hopping energy denoted as $V_{dd\sigma}$, $V_{dd\pi}$ and $V_{dd\delta}$.
Following this approximation as detailed in Ref.~\onlinecite{harrison2012electronic}, one can obtain the Hamiltonian of $d$ orbitals on a Kagome lattice.

We take the $d_{x^2-y^2}$ and $d_{xy}$ orbitals as an example. With only the nearest-neighboring coupling, the tight-binding Hamiltonian reads
\begin{align*}
H = \left( \begin{matrix}
    H_{11} & H_{12} \\
    H_{21} & H_{22} 
    \end{matrix} \right)
\end{align*}
where 1 and 2 indicate $d_{x^2-y^2}$ and $d_{xy}$, separately, and $H_{ij}$ indicates the coupling between the $i$ and $j$ orbital between three sites in a unit cell. Specifically, 
\begin{align*}
H_{11}(\bm{k}) = 2\left( \begin{matrix}
    0 & (t_0+t_1)\cos \bm{k}\cdot\bm{R}_0 & 0 \\
    0 & 0 & (t_0-t_1/2)\cos \bm{k}\cdot\bm{R}_{\phi}\\
    (t_0-t_1/2)\cos \bm{k}\cdot\bm{R}_{-\phi} & 0 & 0\\
    \end{matrix} \right) + h.c.
\end{align*}
where $\bm{k}$ is the momentum, $\phi=2\pi/3$, $\bm{R}_{\theta}=(\cos \theta, \sin \theta)$ is the unit vector pointing along the angle $\theta$, $t_0=\frac{1}{8}(3V_{dd\sigma}+4V_{dd\pi}+V_{dd\delta})$, and 
$t_1=\frac{1}{8}(3V_{dd\sigma}-4V_{dd\pi}+V_{dd\delta})$. Typically, $|t_1| \gg |t_0|$. 
The other matrices are
\begin{align*}
H_{22}(\bm{k}) = 2\left( \begin{matrix}
    0 & (t_0-t_1)\cos \bm{k}\cdot\bm{R}_0 & 0 \\
    0 & 0 & (t_0+t_1/2)\cos \bm{k}\cdot\bm{R}_{\phi}\\
    (t_0+t_1/2)\cos \bm{k}\cdot\bm{R}_{-\phi} & 0 & 0\\
    \end{matrix} \right) + h.c.
\end{align*}
and 
\begin{align*}
H_{12}(\bm{k}) = \sqrt{3} t_1 \left( \begin{matrix}
    0 & 0 & -\cos \bm{k}\cdot\bm{R}_{-\phi} \\
    0 & 0 &  \cos \bm{k}\cdot\bm{R}_{\phi}\\
    -\cos \bm{k}\cdot\bm{R}_{-\phi} & \cos \bm{k}\cdot\bm{R}_{\phi} & 0\\
    \end{matrix} \right).
\end{align*}
with $H_{21}^\dagger=H_{12}$.

\begin{figure}
\includegraphics[scale=0.45]{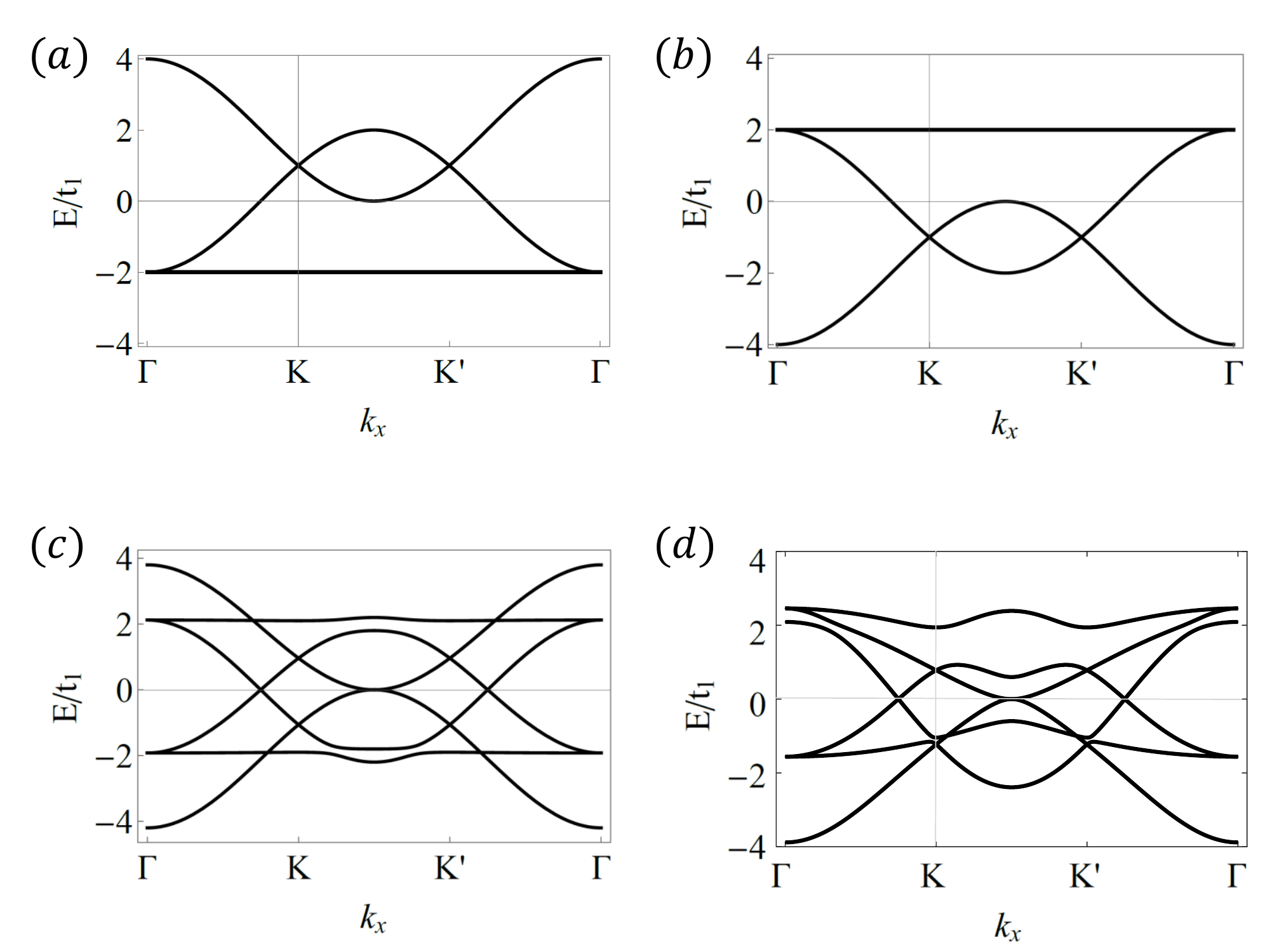} 
\caption{Band structures of $d$ orbitals. (a)-(b) Two decoupled blocks for $e_{g}''$ orbitals with $t_0=0$. (c)  $e_{g}''$ bands with $t_0\neq 0$. (d) Energy bands of $e_g'$ orbitals.}
\label{Fig1}
\end{figure}

Since $|t_1| \gg |t_0|$, we first set $t_0=0$. Such simplification allows one to transform the Hamiltonian into two $3\times 3$ blocks. 
The energy bands for the first and second blocks are plotted in the left and central panels separately in Fig.~\ref{Fig1}.
These band structures are the same as a $s$-orbital kagome lattice that features a flat band and two Dirac cones at $K$ and $K'$ points.

In the presence of a nonzero $t_0$, these two blocks are coupled. The energy bands with $t_0=0.1t_1$ is plotted in the right panel where one can find that the flat bands are no longer flat at $M$ point, i.e., the middle point between $K$ and $K'$. Nevertheless, the dispersions near $K$ and $K'$ are still linear. The presence of $t_0$ does not change the symmetry of the lattice and the Dirac points are robust against the perturbation. A nonzero $t_0$ only modify the energies of the Dirac points and the Dirac velocity.

\end{document}